# Optical Structures for Thermophotovoltaic Emitters: Power Generation from Waste Heat

A dissertation submitted by

**Minsu Oh**

In partial fulfillment of the requirements for the degree of

Doctor of Philosophy

in

Electrical Engineering and Materials Science and Engineering

TUFTS UNIVERSITY

August 2023

Advisor: Prof. Thomas E. Vandervelde

# Abstract

Heat is an inevitable outcome in energy consumption processes, as more than 65% of input energy is wasted as heat. If we can generate electricity from waste heat, it will help minimize the needs for fossil fuels in power plants and can reduce carbon emissions. One way to generate power from heat is through the use of thermophotovoltaics (TPVs), where photons radiated from thermal emitters are converted into electricity. To optimize TPV performance, it is crucial to design emitters such that their emissivity spectrum matches their operating temperature. For example, higher emissivity is needed at shorter (longer) wavelengths at higher (lower) temperatures. Thus, having the ability to create wavelength-selective emitters can enable TPV applications for a wider range of temperatures.

This research focuses on utilizing metamaterials (2D emitters) and planar thin films (1D emitters) to create those emitters. Simulation, fabrication, material property analysis, and radiation measurements were used to characterize the emitters. Based on simulation, metamaterial emitters exhibit engineerable emissivity due to various mechanisms of their optical resonance. Also, large-area fabrication of 1D emitters (78 cm$^2$) was achieved owing to their simple structure, which is required to produce higher TPV power output. Incorporating the characteristics of emitters of each type, their advantages and challenges are discussed. Therefore, the comprehensive results of this research help realize practical implementation of TPV applications.

# Acknowledgements

**Tom:** I can't imagine what my PhD life would have been like if I didn't come to Tufts and have you as my advisor. I was able to learn a lot and spend exciting, fun, and rewarding PhD years with your support and encouragement from the very first day I started my PhD.

**Kevin**: I wouldn't have been able to build my emission measurement apparatus from scratch if it wasn't for you. Your help inspired me with ideas and intuitions. Not to forget, I also had a lot of fun working with you designing the microfabrication lab for the EE-193 as a TA.

I also express my appreciation for my other committee members **Luke** and **Brian**. Your advice led me to a deeper understanding of the subject, which I wouldn't have even thought about if it wasn't for you. I was able to expand my knowledge and make my dissertation stronger, thanks to you.

Another reason why I was able to enjoy doing my research is the **REAP team!**: working together in the cleanroom, helping each other with their expertise and skills, and more importantly, reminding each other as a peer PhD student that you are not alone here.

**Larissa**: you were always willing to help me whether it was giving me insights from a non-technical perspective or providing me with delicious food. I want to get more matcha ice cream with you. **My dear friends**: you guys were (and will always be) my strength since I came to Boston area through summers and winters.

# Contents

# List of Tables

# List of Figures

# 1. INTRODUCTION

## 1.1. Waste Heat and Thermophotovoltaics (TPVs)

According to the Lawrence Livermore National Laboratory in the United States [1], more than 65% of the total input energy in energy consumption processes is wasted as heat, referred to as "waste heat". In 2022, in the United States, approximately $4 \times 10^{15}$ Wh of electricity was generated where the dominant energy sources were fossil fuels such as coal, petroleum, and gas (Table ES1.B on page 13 of Ref. [2]). If thermal energy in waste heat can be converted into electricity, more power can be generated with less fossil fuels, which can reduce carbon emissions as well [3].

One way to generate power from heat is through the use of thermophotovoltaics (TPVs) [4]–[7]. Fig. 1.1 illustrates TPV operation with its key components: heat source, emitter, and photovoltaic cell. While the temperature of the heat source varies from application to application, it generally ranges from room temperature to around 1000°C [8]. TPV emitters absorb thermal energy from the heat source via conduction, convection, and/or radiation [9], [10]. Then, the emitter radiates photons based on its emission spectrum towards the photodiode (or photovoltaic cell). The diode absorbs photons that have energies higher than its bandgap and generates electricity. The radiated power from the emitter at a given temperature is obtained by multiplying its emissivity by the blackbody radiation power at that temperature. Therefore, TPV power output as well as conversion efficiency are dependent on the emitter's emission spectrum.

Compared to a narrowband emitter, a broadband emitter radiates more photons available for power conversion. This leads to a higher power output but can overheat the diode due to the larger number of photons with energies much higher than the bandgap [9]. The overheating may cause a drop in conversion efficiency [11] as well as thermal degradation of the diode [12]. On the other hand, a narrowband emitter can reduce diode overheating at the cost of lower power output [9]. The advantages and disadvantages of each type of emitter are described in Fig. 1.2 –ideal emissivity curves are used for easier description; where the emissivity is one at target wavelengths and zero elsewhere. Therefore, an emission spectrum that is somewhat in-between narrowband and broadband shall be needed to optimize a TPV device [9], [13].

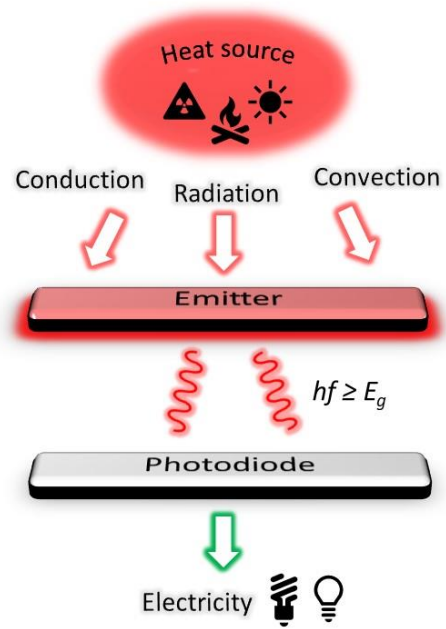

Fig. 1.1. Schematic description of TPV operation. $h$ and $f$ are the Plank's constant and frequency of the radiated light, respectively. $E_g$ is the bandgap energy of the photodiode. Reprinted with permission from Ref. [3].

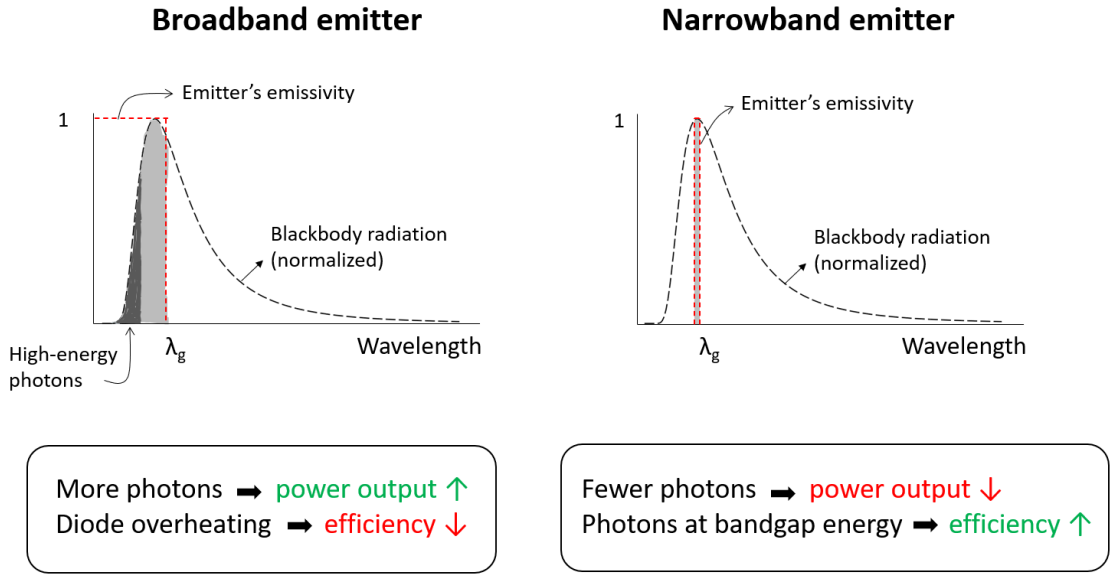

Fig. 1.2. Comparison of a broadband emitter and a narrowband emitter. Gray area is the radiated power from the emitter (darker gray is used to represent high-energy photons). $\lambda_g$ is the wavelength that corresponds to the diode's bandgap. Ideal emissivity curves are used for easier description.

## 1.2. Optical Structures

### 1.2.1. Subwavelength Structures

Upon the interaction between an object and light, every spatial point in the object behaves as a point source of radiation [3], according to Huygens principle [14]. The optical response of an object can be obtained by summing the waves from all those points. This implies that a change in the spatial distribution of those point sources can lead to a different optical response. One example of this is diffraction gratings, where the optical response is dependent on the shape or period of the grating patterns [15]. Another example is subwavelength structures; where the surface pattern dimensions and period are smaller than the wavelength of interest [16]. A subwavelength structure that consists of two different materials is illustrated in Fig. 1.3. In subwavelength structures, compared to diffraction gratings, high-order diffractions become evanescent and only zeroth-order diffraction survives [17]. Optical responses of subwavelength structures [1] can be tuned by changing their surface geometry [18]–[22] and/or constituent materials [23], which is a useful tool for creating TPV emitters with a desired emission spectrum. Subwavelength structures are commonly referred to as metamaterials, metasurfaces [24]–[29], or photonic crystals (PhC) [30]–[32]. In this work, the term "metamaterial (MM)" will be used.

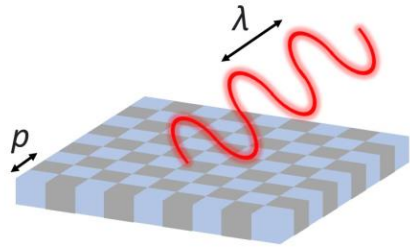

Fig. 1.3. A subwavelength structure consisting of two different materials, represented in blue and gray. $p$ and $\lambda$ are the period and wavelength, respectively, where $p < \lambda$. Reprinted with permission from Ref. [3].

**1.2.2. Thin Film Interference**

When light is incident on an object consisting of thin layers of different materials, the transmitted light reflects at each interface. The reflected waves superpose with one another constructively or destructively, known as thin film interference. The interference then determines the overall optical response of the object, such as reflectance ($R$) or transmittance ($T$), for example. Therefore, optical responses of objects with thin layers can be engineered by changing the layer thicknesses [33]. Fig. 1.4 illustrates the optical interference caused by a thin layer. The structures using optical interference for higher reflectance are sometimes called Bragg reflectors or Fabry-Perot resonators. When layers are optically thick, the transmittance becomes zero. In such case, absorptivity ($A$) [2] of the object is simply given by $A = 1 - R$, where the reflectance (R) can be computed using the transfer matrix method [33], [34]. According to Kirchhoff's law, the emissivity of an object is equal to its absorptivity [35]. Therefore, utilizing Kirchhoff's law, thin film coatings can be used to create TPV emitters with a desired emission spectrum.

---

[1] The optical responses of bulk structures also depend on their geometry. For example, different curvatures of a lens produce different angles of refraction. However, subwavelength structures have more degrees of freedom to manipulate their optical responses than bulk structures. This implies that subwavelength structures can generate responses that bulk structures may not exhibit.

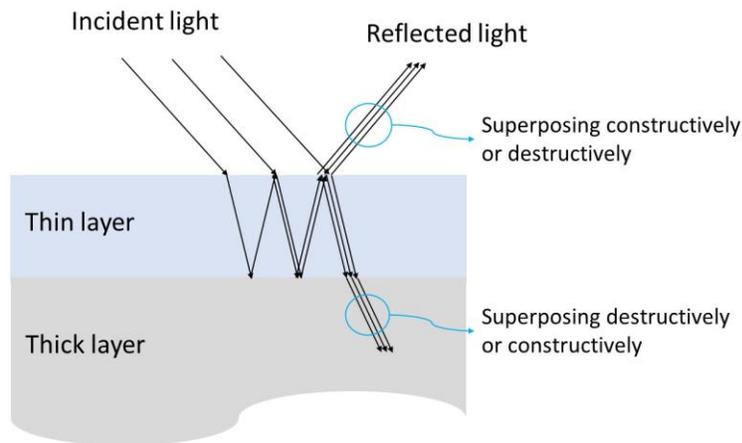

Fig. 1.4. Optical interference caused by a thin layer.

## 1.3. Research Goal

The TPV operation temperature depends on the heat source. This indicates that the emissivity spectrum of an emitter should be designed such that it optimizes the power output and conversion efficiency at the heat source's temperature. Fig. 1.5 shows blackbody radiation spectra at different temperatures with matching emissivity of the emitter to optimize the TPV system. It was the goal of this research to create wavelength-selective, thermally robust emitters for various operation temperatures. Due to the tunability of optical properties, metamaterials (MMs) and thin film interference were used. Materials with a higher melting point such as iridium (Ir), hafnium dioxide (HfO$_2$), chromium (Cr), or silicon (Si) were chosen as the material candidates.

---

[2] The terms "absorptivity" and "absorbance" are used interchangeably in the literature.

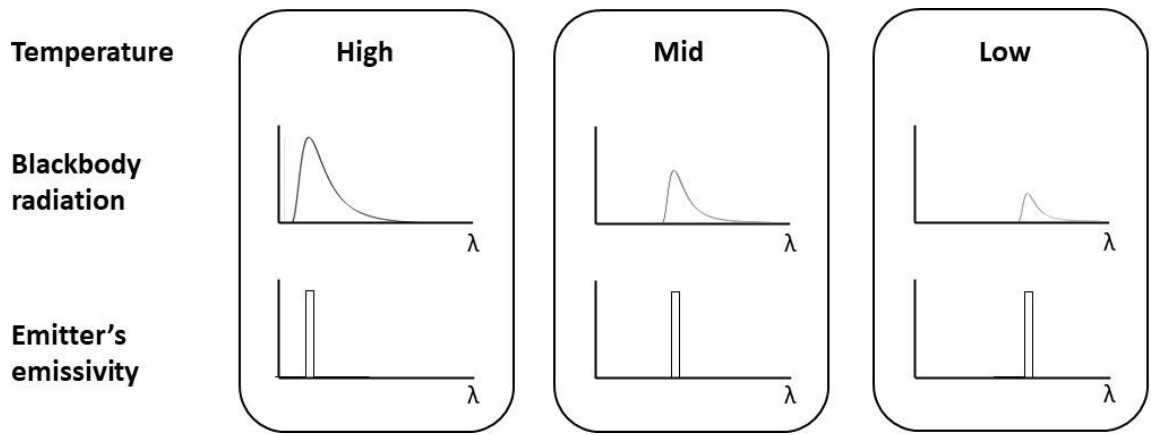

Fig. 1.5. Emissivity peak locations for different temperatures. λ is the wavelength.

# 2. THEORY

## 2.1. Thermal Radiation

This section discusses thermal radiation theories. Important radiometric terminologies are shown in Table 2.1 [36]. These terminologies will be frequently used in this section [3].

Table 2.1. Radiometric terminologies. The subscripts "rb" and "bb" stand for real body and blackbody, respectively. $\lambda$ is the wavelength of the radiated wave.

| Terminology | Symbol | Equation | Unit | Meaning |
|---|---|---|---|---|
| Solid angle | $\Omega$ | | [sr] | |
| Area | $A$ | | [m²] | Area of the radiating surface |
| Projected area | $A_p$ | | [m²] | Projected area of the radiating surface (or area in space on which the radiation is measured) |
| Radiant flux | $\Phi$ | | [W] | Power radiated |
| Radiant intensity | $I$ | $I = \dfrac{d\Phi}{d\Omega}$ | [W sr⁻¹] | Power radiated per unit solid angle |
| Radiant exitance (or radiant emittance) | $M$ | $M = \dfrac{d\Phi}{dA}$ | [W m⁻²] | Power radiated per unit area of the object |
| Spectral exitance (or spectral emittance) | $M_\lambda$ | $M = \int_0^\infty M_\lambda \, d\lambda$ | [W m⁻² m⁻¹] | Power radiated per unit area of the object per unit interval of wavelength |
| Radiance | $L$ | $L = \dfrac{d\Phi}{d\Omega \, dA_p}$ | [W sr⁻¹ m⁻²] | Power radiated per unit solid angle per unit projected area |
| Spectral radiance | $L_\lambda$ | $L = \int_0^\infty L_\lambda \, d\lambda$ | [W sr⁻¹ m⁻² m⁻¹] | Power radiated per unit solid angle per unit projected area per unit interval of wavelength |
| Total emissivity | $\epsilon_{tot}$ | $\epsilon_{tot} = \dfrac{M_{rb}}{M_{bb}}$ | dimensionless | Ratio of the radiant exitance of a real body to that of a blackbody |
| Spectral emissivity | $\epsilon_{spectral}$ or $\epsilon$ | $\epsilon = \dfrac{L_{\lambda,rb}}{L_{\lambda,bb}}$ | dimensionless | Ratio of the spectral radiance of a real body to that of a blackbody |

---

[3] Different terminologies are used in some textbooks. For example, "radiation intensity" [39] or "spectral radiancy" [98] is used instead of "spectral exitance" ($M_\lambda$).

### 2.1.1. Point Source

The radiant intensity $I$ is defined as the radiant flux divided by the solid angle. That is,

Eq. 2.1 $$I = \frac{d\Phi}{d\Omega}$$

, where $\Phi$ and $\Omega$ are the radiant flux and solid angle, respectively. The notation "$d$" is used here to represent infinitesimal amount of each parameter. The radiant intensity $I$ is a constant regardless of solid angle if the radiation is from a point source [37]. The geometry of point source radiation is illustrated in Fig. 2.1. A point source, however, is an idealized object and does not exist in real world [36]. To characterize radiation of objects in three-dimensional world, parameters such as radiant exitance ($M$) or radiance ($L$) are used instead of radiant intensity ($I$) by taking the object's surface area into account.

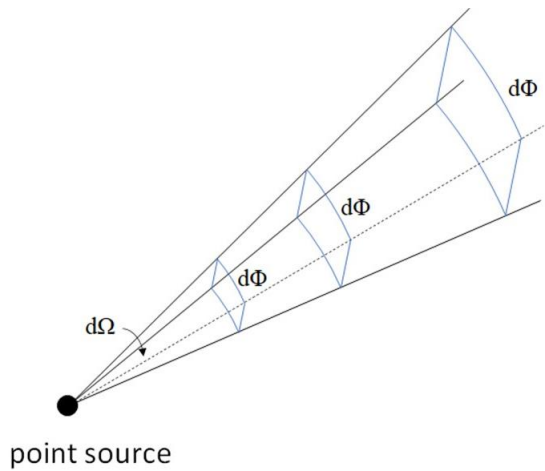

Fig. 2.1. Geometry of radiation from a point source.

### 2.1.2. Blackbody

A blackbody is a theoretical object that absorbs all radiation energy incident on it and re-radiates all of the absorbed energy, at thermal equilibrium [37], [38]. By this definition, the radiation energy from a blackbody is independent of polarization. This indicates that electromagnetic waves radiated from a blackbody are unpolarized. Fig. 2.2 illustrates the radiation emitted from one side of a flat, infinitesimally small blackbody surface with the surface area $dA$; referred to as differential surface element.

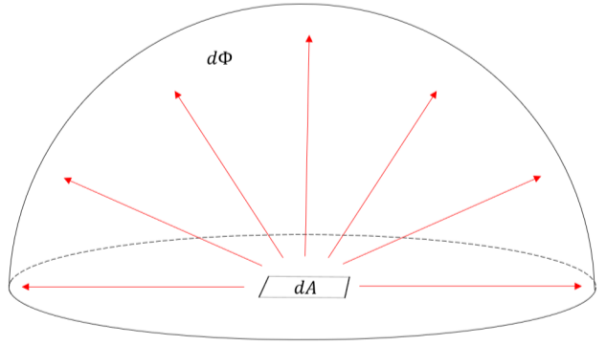

Fig. 2.2. Radiation geometry for a flat, infinitesimally small blackbody surface. The red arrows represent radiation. The radiant flux $d\Phi$ here is the total power radiated from the "top side" of the surface into all space; upper hemisphere. Whether there is radiation from the bottom side of the surface into the lower hemisphere does not change the radiant flux from the top surface.

The radiant exitance from the blackbody ($M_{bb}$) is given by the radiant flux or power ($d\Phi$) radiated into all space divided by the blackbody's surface area ($dA$) [4]. That is,

Eq. 2.2
$$M_{bb} = \frac{d\Phi}{dA}$$

$$= \int_0^\infty M_{\lambda,bb} \, d\lambda$$

$$= \sigma_s T^4$$

, where $M_{\lambda,bb}$ is the spectral exitance, $\lambda$ is the wavelength of the radiated wave, $\sigma_s$ is the Stefan-Boltzmann constant, and $T$ is temperature. Eq. 2.2 is also called Stefan-Boltzmann law. The surface area of an object is given as the integration of the infinitesimally small surface

---

[4] If the radiating surface is spherical, then it radiates energy into both upper and lower hemispheres in the space.

element over all surface; $\int dA \text{ over all surface}$. Thus, from Eq. 2.2, it follows that the total power radiated from a blackbody surface with area *A* is simply

Eq. 2.3 $$P_{bb,tot} = A\sigma_s T^4$$

Moreover, the spectral exitance $M_{\lambda,bb}$ in Eq. 2.2 is the power radiated per unit area of the blackbody in the wavelength interval from $\lambda$ to $\lambda + d\lambda$. According to Plank's law, the spectral exitance of a blackbody is given by

Eq. 2.4 $$M_{\lambda,bb} = \frac{2\pi h c^2}{\lambda^5 \left[ exp\left(\frac{hc}{\lambda k_B T}\right) - 1 \right]}$$

Note that $M_{bb}$ and $M_{\lambda,bb}$ do not provide information about radiated power in relation to radiation angle [37]. For the calculation of radiated power per unit solid angle, radiance ($L$) can be used. Fig. 2.3 shows the geometry for deriving the radiance of an infinitesimally small blackbody surface.

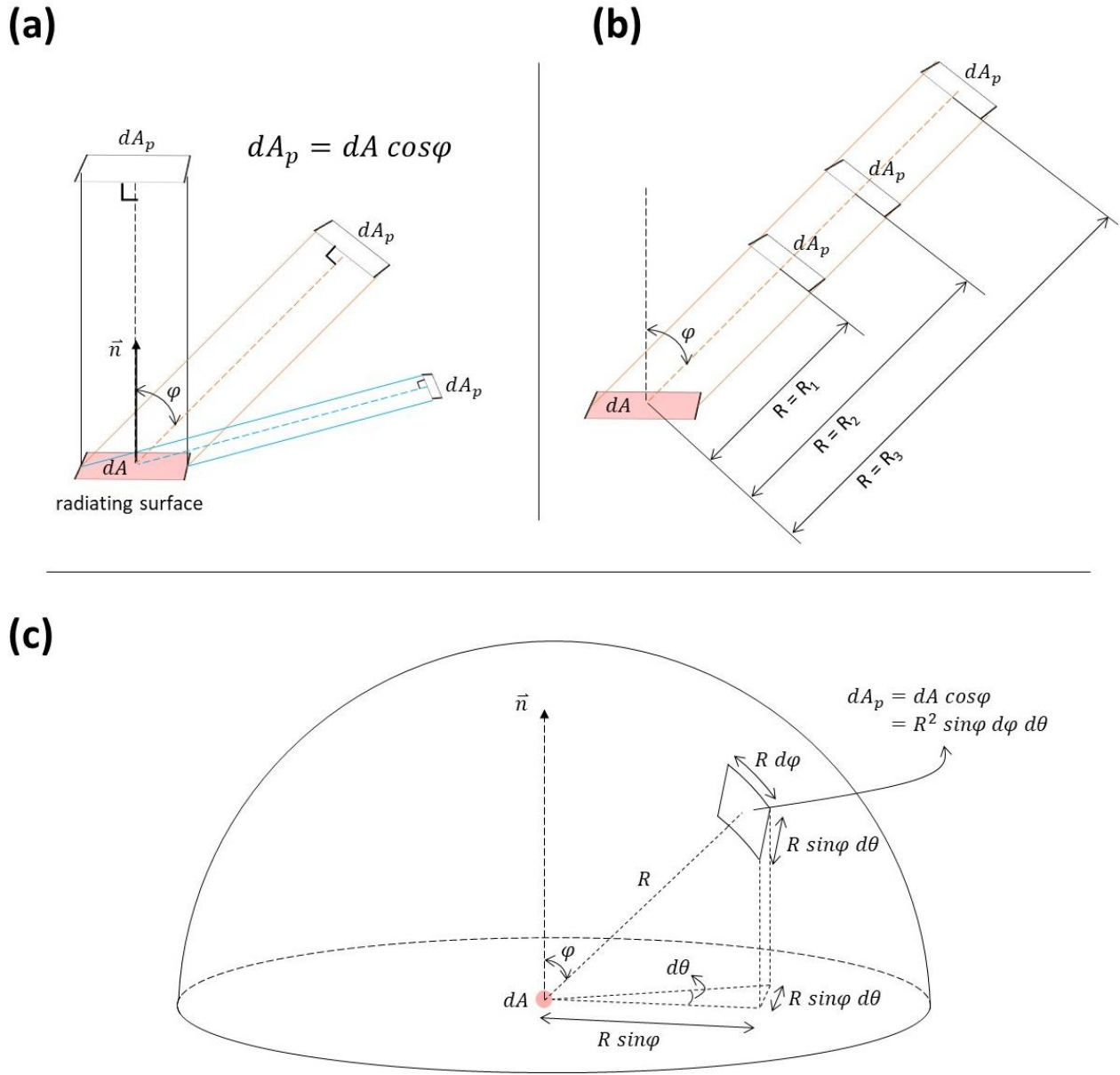

Fig. 2.3. Geometry for the radiance of an infinitesimally small blackbody surface. **(a)**: A radiating surface with surface area $dA$ and its projected area $dA_p$ at various viewing angles. $\vec{n}$ is a vector perpendicular to the radiating surface. **(b):** Projected area at a given viewing angle. The projected area is independent of the distance $R$ at a given viewing angle. **(c)**: Projected area in terms of spherical coordinate parameters. <u>It is assumed that the blackbody surface is far enough from the projected area so that it can be treated like a point source.</u>

The radiance of a blackbody is defined by

$$\text{Eq. 2.5} \qquad L_{bb} = \frac{d\Phi}{d\Omega \, dA_p}$$

The projected area of the blackbody surface in Fig. 2.3 is

$$\text{Eq. 2.6} \qquad dA_p = dA \cos\varphi$$

Inserting Eq. 2.6 into Eq. 2.5,

$$\text{Eq. 2.7} \qquad L_{bb} = \frac{d\Phi}{d\Omega \, dA_p} = \frac{d\Phi}{dA} \frac{1}{d\Omega \cos\varphi}$$

It follows that

$$\text{Eq. 2.8} \qquad \frac{d\Phi}{dA} = L_{bb} \, d\Omega \cos\varphi$$

The radiant exitance of the blackbody surface ($M_{bb}$) in Fig. 2.3 can be obtained by integrating Eq. 2.8 over the (upper) hemisphere in space. That is,

$$\text{Eq. 2.9} \qquad M_{bb} = \frac{1}{dA} \int d\Phi = \int L_{bb} \, d\Omega \cos\varphi \qquad \text{over the hemisphere}$$

Assume that the projected area is far enough from the blackbody surface and the blackbody surface can be treated like a point source. Based on this assumption, the viewing angle $\varphi$ shall be the same at any location on the projected area. This allows for the expression of the projected area in terms of spherical coordinate parameters, illustrated in Fig. 2.3(c), as:

$$\text{Eq. 2.10} \qquad dA_p = R^2 \sin\varphi \, d\varphi \, d\theta$$

Thus, the solid angle for the projected area can be written as

Eq. 2.11 $$d\Omega \equiv \frac{dA_p}{R^2} = sin\varphi \, d\varphi \, d\theta$$

Insert Eq. 2.11 into Eq. 2.9 and integrate it over the hemisphere. For a flat blackbody surface [5], we obtain

Eq. 2.12 $$M_{bb} = \int_0^{2\pi} \int_0^{\pi/2} (L_{bb} \, sin\varphi \, cos\varphi) \, d\varphi \, d\theta = \pi L_{bb}$$

Eq. 2.13 $$L_{bb} = \frac{\sigma_s T^4}{\pi}$$

Moreover, the blackbody radiance $L_{bb}$ can be obtained by integrating the spectral radiance $L_{\lambda,bb}$ over all wavelength intervals. That is,

Eq. 2.14 $$L_{bb} = \int_0^\infty L_{\lambda,bb} \, d\lambda$$

According to Plank's law, the spectral radiance for a flat blackbody surface is given by

Eq. 2.15 $$L_{\lambda,bb} = \frac{2hc^2}{\lambda^5 \left[exp\left(\frac{hc}{\lambda k_B T}\right) - 1\right]}$$

Fig. 2.4 shows the normalized blackbody radiation, either $M_{\lambda,bb}$ or $L_{\lambda,bb}$, at several temperatures as a function of wavelength. The wavelength at which radiation energy peaks ($\lambda_{peak}$) varies with temperature:

Eq. 2.16 $$\lambda_{peak} \, [\mu m] \approx \frac{2898}{T \, [K]}$$

---

[5] It seems that most textbooks spend majority of their efforts in discussing radiation from a "flat" surface.

, where the wavelength is in *μm* and the temperature is in *K*. For example, $\lambda_{peak}$ = 1 *μm* at 2898*K*. Eq. 2.16 is also known as Wien's displacement law.

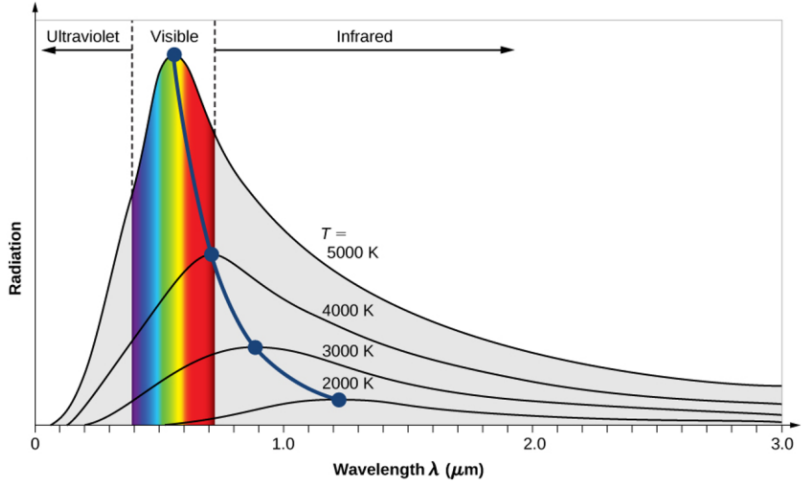

Fig. 2.4. Normalized blackbody radiation spectra at various temperatures [39] (access for free at openstax.org). The wavelength at which radiation energy peaks is indicated with a solid circle at each temperature.

**2.1.3. Real Body**

In this section, radiometric properties of real bodies (or real objects) are compared to those of a blackbody. The parameters relating to real bodies will be denoted with a subscript "rb", which stands for "real body". The total emissivity of a real body ($\epsilon_{tot}$) is defined as the ratio of the total power radiated per unit area of the real body ($M_{rb}$) to that of a blackbody ($M_{bb}$) at a given temperature: $\epsilon_{tot} = \frac{M_{rb}}{M_{bb}}$, where $M_{rb} \leq M_{bb}$ [6]. Similarly, spectral emissivity of a real body ($\epsilon_{spectral}$) is the ratio of the spectral radiance of the real body ($L_{\lambda,rb}$) to that

---

[6] There are exceptions where the radiation energy from a real body can exceed the blackbody limit (i.e. $M_{rb} > M_{bb}$). This can happen at a distance as small as a few nano or micrometers (or subwavelength distances) from the radiating surface. The phenomenon is due to the energy transfer by evanescent waves [99]. One application that utilizes this phenomenon is near-field thermophotovoltaics [100]. In this work, however, near-field thermophotovoltaics is not discussed.

of a blackbody ($L_{\lambda,bb}$) at a given temperature [7]: $\epsilon_{spectral} = \frac{L_{\lambda,rb}}{L_{\lambda,bb}}$. As shown in Eq. 2.15, $L_{\lambda,bb}$ is a constant for a given wavelength at a given temperature. However, $L_{\lambda,rb}$ may vary depending on radiation angle and/or polarization of the wave, although the wavelength and temperature are fixed. Hence, spectral emissivity of a real body is generally a function of wavelength, temperature, as well as radiation angle and polarization [36]. The radiometric parameters for real body are correlated with blackbody parameters as shown below:

Eq. 2.17 $\qquad M_{rb}(T) = \epsilon_{tot}(T)\, M_{bb}(T) = \epsilon_{tot}(T)\, \sigma_s T^4$

Eq. 2.18 $\qquad L_{\lambda,rb}\,(\lambda, T, \varphi, pol) = \epsilon_{spectral}(\lambda, T, \varphi, pol)\, L_{\lambda,bb}(\lambda, T)$

Both total emissivity and spectral emissivity are dimensionless, and their values range from zero to one. The spectral emissivity is oftentimes referred to as just "emissivity" in the literature [8]. The spectral emissivity in this work will also be referred to as "emissivity" with the symbol "$\epsilon$" without any subscript unless otherwise specified. Fig. 2.5 compares radiation spectra of a blackbody and a real body at a given temperature.

---

[7] For spectral emissivity, it is assumed that the radiation is measured far enough from the object so that the radiation angle is same at any point on the detecting area. This assumption was also used in Fig. 2.3(c).

[8] In literature on radiometry, many different terminologies are used that refer to the same type of emissivity. It seems that there is no universal agreement on which term should be used.

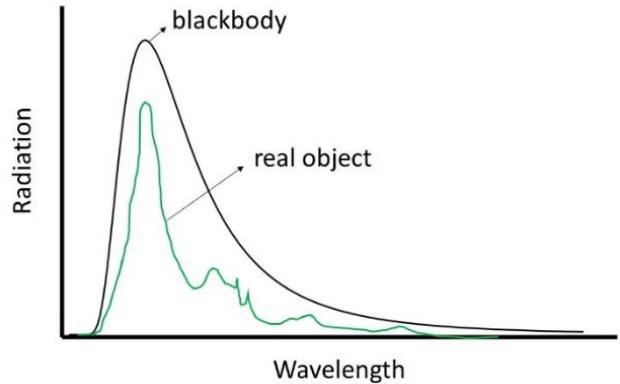

Fig. 2.5. Radiation spectra of a blackbody and a real body at a given temperature.

**2.1.4. Kirchhoff's Law**

According to Kirchhoff's law, the emissivity ($\epsilon$) of an object is equal to its absorptivity (*A*) [9] at a given temperature, wavelength, propagation angle, and polarization at thermal equilibrium [35], [36]. Compared to emissivity, absorptivity can be measured relatively easily by commercially available instruments such as ellipsometers or spectrophotometers. Thus, emissivity of TPV emitters is sometimes reported based on their measured absorptivity in the literature. However, one should be careful when using absorptivity as emissivity because Kirchhoff's law only applies when the emitted and absorbed waves are identical. For example, if the emitted and absorbed waves have different optical coherence, then the emissivity may not be equal to absorptivity. Structures with planar, thin layers are examples where their optical responses can be significantly impacted by the (temporal) coherence of light. When measuring optical properties of such structures by an ellipsometer, highly coherent light is used. However, thermally radiated waves are generally incoherent or partly coherent [40],

---

[9] The terms "absorptivity" and "absorbance" are used interchangeably in the literature. Both terms are dimensionless and range from zero to one.

[41]. Therefore, for emitters whose performance is based on optical interference between thin layers, their emissivity and measured absorptivity may differ.

## 2.2. Electromagnetic Waves

Maxwell's equations are a powerful mathematical tool that can explain and predict interactions between materials and light. This section discusses Maxwell's equations as well as light-matter interactions. Some important electromagnetic parameters are shown in Table 2.2 [42], which will frequently be used in this section.

Table 2.2. Electromagnetic parameters. Vectors are denoted with an arrow hat. $\omega$ is the angular frequency of the wave.

| Terminology | Symbol | Equation |
|---|---|---|
| Electric field | $\vec{E}$ | |
| Electric displacement field | $\vec{D}$ | $\vec{D} = \varepsilon_o \vec{E} + \vec{P} = \varepsilon_r \vec{E}$ |
| Magnetic B field | $\vec{B}$ | |
| Magnetic H field | $\vec{H}$ | $\vec{H} = \dfrac{\vec{B}}{\mu_o} - \vec{M}$ |
| Electric susceptibility | $\chi_e$ | |
| Magnetic susceptibility | $\chi_m$ | |
| Vacuum permittivity | $\varepsilon_o$ | |
| Vacuum permeability | $\mu_o$ | |
| Relative dielectric permittivity | $\varepsilon_r$ | $\varepsilon_r = 1 + \chi_e$ |
| Relative permeability | $\mu_r$ | $\mu_r = 1 + \chi_m$ |
| Dielectric permittivity | $\varepsilon$ | $\varepsilon = \varepsilon_o \varepsilon_r$ |
| Effective permittivity | $\varepsilon_{eff}$ | $\varepsilon_{eff} = \varepsilon + i\dfrac{\sigma}{\omega}$ |
| Relative effective permittivity | $\varepsilon_{r,eff}$ | $\varepsilon_{r,eff} = \dfrac{\varepsilon_{eff}}{\varepsilon_o}$ |
| Permeability | $\mu$ | $\mu = \mu_o \mu_r$ |
| Polarization | $\vec{P}$ | $\vec{P} = \varepsilon_o \chi_e \vec{E}$ for linear materials [10] |
| Bound charge density (volumetric) | $\rho_b$ | $\rho_b = -\nabla \cdot \vec{P}$ |
| Free charge density (volumetric) | $\rho_f$ | |
| Total charge density (volumetric) | $\rho_{tot}$ | $\rho_{tot} = \rho_b + \rho_f$ |
| Electrical conductivity | $\sigma$ | |
| Magnetization | $\vec{M}$ | $\vec{M} = \chi_m \vec{H}$ |
| Polarization current density | $\vec{J}_p$ | $\vec{J}_p = \dfrac{\partial \vec{P}}{\partial t}$ |
| Bound current density | $\vec{J}_b$ | $\vec{J}_b = \nabla \times \vec{M}$ |
| Free current density | $\vec{J}_f$ | $\vec{J}_f = \sigma \vec{E}$ |
| Total current density | $\vec{J}_{tot}$ | $\vec{J}_{tot} = \vec{J}_p + \vec{J}_b + \vec{J}_f$ |

---

[10] For nonlinear materials, the electric susceptibility can be written as $\chi_e = \chi_1 + \chi_2 E + \chi_3 E^2 + \cdots$, and the polarization as $P = \varepsilon_o \chi_1 E + \varepsilon_o \chi_2 E^2 + \varepsilon_o \chi_3 E^3 + \cdots$ (Chapter 24 of Ref. [54]). In the expression of polarization, the first term $\varepsilon_o \chi_1 E$ describes the linear response of the material, and other terms nonlinear responses. The second ($\varepsilon_o \chi_2 E^2$) and third ($\varepsilon_o \chi_3 E^3$) terms are related to Pockels effect and Kerr effect, respectively.

### 2.2.1. Maxwell's Equations in General Form

The (differential form) Maxwell's equations in matter are given below in terms of vacuum permittivity ($\varepsilon_o$) and vacuum permeability ($\mu_o$) [43].

Eq. 2.19    $(a)\ \nabla \cdot \vec{E} = \frac{\rho_{tot}}{\varepsilon_o}$    $(c)\ \nabla \times \vec{E} = -\frac{\partial \vec{B}}{\partial t}$

$(b)\ \nabla \cdot \vec{B} = 0$    $(d)\ \nabla \times \vec{B} = \varepsilon_o \mu_o \frac{\partial \vec{E}}{\partial t} + \mu_o \vec{J}_{tot}$

, where $\rho_{tot} = \rho_b + \rho_f$, $\vec{J}_{tot} = \vec{J}_p + \vec{J}_b + \vec{J}_f$, and $t$ is time. <u>The vector parameters are generally a function of both position and time</u>. Each of the four Maxwell's equations has their own unique title [43]. Eq. 2.19(a) is the Gauss's law for electric fields. Eq. 2.19(b) is the Gauss's law for magnetic fields. Eq. 2.19(c) is the Faraday's law. Eq. 2.19(d) is the Ampere-Maxwell law. The Maxwell's equations in vacuum can also be obtained by setting $\rho_{tot} = \vec{J}_{tot} = 0$ in Eq. 2.19. This implies that the **motion of charges** determines the material's interaction with electromagnetic fields. According to Eq. 2.19(c), a variation of magnetic field in time induces an electric field nearby. Similarly, according to Eq. 2.19(d), a variation of electric field in time and/or electric current induces a magnetic field nearby. If the induced magnetic (electric) field also varies with time, then it will induce another electric (magnetic) field nearby, and so on. This indicates that electric and magnetic fields propagate through space by continuously inducing each other, as illustrated in Fig. 2.6. In Section 2.2.3, we will find a mathematical expression of electric and magnetic fields to describe electromagnetic waves.

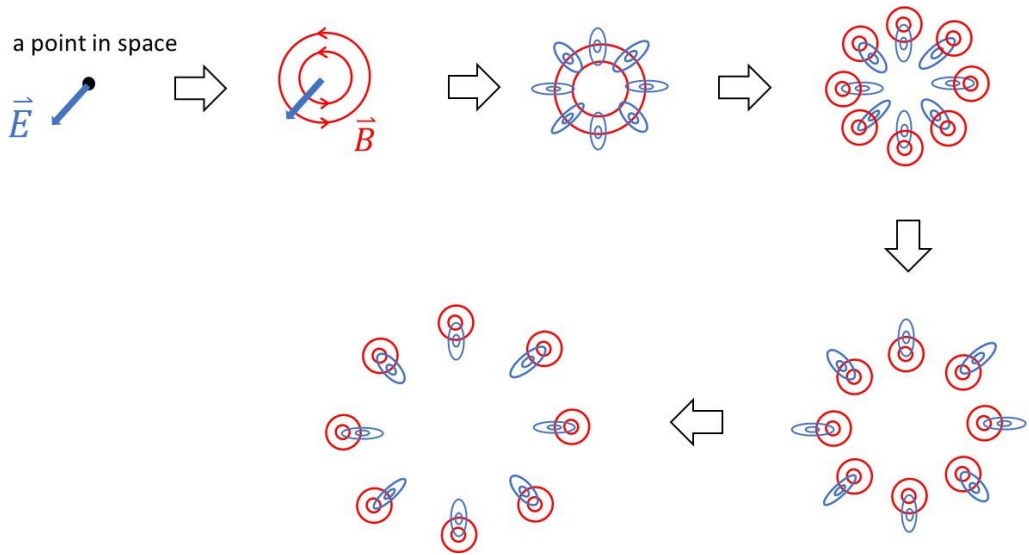

Fig. 2.6. Electromagnetic wave propagation from a point source. The electric (blue) and magnetic (red) fields are drawn in different colors. The illustrated is the wave propagation on the plane of the paper, because it is hard to visualize three-dimensional (3D) propagation on a 2D paper. A point source of radiation, however, produces a spherical wave. Reprinted with permission from Ref. [44].

Eq. 2.19(a) and Eq. 2.19(d) can also be expressed in terms of electric displacement field ($\vec{D}$) and magnetic H field ($\vec{H}$) instead of $\vec{E}$ and $\vec{B}$, respectively. To replace $\vec{E}$ with $\vec{D}$, let us write the total electric charge density ($\rho_{tot}$) as

Eq. 2.20 $\qquad \rho_{tot} = \rho_b + \rho_f = \left(-\nabla \cdot \vec{P}\right) + \rho_f$

By inserting Eq. 2.20 into Eq. 2.19(a),

Eq. 2.21 $\qquad \nabla \cdot \vec{E} = \frac{\left(-\nabla \cdot \vec{P}\right) + \rho_f}{\varepsilon_o}$

By multiplying $\varepsilon_o$ to both sides of Eq. 2.21,

Eq. 2.22 $\qquad \nabla \cdot \varepsilon_o \vec{E} + \nabla \cdot \vec{P} = \rho_f$

$\qquad \rightarrow \quad \nabla \cdot \left(\varepsilon_o \vec{E} + \vec{P}\right) = \rho_f$

Since $\vec{D} \equiv \varepsilon_o \vec{E} + \vec{P}$, Eq. 2.22 can be written as

Eq. 2.23 $$\nabla \cdot \vec{D} = \rho_f$$

To replace $\vec{B}$ with $\vec{H}$, write the total electric current density ($\vec{J}_{tot}$) as

Eq. 2.24 $$\vec{J}_{tot} = \vec{J}_p + \vec{J}_b + \vec{J}_f$$

$$= \left(\frac{\partial \vec{P}}{\partial t}\right) + (\nabla \times \vec{M}) + (\sigma \vec{E})$$

By inserting Eq. 2.24 into Eq. 2.19(d),

Eq. 2.25 $$\nabla \times \vec{B} - \mu_o(\nabla \times \vec{M}) = \varepsilon_o \mu_o \frac{\partial \vec{E}}{\partial t} + \mu_o \frac{\partial \vec{P}}{\partial t} + \mu_o \sigma \vec{E}$$

Dividing both sides of Eq. 2.25 by $\mu_o$,

Eq. 2.26 $$\nabla \times \frac{\vec{B}}{\mu_o} - (\nabla \times \vec{M}) = \varepsilon_o \frac{\partial \vec{E}}{\partial t} + \frac{\partial \vec{P}}{\partial t} + \sigma \vec{E}$$

$$\rightarrow \nabla \times \left(\frac{\vec{B}}{\mu_o} - \vec{M}\right) = \frac{\partial \vec{D}}{\partial t} + \sigma \vec{E}$$

Since $\vec{H} \equiv \frac{\vec{B}}{\mu_o} - \vec{M}$, Eq. 2.26 can be written as

Eq. 2.27 $$\nabla \times \vec{H} = \frac{\partial \vec{D}}{\partial t} + \sigma \vec{E}$$

Therefore, the Maxwell's equations in Eq. 2.19 can also be written by

Eq. 2.28 $\quad$ (a) $\nabla \cdot \vec{D} = \rho_f \quad$ (c) $\nabla \times \vec{E} = -\frac{\partial \vec{B}}{\partial t}$

$\quad\quad\quad\quad\quad\quad\quad$ (b) $\nabla \cdot \vec{B} = 0 \quad$ (d) $\nabla \times \vec{H} = \frac{\partial \vec{D}}{\partial t} + \sigma \vec{E}$

Note that Eq. 2.19 and Eq. 2.28 apply to both isotropic and anisotropic media [45].

**2.2.2. Maxwell's Equations in Isotropic Media**

Isotropic media are those in which the electromagnetic properties, such as permittivity and permeability, are independent of the incident field's orientation. If a material is isotropic, then its electromagnetic properties can be treated as a scalar constant. In "anisotropic" media, however, such properties should be treated as a tensor [46]. Thus, the mathematical analysis of electromagnetic phenomena in isotropic materials is relatively much easier than that for anisotropic media [45]. The Maxwell's equations for isotropic materials [11] are shown in Eq. 2.29. Isotropic materials are assumed throughout this work unless otherwise specified.

Eq. 2.29

$$(a) \quad \nabla \cdot \vec{E} = \frac{\rho_{tot}}{\varepsilon_o} = \frac{\rho_f}{\varepsilon} \qquad (c) \quad \nabla \times \vec{E} = -\frac{\partial \vec{B}}{\partial t}$$

$$(b) \quad \nabla \cdot \vec{B} = 0 \qquad (d) \quad \nabla \times \vec{B} = \varepsilon\mu \frac{\partial \vec{E}}{\partial t} + \mu\sigma\vec{E}$$

**2.2.3. Sinusoidal Solution to Maxwell's Equations**

"Vector" Laplacian ($\nabla_v^2$) provides a way to solving Maxwell's equations [12]. The vector Laplacian is defined such that $\nabla_v^2 \vec{V} \equiv \nabla(\nabla \cdot \vec{V}) - \nabla \times (\nabla \times \vec{V})$, where the vector quantity $\vec{V} = V\hat{u} = (V_x, V_y, V_z)$. In the Cartesian coordinate, the vector Laplacian can be expressed by [47]:

---

[11] Appendix A discusses charge densities in isotropic media regarding Eq. 2.29(a).

[12] In the literature, there is a very common confusion associated with vector Laplacian, which operates on a vector, and scalar Laplacian, which operates on a scalar. This is because these two operators use the same symbol [47], $\nabla^2$. In this work, $\nabla_v^2$ is used for vector Laplacian and $\nabla_s^2$ for scalar Laplacian to avoid confusion.

Eq. 2.30 $\qquad \nabla_v^2 \vec{V} \equiv \nabla(\nabla \cdot \vec{V}) - \nabla \times (\nabla \times \vec{V})$

$$= \left( \left[ \frac{\partial^2 V_x}{\partial x^2} + \frac{\partial^2 V_x}{\partial y^2} + \frac{\partial^2 V_x}{\partial z^2} \right], \left[ \frac{\partial^2 V_y}{\partial x^2} + \frac{\partial^2 V_y}{\partial y^2} + \frac{\partial^2 V_y}{\partial z^2} \right], \left[ \frac{\partial^2 V_z}{\partial x^2} + \frac{\partial^2 V_z}{\partial y^2} + \frac{\partial^2 V_z}{\partial z^2} \right] \right)$$

$$= (\nabla_s^2 V_x, \ \nabla_s^2 V_y, \ \nabla_s^2 V_z)$$

, where the "scalar" Laplacian ($\nabla_s^2$) is defined such that $\nabla_s^2 = \nabla \cdot \nabla = \left( \frac{\partial}{\partial x}, \frac{\partial}{\partial y}, \frac{\partial}{\partial z} \right) \cdot \left( \frac{\partial}{\partial x}, \frac{\partial}{\partial y}, \frac{\partial}{\partial z} \right) = \left( \frac{\partial^2}{\partial x^2} + \frac{\partial^2}{\partial y^2} + \frac{\partial^2}{\partial z^2} \right)$ [48]. All components of $\vec{V}$ are generally a function of all three perpendicular coordinates such that $\vec{V} = \left( V_x(x,y,z), \ V_y(x,y,z), \ V_z(x,y,z) \right)$.

Applying the vector Laplacian to the magnetic field,

Eq. 2.31 $\qquad \nabla_v^2 \vec{B} = \nabla(\nabla \cdot \vec{B}) - \nabla \times (\nabla \times \vec{B})$

According to Eq. 2.29(b) – (d),

Eq. 2.32 $\qquad \nabla_v^2 \vec{B} = -\nabla \times \left( \varepsilon\mu \frac{\partial \vec{E}}{\partial t} + \mu\sigma \vec{E} \right)$

$$= -\varepsilon\mu \frac{\partial}{\partial t}(\nabla \times \vec{E}) - \mu\sigma(\nabla \times \vec{E})$$

$$= -\varepsilon\mu \frac{\partial}{\partial t}\left( -\frac{\partial \vec{B}}{\partial t} \right) - \mu\sigma\left( -\frac{\partial \vec{B}}{\partial t} \right)$$

$$= \varepsilon\mu \frac{\partial^2 \vec{B}}{\partial t^2} + \mu\sigma \frac{\partial \vec{B}}{\partial t}$$

Define the propagation vector ($\vec{k}$) and position vector ($\vec{r}$) in the Cartesian coordinate such that

Eq. 2.33 $\qquad$ (a) $\vec{k} = (k_x, \ k_y, \ k_z) = k\hat{u}_k$ $\qquad$ (b) $\vec{r} = (x, \ y, \ z)$

, where $\hat{u}_k$ is the unit vector. Now, suppose a uniform [13], sinusoidal plane wave of a magnetic field that extends to the infinite of the space [14]. Using the propagation vector, the magnetic field can be written by [15]

Eq. 2.34
$$\vec{B} = \vec{B}_o e^{i(\vec{k}\cdot\vec{r} - \omega t)}$$
$$= (B_x, \ B_y, \ B_z)$$
$$= (B_{ox}, \ B_{oy}, \ B_{oz}) e^{i(k_x x + k_y y + k_z z - \omega t)}$$

, where $\vec{B}_o$ is the amplitude vector, $\omega$ is the angular frequency of the wave, and $t$ is time. Using the magnetic field expression in Eq. 2.34, the lefthand side of Eq. 2.32 can be written by

Eq. 2.35
$$\nabla_v^2 \vec{B} = (\nabla_s^2 B_x, \ \nabla_s^2 B_y, \ \nabla_s^2 B_z)$$
$$= \left(B_{ox} \nabla_s^2 \left[e^{i(k_x x + k_y y + k_z z - \omega t)}\right], B_{oy} \nabla_s^2 \left[e^{i(k_x x + k_y y + k_z z - \omega t)}\right], B_{oz} \nabla_s^2 \left[e^{i(k_x x + k_y y + k_z z - \omega t)}\right]\right)$$
$$= (B_{ox}, B_{oy}, B_{oz}) \nabla_s^2 \left[e^{i(k_x x + k_y y + k_z z - \omega t)}\right]$$
$$= \vec{B}_o \ i^2 (k_x^2 + k_y^2 + k_z^2) \ e^{i(k_x x + k_y y + k_z z - \omega t)}$$
$$= -\vec{B}_o \ k^2 \ e^{i(\vec{k}\cdot\vec{r} - \omega t)}$$
$$= -k^2 \vec{B}$$

---

[13] Uniform waves are those where the phase propagation and energy loss occur in the same direction. Nonuniform waves are those where they occur in different directions. For a nonuniform wave, therefore, the propagation vectors associated with phase propagation ($\vec{k}_R$) and energy loss ($\vec{k}_I$) directions are not parallel to each other [101], [102], where the relationship $\vec{k}_R + i\vec{k}_I = \vec{k} = k\hat{u}_k$ can still be used. In such case, the unit vector $\hat{u}_k$ is a complex-valued vector so that it maintains the equality [103].

[14] If a wave is confined in a finite space, such as waveguides, then the boundary conditions of electric and magnetic fields need to be incorporated when solving the Maxwell's equations.

[15] The divergence of magnetic fields is always zero. Thus, $\nabla \cdot \left\{\vec{B}_o e^{i(\vec{k}\cdot\vec{r} - \omega t)}\right\} = \nabla \cdot (\vec{B}_R + i\vec{B}_I) = 0$, where $\nabla \cdot \vec{B}_R = \nabla \cdot \vec{B}_I = 0$ [104].

By equating the righthand sides of Eq. 2.32 and Eq. 2.35 and using the magnetic field form in Eq. 2.34, one can obtain

Eq. 2.36 $$k^2 = \varepsilon\mu\omega^2 + i\mu\sigma\omega$$

Now, apply the vector Laplacian to the electric field,

Eq. 2.37 $$\nabla_v^2 \vec{E} = \nabla(\nabla \cdot \vec{E}) - \nabla \times (\nabla \times \vec{E})$$

Inserting Eq. 2.29(a), Eq. 2.29(c), and Eq. 2.29(d) into Eq. 2.37,

Eq. 2.38 $$\nabla_v^2 \vec{E} = \nabla\left(\frac{\rho_{tot}}{\varepsilon_o}\right) + \nabla \times \frac{\partial \vec{B}}{\partial t}$$

$$= \nabla\left(\frac{\rho_{tot}}{\varepsilon_o}\right) + \frac{\partial}{\partial t}(\nabla \times \vec{B})$$

$$= \nabla\left(\frac{\rho_{tot}}{\varepsilon_o}\right) + \frac{\partial}{\partial t}\left(\varepsilon\mu\frac{\partial \vec{E}}{\partial t} + \mu\sigma\vec{E}\right)$$

$$= \nabla\left(\frac{\rho_{tot}}{\varepsilon_o}\right) + \varepsilon\mu\frac{\partial^2 \vec{E}}{\partial t^2} + \mu\sigma\frac{\partial \vec{E}}{\partial t}$$

If the total charge density is zero ($\rho_{tot} = 0$) [16], Eq. 2.38 reduces to

Eq. 2.39 $$\nabla_v^2 \vec{E} = \varepsilon\mu\frac{\partial^2 \vec{E}}{\partial t^2} + \mu\sigma\frac{\partial \vec{E}}{\partial t}$$

Similar to the magnetic field above, suppose a sinusoidal plane wave of an electric field that extends to the infinite of the space. That is,

---

[16] This assumption plays a crucial role in solving the Maxwell's equations for electric fields. If the total charge density is not zero, then the derivations based on this assumption may not be true anymore, unless the gradient of the total charge density is still zero ($\nabla\rho_{tot} = 0$). Furthermore, using the assumption, the divergence of electric fields is $\nabla \cdot \vec{E} = \nabla \cdot \{\vec{E}_o e^{i(\vec{k}\cdot\vec{r}-\omega t)}\} = \nabla \cdot (\vec{E}_R + i\vec{E}_I) = 0$. Thus, for both real and imaginary parts of the field: $\nabla \cdot \vec{E}_R = \nabla \cdot \vec{E}_I = 0$.

Eq. 2.40
$$\vec{E} = \vec{E}_o e^{i(\vec{k}\cdot\vec{r}-\omega t)}$$
$$= (E_x,\ E_y,\ E_z)$$
$$= (E_{ox},\ E_{oy},\ E_{oz}) e^{i(k_x x + k_y y + k_z z - \omega t)}$$

The electric field in the above expression can be used to solve Eq. 2.39, which produces the same result as Eq. 2.36 for the electric field. In this section, we have identified a (sinusoidal) solution to Maxwell's equations for both electric and magnetic fields. This suggests that a sinusoidal electric field induces a sinusoidal magnetic field and vice versa. Additional mathematical details about how electric and magnetic fields are correlated are provided in Eq. A5 in Appendix B.

### 2.2.4. Effective Permittivity

Eq. 2.41 defines a parameter that combines dielectric permittivity ($\varepsilon$), which arises due to bound electrons, and electrical conductivity ($\sigma$), which arises due to free electrons. The parameter is referred to as "effective" permittivity ($\varepsilon_{eff}$). Fig. 2.7 illustrates bound and free electrons in a material and how they each relate to the electric properties.

Eq. 2.41
$$\varepsilon_{eff} \equiv \varepsilon + i\frac{\sigma}{\omega}$$
$$= \varepsilon_{eff,r}\, \varepsilon_o$$

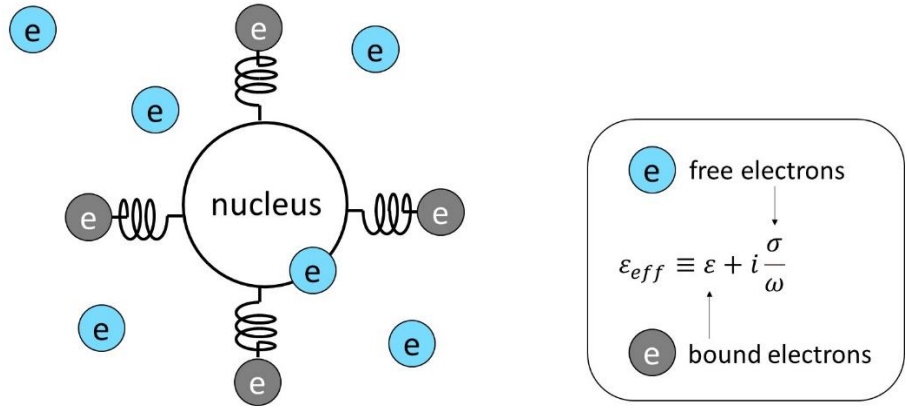

Fig. 2.7. Bound and free electrons in a material.

The dielectric permittivity ($\varepsilon$) and electrical conductivity ($\sigma$) are generally complex-valued numbers; $\varepsilon = \varepsilon' + i\varepsilon''$ and $\sigma = \sigma' + i\sigma''$, which will be discussed in Section 2.3. Therefore, the terms $\varepsilon$ and $\frac{\sigma}{\omega}$ in Eq. 2.41 are not the real and imaginary parts of the effective permittivity; $\varepsilon \neq Re(\varepsilon_{eff})$ and $\frac{\sigma}{\omega} \neq Im(\varepsilon_{eff})$. The real and imaginary parts of the effective permittivity shall be given by

Eq. 2.42
$$\varepsilon_{eff} = \varepsilon_{eff}' + i\varepsilon_{eff}''$$
$$= \left(\varepsilon' - \frac{\sigma''}{\omega}\right) + i\left(\varepsilon'' + \frac{\sigma'}{\omega}\right)$$

, where $Re[\varepsilon_{eff}] = \varepsilon' - \frac{\sigma''}{\omega}$ and $Im[\varepsilon_{eff}] = \varepsilon'' + \frac{\sigma'}{\omega}$. Given that the electric and magnetic fields are sinusoidal, Eq. 2.32 can be written in a simpler manner using the effective permittivity. That is,

Eq. 2.43
$$\nabla_v^2 \vec{B} = \varepsilon\mu \frac{\partial^2 \vec{B}}{\partial t^2} + \mu\sigma \frac{\partial \vec{B}}{\partial t}$$
$$= \varepsilon_{eff}\mu \frac{\partial^2 \vec{B}}{\partial t^2}$$

The same is true for Eq. 2.39, which produces

$$\text{Eq. 2.44} \qquad \nabla_v^2 \vec{E} = \varepsilon\mu \frac{\partial^2 \vec{E}}{\partial t^2} + \mu\sigma \frac{\partial \vec{E}}{\partial t}$$

$$= \varepsilon_{eff}\mu \frac{\partial^2 \vec{E}}{\partial t^2}$$

In addition, the Ampere-Maxwell law in Eq. 2.29(d) can also be written as $\nabla \times \vec{B} = \varepsilon\mu \frac{\partial \vec{E}}{\partial t} + \mu\sigma\vec{E} = \varepsilon_{eff}\mu \frac{\partial \vec{E}}{\partial t}$. Thus, the Maxwell's equations in terms of effective permittivity are:

$$\text{Eq. 2.45} \qquad (a)\ \nabla \cdot \vec{E} = \frac{\rho_b + \rho_f}{\varepsilon_o} = \frac{\rho_f}{\varepsilon} \qquad (c)\ \nabla \times \vec{E} = -\frac{\partial \vec{B}}{\partial t}$$

$$(b)\ \nabla \cdot \vec{B} = 0 \qquad (d)\ \nabla \times \vec{B} = \varepsilon_{eff}\mu \frac{\partial \vec{E}}{\partial t}$$

The "effective permittivity" is commonly used when discussing optical properties of conductive materials, where $\sigma \neq 0$, without the notation "effective" or subscript "eff" in the literature. As this may cause confusion with dielectric permittivity, it is important to tell based on the context.

**2.2.5. Wave Equation and Speed of Light in Vacuum**

Eq. 2.43 and Eq. 2.44 appear to be in a similar form to that of a wave equation, which is given by

$$\text{Eq. 2.46} \qquad \nabla_s^2 d = \frac{1}{v^2}\frac{\partial^2 d}{\partial t^2}$$

, where $d$ is the displacement of the oscillation, and $v$ is the propagation speed of the wave (p. 399 of Ref. [42]). This hints that it might be possible to obtain the propagation speed of an electromagnetic wave by setting $\frac{1}{v^2} = \varepsilon_{eff}\mu$ in Eq. 2.46. To check this, we will write Eq. 2.43 and Eq. 2.44 in terms of "scalar" Laplacian, since Eq. 2.46 is based on scalar Laplacian and align the wave such that $\vec{B} = (B_x, 0, 0) = (B, 0, 0)$. Then, the lefthand and righthand sides of Eq. 2.43 can be written by $\nabla_v^2 \vec{B} = (\nabla_s^2 B, 0, 0)$ and $\varepsilon_{eff}\mu \frac{\partial^2 \vec{B}}{\partial t^2} = \varepsilon_{eff}\mu \left(\frac{\partial^2 B}{\partial t^2}, 0, 0\right)$, respectively. This leads to

Eq. 2.47
$$\nabla_s^2 B = \varepsilon_{eff}\mu \frac{\partial^2 B}{\partial t^2}$$

Due to the symmetry between Eq. 2.46 and Eq. 2.47, it might be tempting to conclude that the speed of light ($v$) can be computed by setting $\frac{1}{v^2} = \varepsilon_{eff}\mu$. However, $B$ and $\varepsilon_{eff}\mu$ are generally complex numbers, and $d$ and $v$ are purely real. For this reason, using $\frac{1}{v^2} = \varepsilon_{eff}\mu$ to calculate the speed of light in materials may lead to some misunderstandings [17]. Nevertheless, the speed of light in vacuum can be calculated by using $\frac{1}{v^2} = \varepsilon_{eff}\mu$, since both $\varepsilon_{eff}$ and $\mu$ are purely real numbers for vacuum: Take the real parts of the lefthand and righthand sides of Eq. 2.47:

Eq. 2.48
$$Re[\nabla_s^2 B] = Re\left[\varepsilon_{eff}\mu \frac{\partial^2 B}{\partial t^2}\right]$$

$$= \varepsilon_o \mu_o \frac{\partial^2 Re[B]}{\partial t^2}$$

$$\rightarrow \nabla_s^2 B' = \varepsilon_o \mu_o \frac{\partial^2 B'}{\partial t^2}$$

---

[17] Appendix C discusses complex-valued speed of light.

, where $B' = Re[B]$. Therefore, the speed of light in vacuum (*c*) is given by

$$\text{Eq. 2.49} \qquad c = v = \sqrt{\frac{1}{\varepsilon_o \mu_o}} \approx 3 \times 10^8 \; m/s$$

Recall that to derive the speed of light in vacuum, we had aligned an axis of the Cartesian coordinate parallel to the field. This was to avoid long mathematical derivation. The speed of light, however, is independent of how we "mathematically" set the coordinate. Even if we choose different coordinate orientations, Eq. 2.49 is still obtained (see Appendix D). This is also true for the electric field.

### 2.2.6. Dispersion Equation and Speed of Light in Matter

Define a term such that

$$\text{Eq. 2.50} \qquad n^2 \equiv \varepsilon_{eff,r} \, \mu_r$$

$$= \frac{\varepsilon_{eff} \, \mu}{\varepsilon_o \mu_o}$$

The term "*n*" hereafter will be refered to as refractive index. Using the effective permittivity and refractive index, Eq. 2.36 can be written as

$$\text{Eq. 2.51} \qquad k^2 = \varepsilon \mu \omega^2 + i \mu \sigma \omega$$

$$= \omega^2 \varepsilon_{eff} \mu$$

$$= \frac{n^2 \omega^2}{c^2}$$

Eq. 2.51 is commonly called dispersion equation. An electric field wave propagating along the z-axis can be written by

Eq. 2.52
$$\vec{E} = \vec{E}_o e^{i(kz-\omega t)}$$
$$= \vec{E}_o e^{i\left((k'+ik'')\cdot z - \omega t\right)}$$
$$= \vec{E}_o \left(e^{-k''z}\right) e^{i(k'z - \omega t)}$$

, where $k'$ and $k''$ are the real and imaginary parts of the propagation constant $k$, respectively. The phase velocity (or propagation speed) of the electric field wave in Eq. 2.52 is given by

Eq. 2.53
$$v = \frac{\omega}{k'}$$
$$= \frac{1}{Re\left(\sqrt{\varepsilon_{eff}\mu}\right)}$$
$$= \frac{c}{n'}$$

, where $n'$ is the real part of the refractive index ($n''$ is used for the imaginary part). Eq. 2.53 is also true for the magnetic field. It can be inferred that $n'$ being smaller than one ($n' < 1$) indicates the speed of light in matter faster than the speed of light in vacuum ($v > c$) [18].

---

[18] **Violation of theory of relativity?**
For some materials at some wavelengths, the real part of their refractive index is smaller than one [105]. There are ongoing debates in the literature about the speed of light being faster than $c$, regarding the theory of relativity. For example, the following statements are found:

1. "Signals that carry information do not travel faster than $c$" (p. 25 of Ref. [106]).
2. "It is actually the group velocity, not phase velocity, at which information-carrying signals travel" [106], [107].

Ref. [106] also says that the refractive index being smaller than one is not a violation of the theory of relativity because the group velocity is actually smaller than $c$. As opposed to the statements above, some researchers have demonstrated that group velocities can still be faster than $c$ [108]–[110]. Regardless, one phenomenon that occurs due to the real part of refractive index smaller than one is total external reflection when an x-ray is incident on a material from air at certain angles [106], [111].

### 2.2.7. Loss Tangent

The (electric) loss tangent refers to the ratio of the lossy term to the lossless term in the expression of $\nabla \times \vec{H}$. I will use this definition, p. 334 of Ref. [49] and p. 252 of Ref. [50], in deriving the general form of loss tangent, using complex-valued electrical conductivity [19]. As we will see towards the end of this section, the loss tangent derived in this work is slightly different from the one that is commonly seen in the literature. To begin the derivation, write the curl of H field:

Eq. 2.54
$$\nabla \times \vec{H} = \frac{\partial \vec{D}}{\partial t} + \vec{J}_f$$
$$= \varepsilon_{eff} \frac{\partial \vec{E}}{\partial t}$$
$$= \varepsilon_{eff}(-i\omega \vec{E})$$
$$= (\varepsilon_{eff}' + i\varepsilon_{eff}'')(-i\omega \vec{E})$$
$$= (\omega \varepsilon_{eff}'')\vec{E} + i(-\omega \varepsilon_{eff}')\vec{E}$$

The terms $\omega \varepsilon_{eff}''$ and $-\omega \varepsilon_{eff}'$ relate to optical energy loss and energy storage, respectively (p. 252 of Ref. [50]). Thus, the loss tangent is given by

Eq. 2.55
$$\tan \delta = \frac{\omega \varepsilon_{eff}''}{-\omega \varepsilon_{eff}'}$$
$$= \frac{\varepsilon_{eff}''}{-\varepsilon_{eff}'}$$
$$= \frac{\sigma' + \omega \varepsilon''}{\sigma'' - \omega \varepsilon'}$$

---

[19] The electrical conductivity ($\sigma$) is in general a frequency-dependent complex-valued number. This is discussed in detail in Section 2.3.2.

The term $\varepsilon_{eff}''$ in the numerator in Eq. 2.55 cannot be negative and is always $\geq 0$ [20]. Hence, the sign of loss tangent is determined by the terms in the denominator; $\sigma''$ and $\omega\varepsilon'$.

Note that loss tangent in most textbooks or published literature [49], [51]–[53] is given by $tan\,\delta = \frac{\sigma' + \omega\varepsilon''}{\omega\varepsilon'}$, ignoring the imaginary part of the electrical conductivity. The term in the denominator is also $\omega\varepsilon'$ instead of $-\omega\varepsilon'$. This is because a different sign convention was used when deriving the loss tangent. In that sign convention, the "-" sign replaces the "+" sign in expressions of electromagnetic parameters. The loss tangent derivation based on that sign convention is shown below.

$$\vec{E} = \vec{E}_o e^{-i(\vec{k}\cdot\vec{r} - \omega t)}$$

$$\varepsilon_{eff} = \varepsilon - i\frac{\sigma}{\omega}$$

$$= \varepsilon_{eff}' - i\varepsilon_{eff}''$$

$$= \left(\varepsilon' - \frac{\sigma''}{\omega}\right) - i\left(\varepsilon'' + \frac{\sigma'}{\omega}\right)$$

Thus,

$$\nabla \times \vec{H} = \varepsilon_{eff}\frac{\partial \vec{E}}{\partial t}$$

$$= \varepsilon_{eff}(+i\omega\vec{E})$$

---

[20] For plane waves in materials, the electric field amplitude is proportional to $e^{-\vec{k}_I \cdot \vec{r}}$. Since the wave cannot gain energy more than what it had, one can infer $|\vec{k}_I| = \frac{n''\omega}{c} \geq 0$, which suggests $n'' \geq 0$. From the relationship $n^2 = \varepsilon_{eff,r}$, one can obtain $\varepsilon_{eff,r}'' = 2n'n''$, where $n' \geq 0$. Therefore, $\varepsilon_{eff}'' \geq 0$.

$$= \left(\omega\varepsilon_{eff}''\right)\vec{E} + i\left(+\omega\varepsilon_{eff}'\right)\vec{E}$$

$$\tan\delta = \frac{\omega\varepsilon_{eff}''}{+\omega\varepsilon_{eff}'}$$

$$= \frac{\sigma' + \omega\varepsilon''}{-\sigma'' + \omega\varepsilon'}$$

If we ignore the imaginary part of the electrical conductivity, the loss tangent becomes

$$\tan\delta = \frac{\sigma' + \omega\varepsilon''}{\omega\varepsilon'}$$

Notice that the absolute value of loss tangent is independent of sign convention:

Eq. 2.56 $\qquad |\tan\delta| = \left|\frac{\sigma'+\omega\varepsilon''}{\sigma''-\omega\varepsilon'}\right|$

$$= \left|\frac{\sigma'+\omega\varepsilon''}{-\sigma''+\omega\varepsilon'}\right|$$

## 2.3. Optical Dispersion of Materials

In Section 2.2, light-matter interactions were discussed in terms of material properties such as dielectric permittivity ($\varepsilon$), permeability ($\mu$), and conductivity ($\sigma$) based on Maxwell's equations. Generally, these properties are dependent on the wave frequency. The frequency-dependent variation of material properties is referred to as optical dispersion. In this section, Lorentz, Cauchy's, Gaussian, and Drude models are discussed. Each model is useful to explain the optical dispersion of different types of materials. <u>Non-magnetic materials are assumed such that the permeability of materials is equal (or almost equal) to the vacuum permeability at all frequencies; $\mu = \mu_o$</u>. This assumption leads to a simpler expression of refractive index in Eq. 2.50:

$$\text{Eq. 2.57} \qquad n^2 = \varepsilon_{eff,r}$$

**2.3.1. Dielectric Materials**

Dielectric, or non-conductive, materials are those where there is no free charge, so both free charge density and electrical conductivity are zero ($\rho_f = \sigma = 0$). In isotropic materials, if the free charge density is zero, then the bound charge density is also zero –this is discussed in detail in Appendix A. Therefore, all derivations in Eq. 2.39 through Eq. 2.57 are valid for dielectric materials. Moreover, the effective permittivity of a dielectric material is equal to its dielectric permittivity; $\varepsilon_{eff} = \varepsilon$. When an electric field is incident on a dielectric material, the atoms (or molecules) make electric dipoles, which results in polarization. The polarization behavior in response to the incident field determines the permittivity of the dielectric material. The permittivity then determines the refractive index. In this section, the permittivity of dielectric materials is described based on Lorentz, Cauchy's, or Gaussian model.

a. **Lorentz Dispersion**

Let us assume that all bound electrons in the material react the same way in response to an incident field. Then, its relative permittivity based on Lorentz model [42] is given by

Eq. 2.58
$$\varepsilon_{r,eff} = \varepsilon_r$$
$$= \varepsilon_r' + i\varepsilon_r''$$
$$= 1 + \left(\frac{N_L q^2}{m_e \varepsilon_o}\right) \frac{1}{\omega_0^2 - \omega^2 - i\omega\gamma_L}$$
$$= n^2$$

, where $N_L$ is the number of dipoles created per unit volume, $q$ is the magnitude of the electron charge, $m_e$ is the electron mass, $\omega_o$ is the angular resonance frequency of the dipole, $\omega$ is the angular frequency of the incident wave, and $\gamma_L$ is the damping (or collision) frequency of the dipole. It is important to note that the permittivity and refractive index in Eq. 2.58 are complex-valued and frequency-dependent. The imaginary part of the refractive index is associated with optical energy extinction (or absorption) in the material –the larger the imaginary part, the higher the extinction– as can be inferred from Eq. 2.52. Fig. 2.8 shows Lorentz dispersion curves for permittivity and refractive index; where the dielectric material has a single resonance frequency and a single damping frequency. The Cauchy's regime is indicated in an area shaded in yellow, which will be discussed in the next section. At the resonance frequency, $\omega_o$, the imaginary part of the permittivity peaks and its real part goes through an inflection point. Dielectric materials in the real world, however, tend to have multiple dipole resonances, resulting in many different resonance ($\omega_j$) and damping ($\gamma_{L,j}$) frequencies. Thus, in such case, the relative permittivity is modified as

Eq. 2.59
$$\varepsilon_r = \varepsilon_r' + i\varepsilon_r''$$
$$= 1 + \frac{N_L q^2}{m_e \varepsilon_o} \sum_j \frac{f_j}{\omega_j^2 - \omega^2 - i\omega\gamma_{L,j}}$$
$$= n^2$$

, where the fraction term $f_j$ ranges from 0 to 1 and $\sum f_j = 1$.

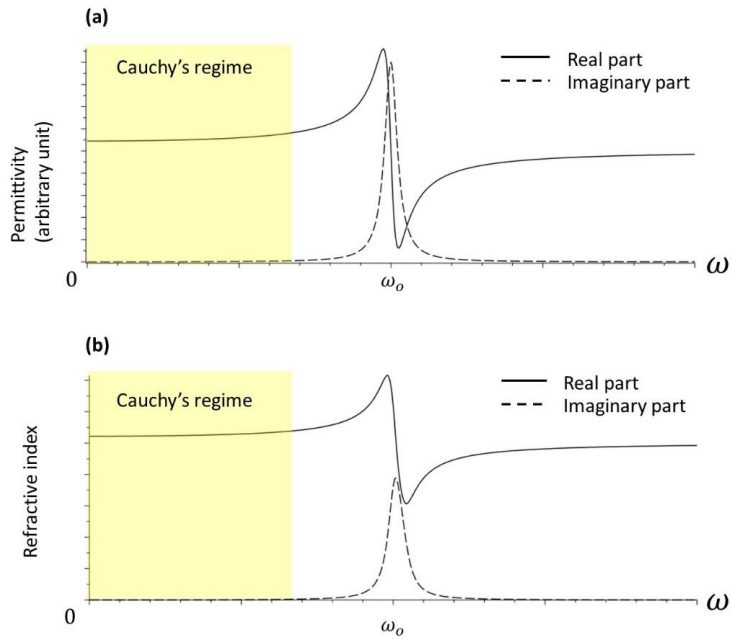

Fig. 2.8. Lorentz dispersion curves for permittivity and refractive index. The material is assumed to have a single resonance frequency and a single damping frequency. Cauchy's regime is shaded in yellow. Note that the real part of the dielectric permittivity can be a negative number depending on the ratio between resonance frequency, damping frequency, and plasma frequency. The plasma frequency ($\omega_p$) of a dielectric material can be written by $\omega_p = \sqrt{\frac{N_L q^2}{m_e \varepsilon_o}}$.

### b. Cauchy's Dispersion

A special case of Lorentz dispersion is when $\omega \ll \omega_o$ and $\omega_o^2 - \omega^2 \gg \omega \gamma_L$ in Eq. 2.58; where the material is transparent (or non-absorbing) to the incident wave. The frequency regime where those two conditions are satisfied are indicated in Fig. 2.8 in the yellow shade: Cauchy's regime. In the Cauchy's regime, Eq. 2.58 reduces to

Eq. 2.60
$$\varepsilon_r = Re(\varepsilon_r)$$
$$= 1 + \left(\frac{N_L q^2}{m_e \varepsilon_o}\right) \frac{1}{\omega_0^2 - \omega^2}$$
$$= n^2$$

By applying the binomial series expansion on the term $\frac{1}{\omega_0^2 - \omega^2}$ in Eq. 2.60, the relative permittivity of a dielectric material within the Cauchy's regime can be written as [54]

Eq. 2.61
$$(a)\ \varepsilon_r = Re(\varepsilon_r)$$
$$= 1 + \frac{N_L q^2}{m_e \varepsilon_o \omega_0^2}\left(1 + \frac{\omega^2}{\omega_o^2} + \frac{\omega^4}{\omega_o^4} + \cdots\right)$$
$$= n^2$$

$$(b)\ \sqrt{\varepsilon_r} = Re(\sqrt{\varepsilon_r})$$
$$= A + \frac{B}{\lambda^2} + \frac{C}{\lambda^4} + \cdots$$
$$= n$$

, where A, B, and C are positive constants and called Cauchy's coefficients. Typically, deploying the first two coefficients, A and B, is enough to describe the material's dispersion within the Cauchy's regime.

c.  **Gaussian Dispersion**

The Gaussian dispersion is similar to Lorentz dispersion [34]. One difference between these two models is that the imaginary part of the permittivity in the Gaussian model decreases to zero more quickly than that in the Lorentz model [34], as the wave frequency moves away from the resonance frequency. This is shown in Fig. 2.9. One benefit of using Gaussian model is that it can describe transparent materials at frequencies where the Lorentz model fails to describe due to the imaginary part of the permittivity being not small enough [34].

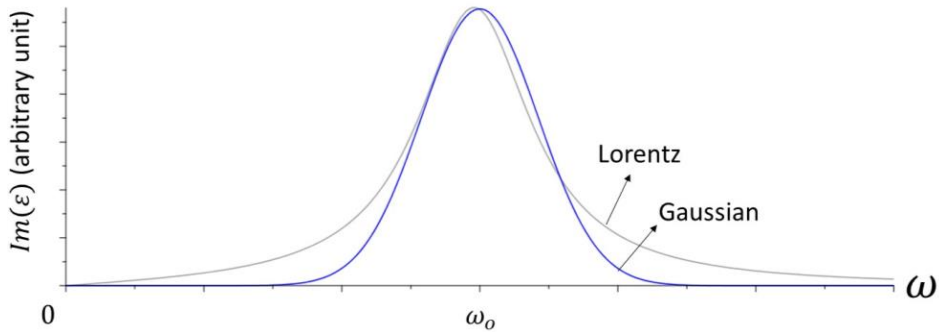

Fig. 2.9. Imaginary parts of the permittivity based on Lorentz and Gaussian dispersion models.

**2.3.2. Conductive Materials**

Conducive materials are those where there are free charges and, therefore, their electrical conductivity is greater than zero. Unlike dielectric materials, the effective permittivity of conductive materials is influenced by their conductivity. In response to an incident field, free charges might get locally accumulated inside the conductor, which creates a positive or negative net charge density at that location. This could make it more complicated to solve Eq. 2.38. However, we will assume that free charges do not build up in the volume of the conductor, except at the edge. This is visualized in Fig. 2.10. The following derivations for conductive materials, therefore, may not work for their edges. However, our interest is in the volume of the material, not the edge.

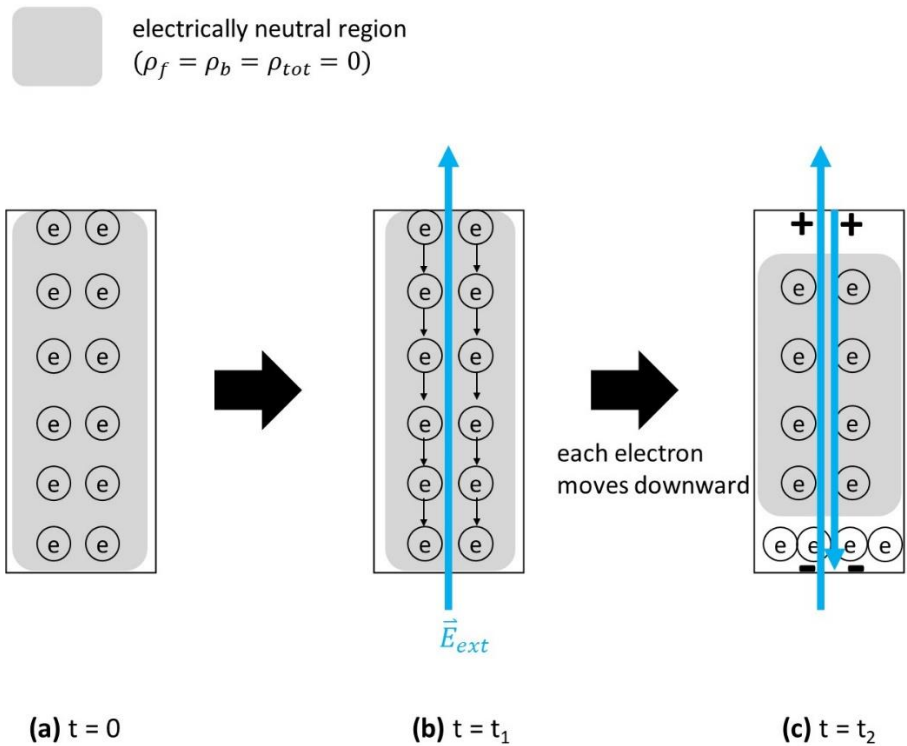

Fig. 2.10. Motion of free electrons in a conductor in response to an external electric field ($\vec{E}_{ext}$). (a): An electrically neutral state without external electric fields. (b): An external electric field is just incident on the conductor. (c): Electrons move due to the external electric field. It is assumed that there are many enough free electrons in the conductor so that the electric field due to the surface charges can cancel the external electric field. The electrically neutral region is shaded in gray. $t$ is time ($0 < t_1 < t_2$).

The frequency-dependent conductivity ($\sigma$) and DC conductivity ($\sigma_{DC}$) are given by (p. 544 of Ref. [54]):

Eq. 2.62  (a) $\sigma = \dfrac{\sigma_{DC}}{1-i\frac{\omega}{\gamma_D}}$    (b) $\sigma_{DC} = \dfrac{N_D q^2}{m_e \gamma_D}$

, where $\omega$ is the angular frequency of light, $\gamma_D$ is the damping (or collision) frequency of free charges, $N_D$ is the number of free charges per unit volume, $q$ is the magnitude of electron charge, and $m_e$ is the electron mass. Fig. 2.11 shows the frequency-dependent complex-valued electrical conductivity based on Drude model. The imaginary part goes to zero at $\omega = 0$ at which $\sigma = \sigma_{DC}$.

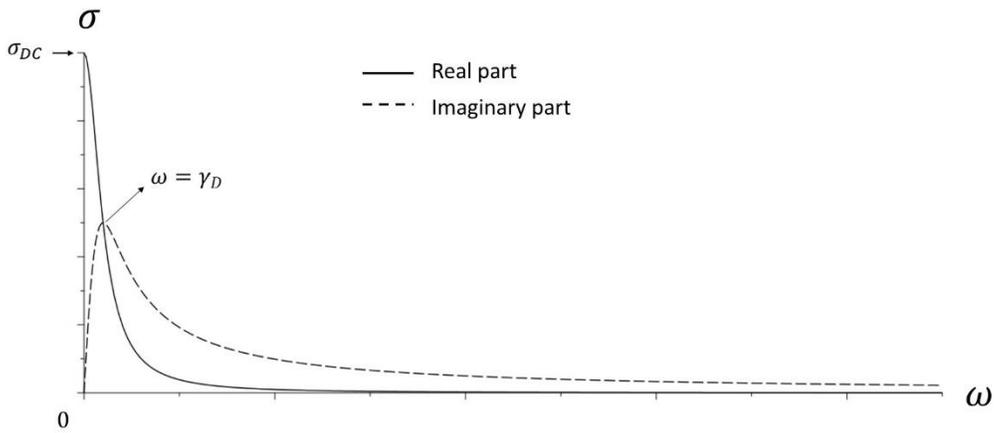

Fig. 2.11. Electrical conductivity based on Drude model.

Let us define a parameter $\omega_p$, which is commonly called plasma frequency (p. 546 of Ref. [54]). The plasma frequency for conductors is defined such that

Eq. 2.63  $\omega_p^2 \equiv \dfrac{N_D q^2}{m_e \varepsilon_o}$

$= \dfrac{\gamma_D \sigma_{DC}}{\varepsilon_o}$

$= \gamma_D \sigma_{DC} c^2 \mu_0$

According to Drude model [55], [56], the relative effective permittivity of conductors can be written as

Eq. 2.64
$$\varepsilon_{r,eff} = \varepsilon_{r,eff}' + i\varepsilon_{r,eff}''$$
$$= \varepsilon_r - \frac{\omega_p^2}{\omega^2 + i\omega\gamma_D}$$
$$= n^2$$

Note that $\varepsilon_r$ is the relative "dielectric" permittivity that arises due to bound electrons in the conductor, which is also generally a complex number. The $\varepsilon_r$ can be described by Lorentz or Gaussian dispersion model. Combining Lorentz and Drude dispersions, the relative effective permittivity of conductors is expressed as

Eq. 2.65
$$\varepsilon_{r,eff} = \left(1 + \frac{N_L q^2}{m_e \varepsilon_o} \sum_j \frac{f_j}{\omega_j^2 - \omega^2 - i\omega\gamma_{L,j}}\right) - \left(\frac{\omega_p^2}{\omega^2 + i\omega\gamma_D}\right)$$
$$= n^2$$

This equation suggests that the refractive index of conductors converges to 1 at high frequencies; say $\omega \gg \omega_j$, $\omega \gg \gamma_{L,j}$, and $\omega \gg \omega_p$. Dispersion curves of the effective permittivity and refractive index of conductors are shown in Fig. 2.12, where the Lorentz dispersion effect is assumed to be negligible; $\varepsilon_r = 1$. In some literature, the righthand side of Eq. 2.64 is written with a purely real term $\varepsilon_\infty$, called epsilon infinity, instead of $\varepsilon_r$. That is,

Eq. 2.66
$$\varepsilon_{r,eff} = \varepsilon_\infty - \frac{\omega_p^2}{\omega^2 + i\omega\gamma_D}$$
$$= n^2$$

The infinity symbol (∞) implies that the relative effective permittivity converges to $\varepsilon_\infty$ at very high frequencies. The value of epsilon infinity may vary depending on materials or wavelength range. For several metals at infrared wavelengths, their Drude parameters are well-studied in the literature [57].

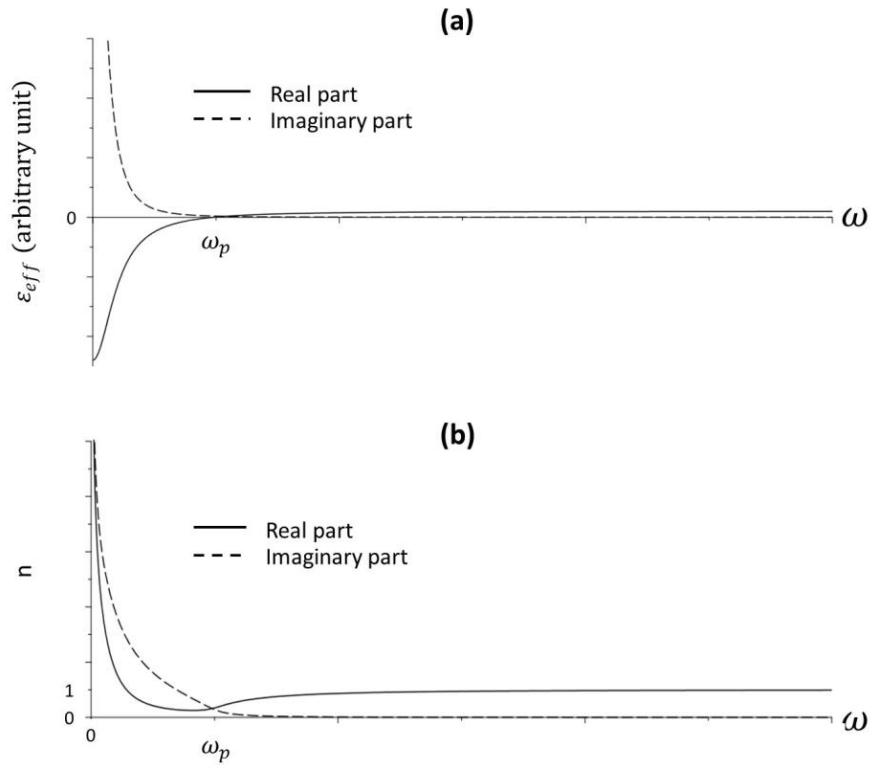

Fig. 2.12. Effective permittivity (a) and refractive index (b) of a conductor. The Lorentz dispersion effect is assumed to be negligible; $\varepsilon_r = 1$.

## 2.4. Stress in Thin Films

In structures consisting of thin layers of different materials, layers can delaminate due to stress. The total stress in a film can be obtained by adding up stresses due to lattice mismatch and CTE mismatch [58], [59], where CTE stands for coefficient of thermal expansion. These stresses are determined by numerous parameters such as Young's modulus, Poisson's ratio, film thickness, lattice misfit, CTE, and temperature [58], [60], [61]. In addition, the bonding energy between different materials affects delamination as well. As there are numerous variables involved, it is difficult to predict if a film will delaminate or not. Nevertheless, the stress due to CTE mismatch can be estimated by the relationship below [60].

$$\text{Eq. 2.67} \qquad d\sigma_f = \frac{E_f}{1-\nu_f}(CTE_s - CTE_f)dT$$

$\sigma_f$ is the stress in the film. $E_f$ and $\nu_f$ are the Young's modulus and Poisson's ratio of the film, respectively. $CTE_s$ and $CTE_f$ are thermal expansion coefficients of the substrate and film, respectively. $T$ is temperature. The positive (negative) stress means tension (compression). As one can see, a lower Young's modulus, lower Poisson's ratio, and smaller CTE mismatch can reduce stress upon temperature changes. Fig. 2.13 shows thermal expansion coefficients of selected materials as a function of temperature. Fig. 2.14 shows Young's moduli of materials at room temperature.

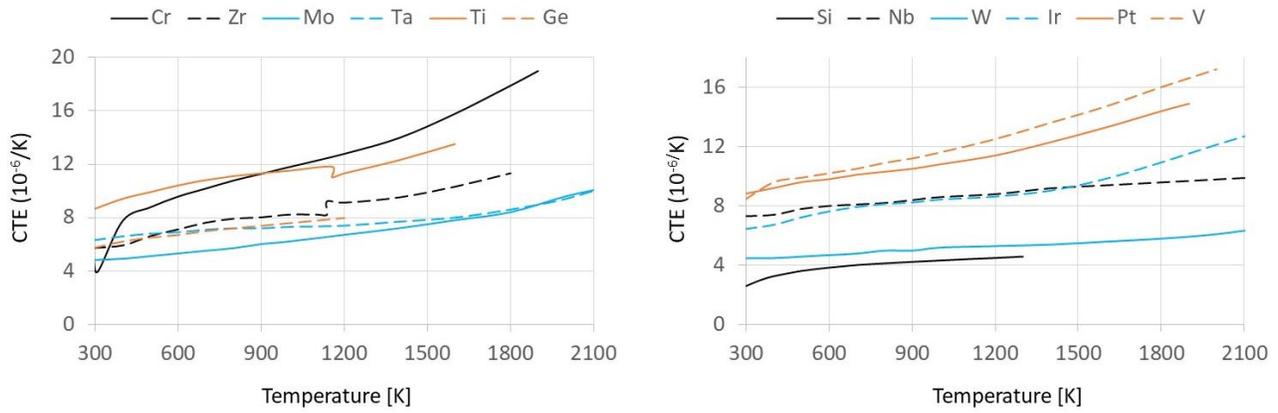

Fig. 2.13. Thermal expansion coefficients (CTEs) vs. temperature for selected materials. Zr and Ti undergo phase transition at near 1137K and 1156K, respectively. The shown Zr and Ti data are for polycrystals, and all other materials have the simple cubic crystal system where the CTE is independent of direction. Plots are generated based on data from Refs. [62], [63].

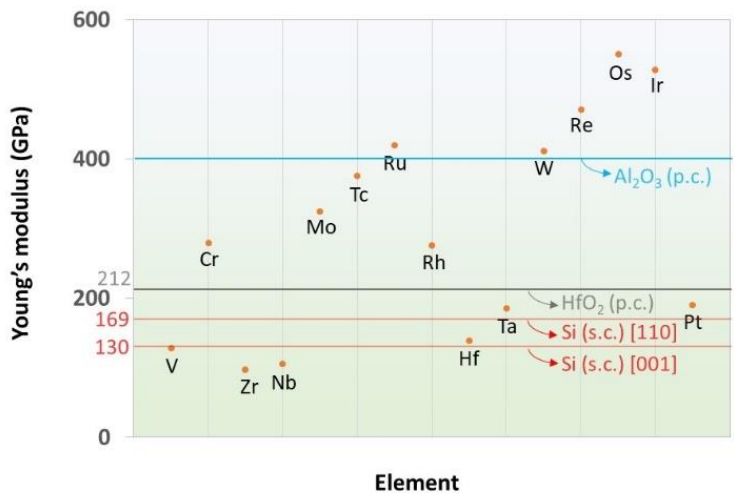

Fig. 2.14. Young's moduli of selected materials at room temperature. Data are shown for polycrystalline metals and some dielectric materials. s.c. and p.c. stand for single crystal and polycrystal, respectively. Reprinted with permission from Ref. [3].

# 3. METHOD

## 3.1. Finite-Element Simulation of Light-Matter Interactions

The optical responses of bulk materials can be predicted via relatively simple equations such as Fresnel reflection/transmission coefficients [64] or absorption coefficient (p. 507 of Ref. [54]). For metamaterials (MMs), however, those equations cannot be directly applied because their geometry needs to be taken into account when solving the Maxwell's equations [21]. Thus, finite-element simulation tools that solve Maxwell's equations, such as CST Microwave Studio, are commonly used for MM research [16], [65]–[67]. Electric and magnetic fields can also be visualized by those simulation tools, which can provide critical insights when analyzing light-matter interactions [10], [67]. In this work, CST Microwave Studio (or simply CST) is used for MM simulation. In CST, optical properties of materials can be defined by using the optical dispersion models introduced in Section 2.3 or by directly inputting measured (effective) permittivity values into the software. Fig. 3.1 shows the software interface for inputting measured optical properties. The effective permittivity can be measured by spectroscopic ellipsometry.

---

[21] Indeed, "effective" coefficients of a MM can be used to calculate its optical properties such as reflectance, transmittance, and/or absorbance [112], [113]. Nevertheless, Maxwell's equations need to be solved first to obtain their effective coefficients.

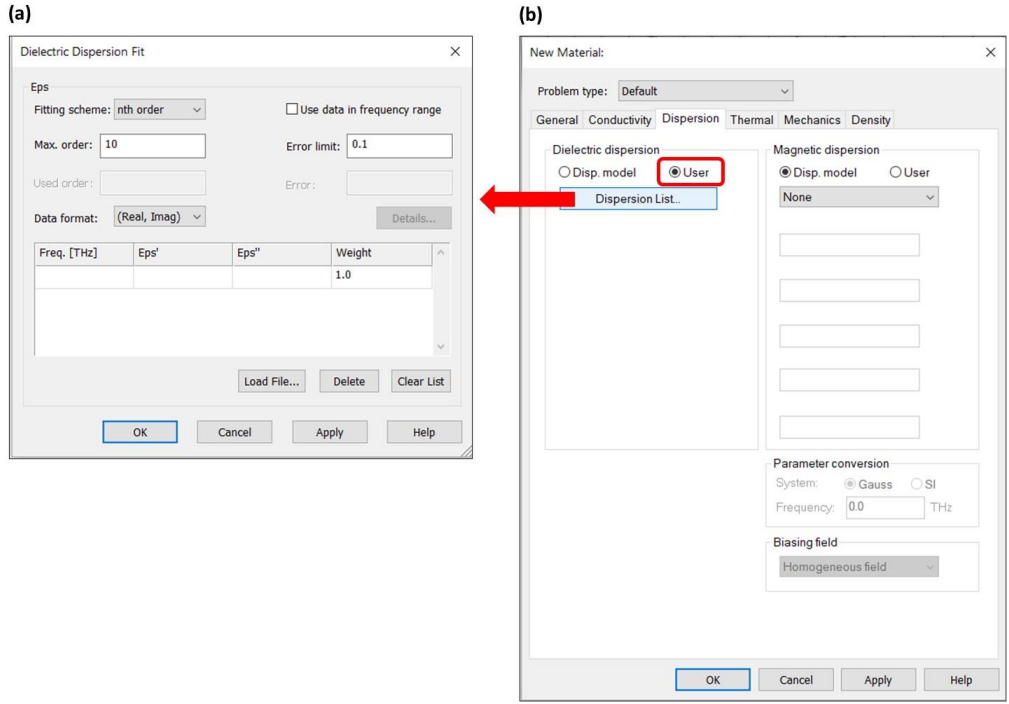

Fig. 3.1. CST Microwave Studio interface for inputting user-defined permittivity. The Eps' and Eps'' are the real and imaginary parts of the relative effective permittivity.

## 3.2. Spectroscopic Ellipsometry

Wavelength-dependent optical properties such as (effective) permittivity or refractive index can be measured by spectroscopic ellipsometry [34]. The measured properties can be used to simulate optical responses of emitters. Fig. 3.2(a) shows basic operation principles of ellipsometry. By measuring differences between the reflected waves of $p$ and $s$ polarizations [22], the ellipsometer determines $\Psi$ and $\Delta$, as shown in Eq. 3.1. The $r_p$ and $r_s$ are reflection coefficients for $p$ and $s$ waves, respectively. To extract information such as film thickness or refractive index, the Ψ and Δ data must be modeled. Fig. 3.3 shows the software interfaces, WVASE, for modeling and data analysis [23]. Fig. 3.3(a) shows modeled layers and their thicknesses. Fig. 3.3(b) shows a window where users can change optical parameters of a GENOSC layer (or general oscillation layer). Fig. 3.3(c) shows physical dispersion models that can be used in a GENOSC layer. Generally speaking, the complexity of modeling and data analysis increases with the number of layers involved. In this work, spectroscopic ellipsometers from J. A. Woollam and their analysis software WVASE are used. The company also provides a heating cell (Linkam Heat Cell) for measurements at elevated temperatures up to approximately 600°C. This cell was used in this work to characterize temperature-dependent optical properties of materials.

---

[22] The $p$ and $s$ refer to the polarization of the electric field.

Eq. 3.1 $$\tan \Psi \, e^{i\Delta} = \frac{r_p}{r_s}$$

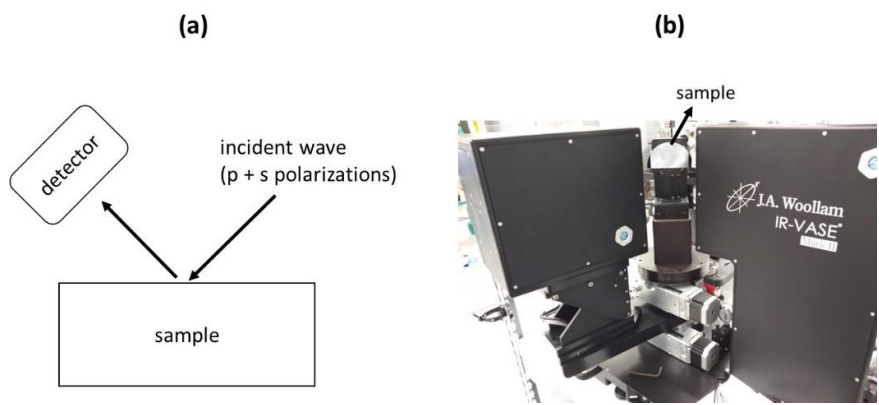

Fig. 3.2. (a): schematic operation of ellipsometry. (b): a spectroscopic ellipsometer (IR-VASE, J. A. Woollam).

---

[23] When fitting the data in WVASE, all optical parameters in the model must "physically" make sense. For example, the amplitude (denoted by "Amp") of a General Oscillator Model must be greater than zero (or zero at the smallest). Similarly, Cauchy's coefficients must also be greater than zero (or zero at the smallest). The software allows negative values of these parameters, but the user has to be careful to not use negative (or unphysical) numbers.

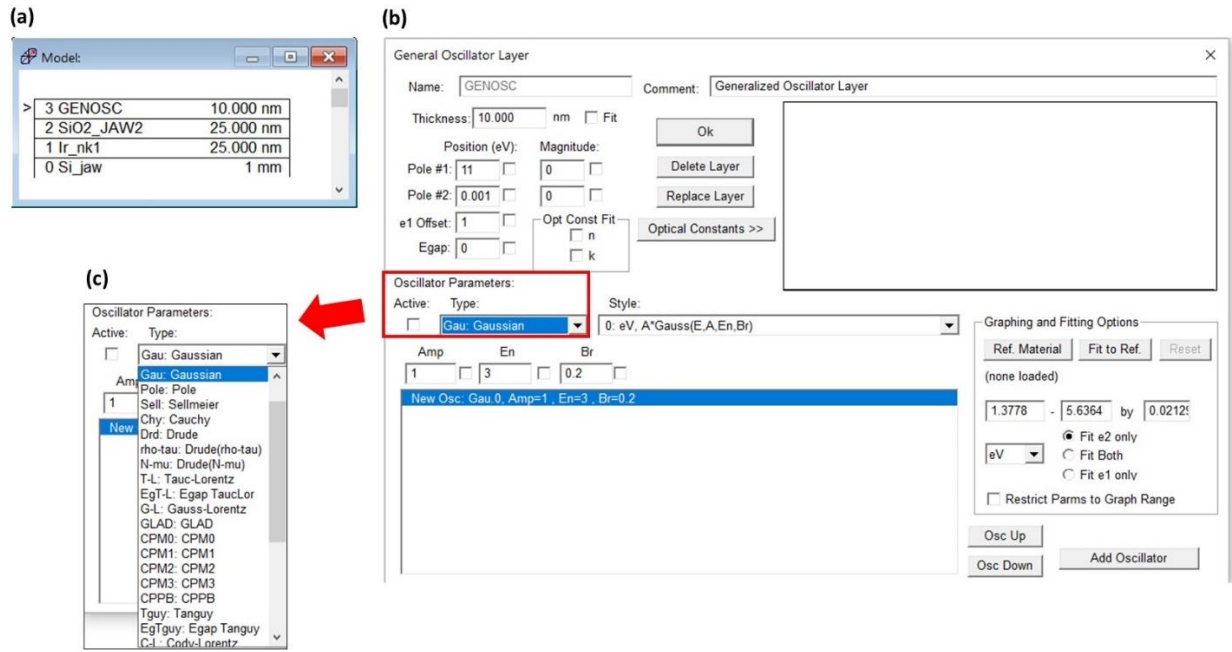

Fig. 3.3. WVASE interfaces for modelling (a) and optical dispersion analysis (b and c). In (a), "GENOSC" is the top layer and "Si_jaw" is the substrate.

## 3.3. Micro/Nanofabrication

### 3.3.1. Thin Film Deposition

In general, optical properties of deposited films depend on deposition technique [68] or deposition parameters [69]. This section reviews several thin film deposition techniques that are commonly used to fabricate optical structures. Optical properties of some deposited films in this work will be discussed in Section 4.2.

a. **Sputtering**

Sputtered layers tend to have better adhesion to the underlying layer or substrate. This is because sputtering ions etch the substrate surface due to the electric potential difference in the chamber, as shown in Fig. 3.4(b). This helps remove surface contamination or surface oxide of the substrate [24] –the solvent cleaning usually removes only oils and grease [25] [70]. Sputtering yield is defined by the number of ejected atoms from the target divided by the number of ions that strike the target. Typically, sputtering yield increases as the atomic mass of the target material increases, as shown in Fig. 3.5(a) [71]. A higher sputtering yield indicates a higher deposition rate. For direct current (DC) sputtering, Fig. 3.4(b), a negative bias is applied on the target. The sputtering ions, which are positive ions such as $Ar^+$, are pulled towards the target and strike its surface. This may knock some atoms out of the target surface. The knocked atoms travel towards the substrate and are deposited there. If the target material is conductive, the sputtering ions at the target's surface can receive free electrons and become electrically neutral. The neutralized Ar atoms are not pulled by the electric force anymore and, therefore, can be moved away. This allows other $Ar^+$ ions in the chamber to keep on striking the target surface. However, if the target material is not conductive, then $Ar^+$ ions can be stuck at the target surface since they cannot receive free electrons. This can prevent other $Ar^+$ ions in the chamber from hitting the target. For this reason, alternating current (AC) sputtering is used for nonconductive targets. This technique is also called radio frequency (RF) sputtering.

---

[24] On top of the surface oxide, there can be a layer of water molecules stuck to the surface. This typically is the case if the substrate cleaning involves a rinse with deionized (DI) water at the last step, even after $N_2$ blow. The water molecules prevent adhesion of atoms (or molecules) to the substrate. To remove water molecules and improve adhesion, the substrate can be heated before deposition.

[25] The typical solvent cleaning of a substrate involves a rinse with Acetone → Isopropyl alcohol (IPA) → DI water. An example of metal adhesion to a Si substrate with and without a DI water rinse is shown in Appendix E.

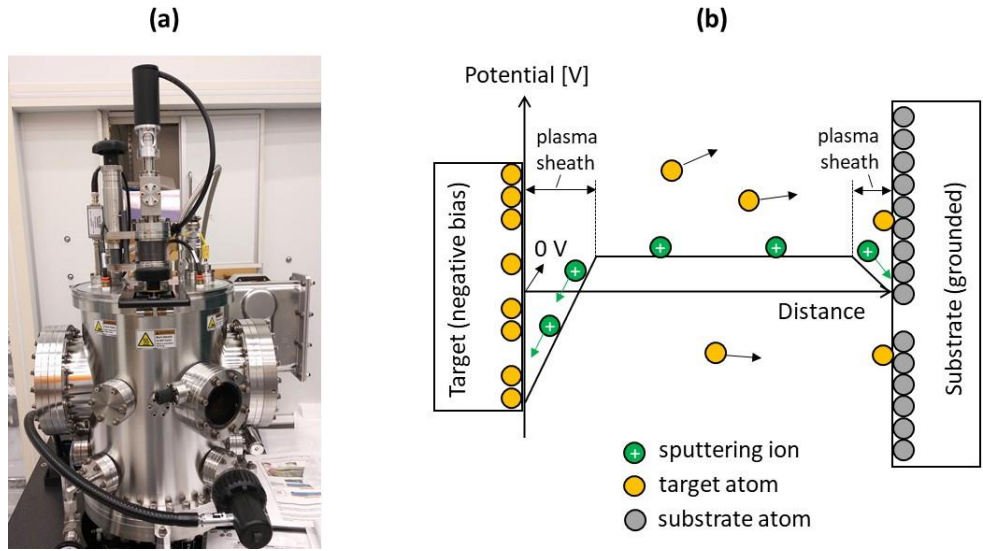

Fig. 3.4. (a): a sputter (ATC Orion-3-HV, AJA International). (b): sputter chamber at a DC bias.

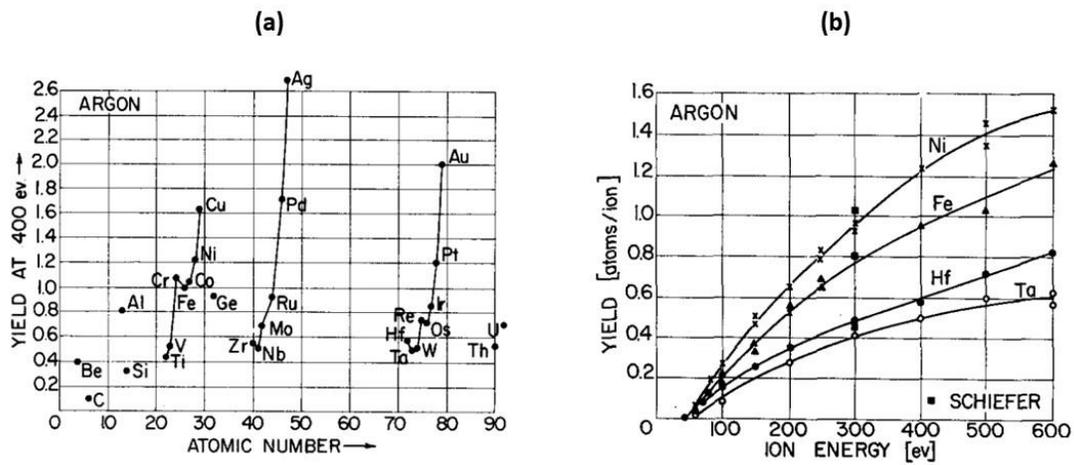

Fig. 3.5. Sputtering yields as a function of atomic number (a) and as a function of sputtering ion energy (b). The sputtering ion is Ar$^+$ for the shown data. Reprinted with permission from Ref. [71].

## b. Chemical Vapor Deposition (CVD) and Atomic Layer Deposition (ALD)

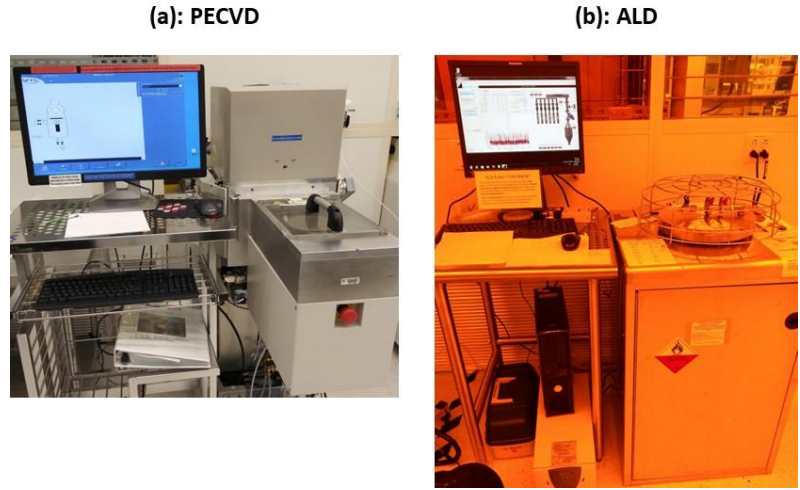

Fig. 3.6. (a): a plasma-enhanced chemical vapor depositor (PECVD, Surface Technology Systems) (b): an atomic layer depositor (S-200, Cambridge NanoTech).

Chemical vapor deposition (CVD) is a technique that uses precursor gases to deposit thin layers. Plasma-enhanced chemical vapor deposition (PECVD) is a type of CVD where chemical reactions are enhanced by a plasma in the chamber, Fig. 3.6(a). The deposition rate of CVD ranges from a few nm/min to a few hundred nm/min, which is much faster than sputtering. Another type of CVD is atomic layer deposition (ALD), Fig. 3.6(b). ALD can deposit thin layers with higher quality; low defect density [72] and uniform thickness, for example. However, ALD is not suitable for depositing thicker layers because its deposition rate is typically as low as a few Å/min. Furthermore, in general, CVD techniques provide higher step coverage compared to physical vapor deposition (PVD) techniques like sputtering or evaporation, as shown in Fig. 3.7. Higher step coverage is useful if the device needs a passivation layer on the surface. However, it can result in slower liftoff since it blocks liftoff chemicals from reaching the underlying layer, photoresist for example.

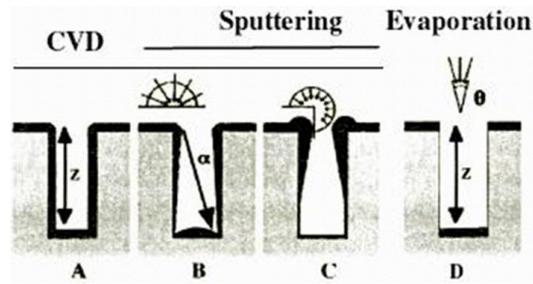

Fig. 3.7. Step coverage comparison of CVD and PVD. Black layers are deposited layers. Reprinted from Ref. [73] (free access).

### 3.3.2. Lithography

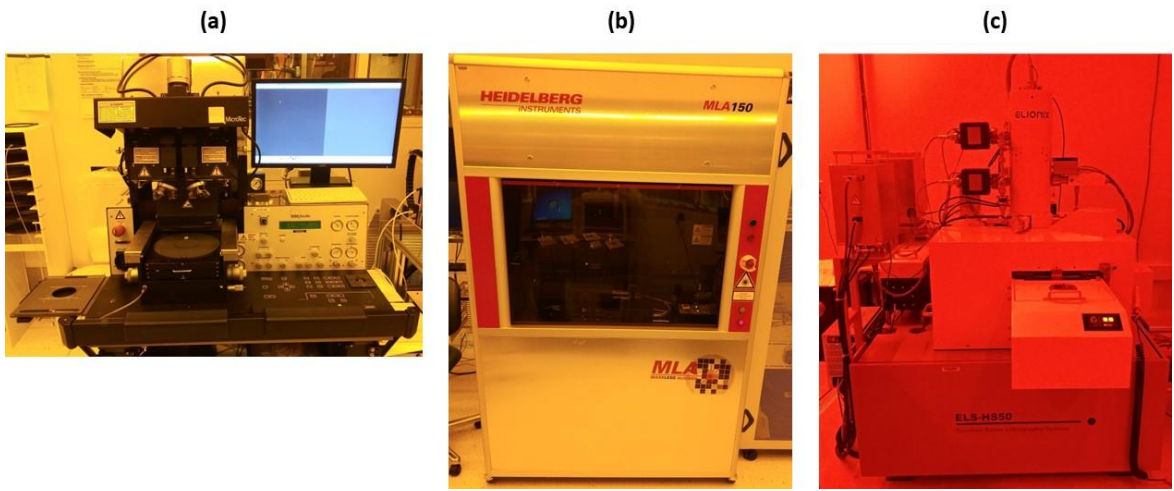

Fig. 3.8. Lithography equipment. (a): conventional photolithography (MA6 Mask Aligner, Suss MicroTec). (b): direct laser beam photolithography (MLA150 Maskless Aligner, Heidelberg Instruments). (c): electron beam lithography (Elionix HS50, Elionix).

Lithography is used to make patterns on the surface. Fig. 3.8 shows some lithography equipment. Photolithography and electron beam lithography use ultraviolet light and electrons to write patterns, respectively. While the resolution of photolithography depends on the wavelength of the light used, it is typically around 1 μm. Electron beam lithography can achieve a much better resolution, tens or hundreds of nm, but it takes longer time to write patterns. Fig. 3.9 compares photolithography processes with a positive resist and a negative resist. For the positive resist, the liftoff process can take much longer time because the deposited layer blocks liftoff chemicals from reaching the resist [26]. This can be avoided by using a negative resist and directional deposition techniques such as PVD. In addition, compared to conventional photolithography that uses a mask, direct laser beam photolithography uses a (UV) laser beam to write patterns without a mask. While the laser beam size determines the resolution, it is typically 1 µm or less. Fig. 3.10 describes the lithography process with a laser beam that has a Gaussian intensity profile.

---

[26] In practice, a sonicator (or ultrasound cleaner) can be used to accelerate the liftoff process. However, if adhesion between the deposited layer and substrate is weak, using a sonicator can cause the layer to peel off.

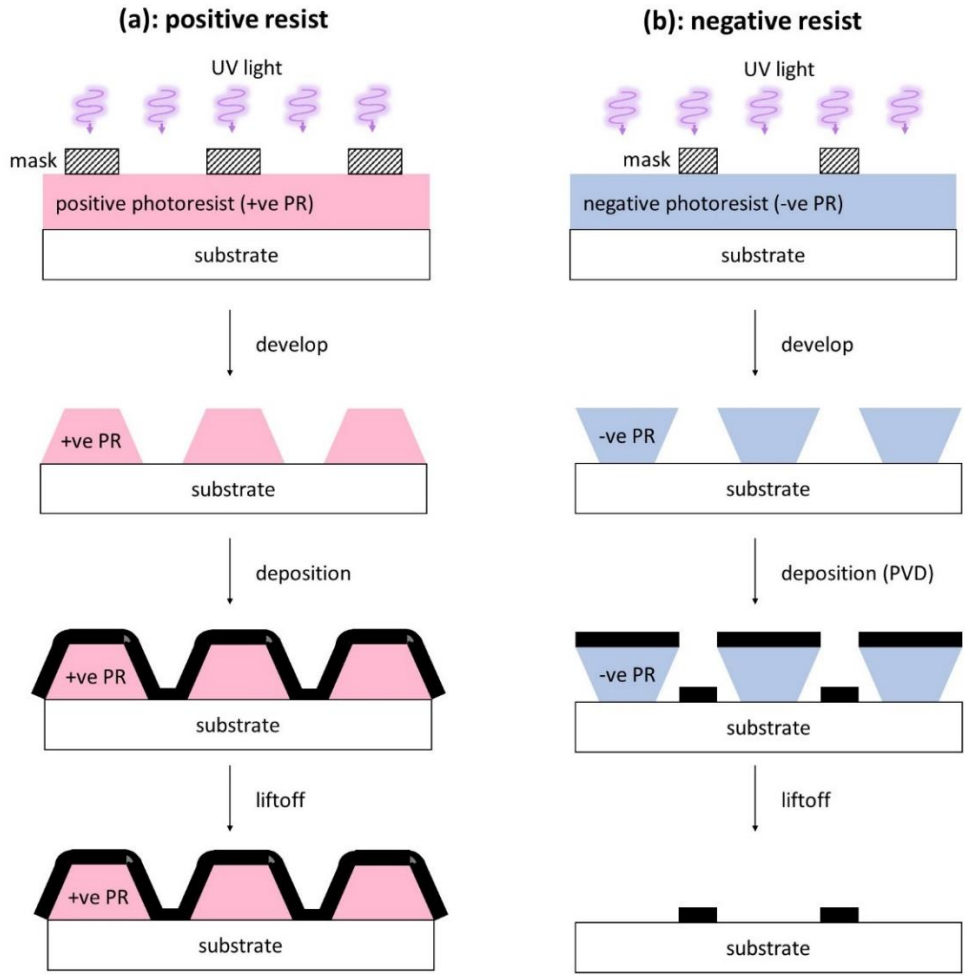

Fig. 3.9. Photolithography processes with a positive resist (a) and a negative resist (b).

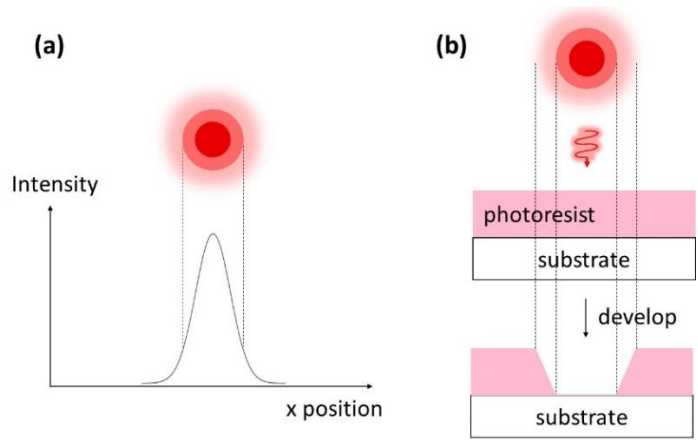

Fig. 3.10. (a): laser beam intensity profile. (b): photolithography with a direct laser beam.

# 4. RESULTS AND DISCUSSION

## 4.1. Emission Measurement Apparatus

### 4.1.1. Motivation

There are many commercialized instruments that can measure spectroscopic absorptivity, such as ellipsometers or spectrophotometers. However, instruments for spectroscopic emissivity are not as common. This motivated me to build a setup to characterize emission properties of TPV emitters at higher temperatures.

### 4.1.2. Results

The emission measurement setup in Fig. 4.1 can be divided into two parts: electrical and optical. The electrical part consists of components to electrically power the heater, shown in Fig. 4.2(a), and to control voltage and current. The electrical part can also be used to test thermal stability of emitters in air. The emitter temperature is measured at its top surface by a thermocouple, shown in Fig. 4.2(b). The optical part consists of components to spectroscopically measure thermal radiation from the emitter. Note that the emitter in the setup is exposed to air. Fig. 4.3(a) shows beam paths for laser and optics alignment. The beam path 1 (red) is a straight laser beam passing through the center of optics components. The beam path 2 (blue) is used to align the emitter to measure its radiation at normal incidence. Fig. 4.3(b) shows the beam path when the emitter is aligned.

The voltage signal from the photodetector is collected and processed by a LabVIEW code that is also custom-written. The LabVIEW interfaces are shown in Fig. 4.4. The method of signal normalization is shown in detail in Appendix J. The measurable wavelength range of the setup is 2 – 4.5 μm with a 0.5 mm-thick quartz window, in Fig. 4.1(b). The range can be extended to 2 – 7 μm if a 2 mm-thick $CaF_2$ window is used. In this work, however, a quartz windows was used because it is stronger to thermal shock [27]. Additional details such as setup diagrams, data signal flow, and LabVIEW codes are shown in Appendix F – H.

---

[27] The $CaF_2$ window broke when used due to thermal shock.

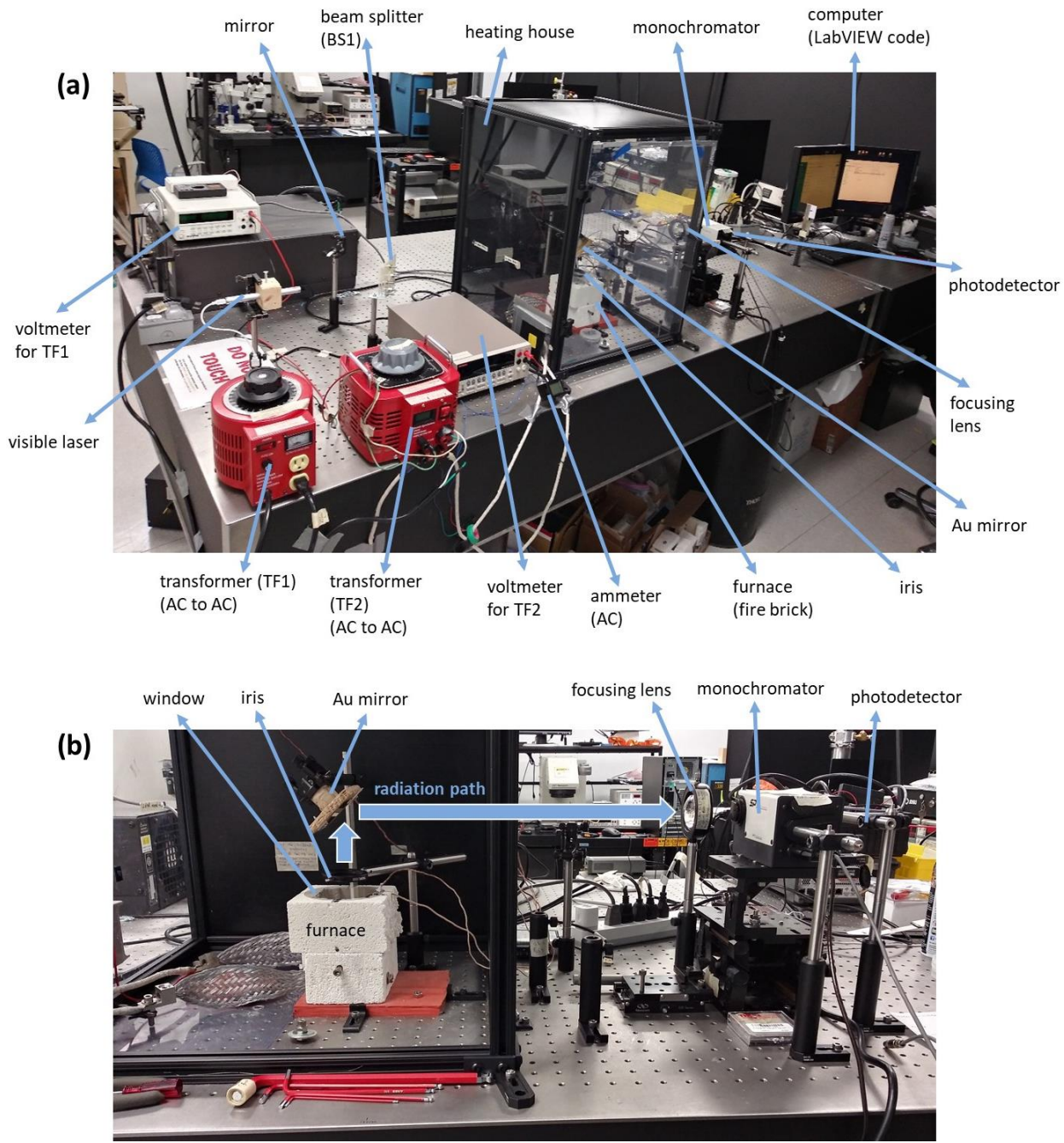

Fig. 4.1. Custom-built emission measurement apparatus. (a): an apparatus picture including electrical and optical components. (b): some optical components and radiation path.

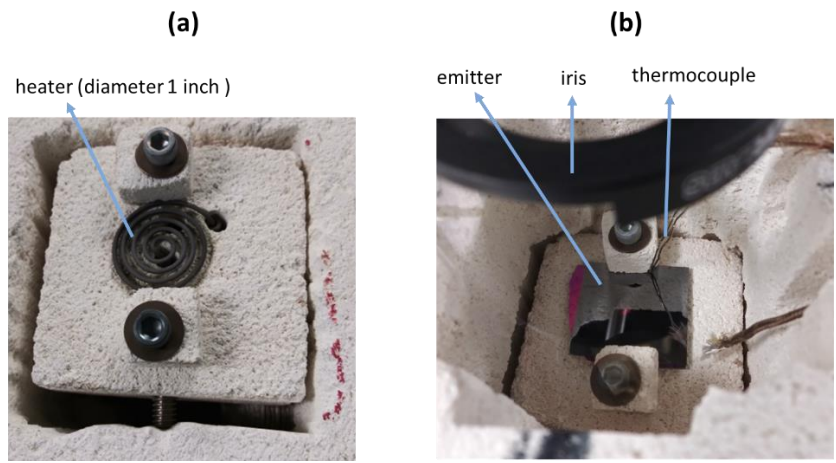

Fig. 4.2. Furnace in the emission measurement apparatus. (a): a heater with a 1 inch diameter. (b): an emitter sample loaded.

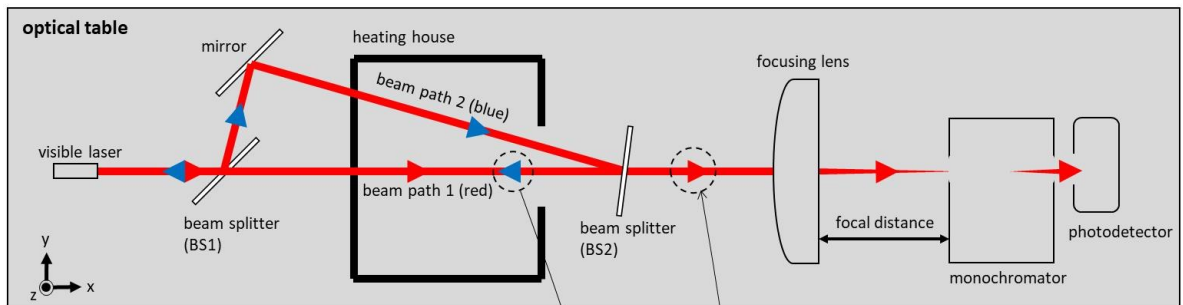

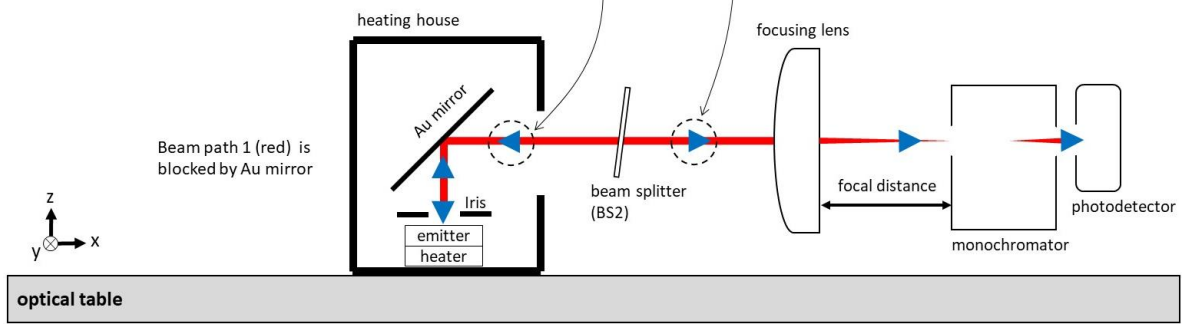

Fig. 4.3. Laser beam paths for optics alignment. Beam path 1 (red) is used for laser and optics alignment. Beam path 2 (blue) is used for emitter alignment.

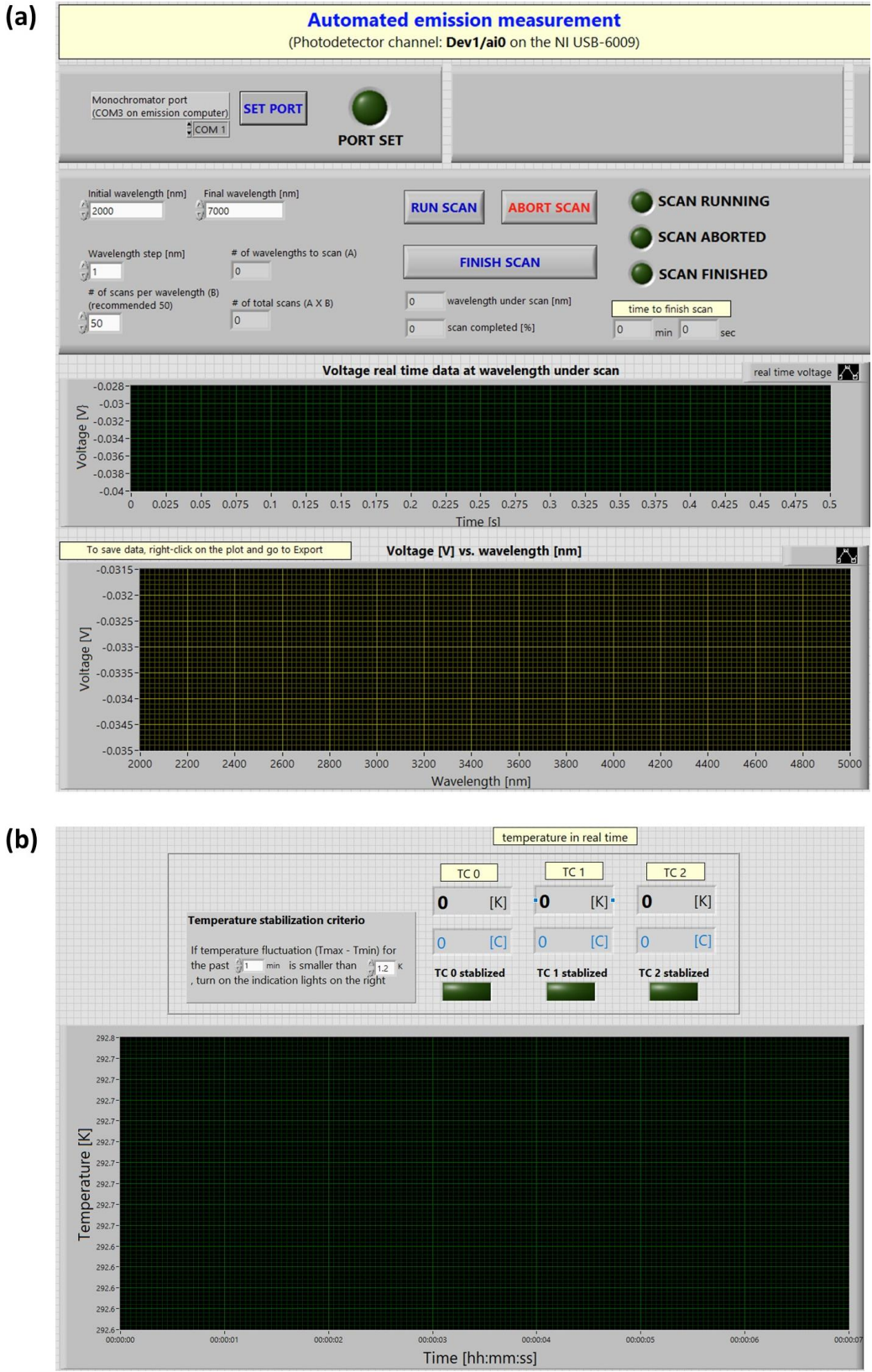

Fig. 4.4. LabVIEW interfaces for spectroscopic emission measurement (a) and temperature monitoring (b).

## 4.2. Optical Properties of Materials

### 4.2.1. Motivation

In order to model or simulate optical properties of TPV emitters, the refractive index of all constituent materials of the emitter must be known. This section reports complex-valued refractive index ($n = n' + in''$) of selected materials. The film thicknesses and optical properties were measured by a profilometer and spectroscopic ellipsometers, respectively. The difference between measured optical data and fit is represented by mean-squared error (MSE). A lower MSE indicates a more accurate fit.

### 4.2.2. Cr

**a. Refractive Index at Room Temperature**

A thin Cr layer was sputter-deposited at two different temperatures: room temperature and 400°C. The sputtering parameters were: Ar pressure 4m Torr, DC power 200W, and chamber pressure before deposition ~$2 \times 10^{-5}$ Torr. The refractive index of both Cr films was measured at room temperature, shown in Fig. 4.5. In the figure, it is evident that the imaginary parts of their refractive index, which are also called extinction coefficients, are quite different. The smaller extinction coefficients of Cr deposited at room temperature are attributed to lower density –i.e. larger content of voids [69]. Therefore, the Cr film deposited at room temperature requires a much higher thickness to be optically thick so that the transmittance is zero. The Cr film deposited at 400°C, however, can be optically thick with a thickness of ~100 nm at the measured wavelengths due to its higher density.

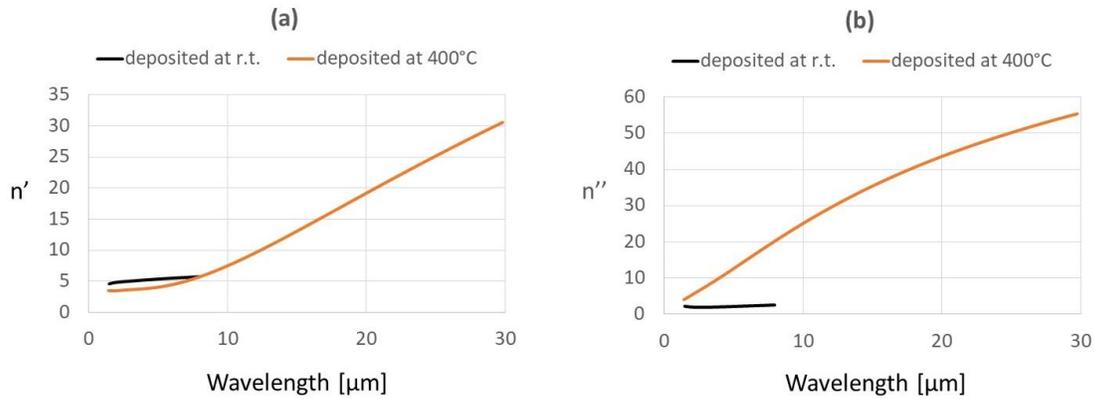

Fig. 4.5. Refractive index of sputtered Cr films, measured at room temperature. (a): real part. (b): imaginary part (also called extinction coefficient). The fit MSE was lower than 1.7 for both samples.

**b. Refractive Index at High Temperatures**

Optical properties of the Cr film that was deposited at 400°C were also measured at various temperatures between 100°C and 600°C. For ellipsometry at elevated temperatures, the heating cell (Linkam Heat Cell) provided by JA Woollam was attached to the ellipsometer. The Cr film was in $N_2$ ambient during measurements. Fig. 4.6 shows the refractive index of Cr measured at 100 – 400°C. At up to 400°C, the Cr surface was clean, smooth, and had a metallic tint. Surface oxidation was negligible since the detected chromium oxide thickness was as small as less than 0.13 nm. However, the Cr surface started to have a yellowish tint at 500°C. At 600°C, it turned purple and then blue; these observations were made by naked eyes through the window in the heating cell. The surface color changes are attributed to oxidation, which can be due to residual oxygen and/or oxygen leak into the cell [28]. Because the analysis software didn't have refractive index data of chromium oxide for the entirety of the measured wavelengths, the refractive index of Cr could not be fit accurately at 500°C and above. The oxidized Cr surface after the measurement at 600°C is shown in Fig. 4.7.

---

[28] To prevent or minimize oxidation, a higher $N_2$ pressure can be used.

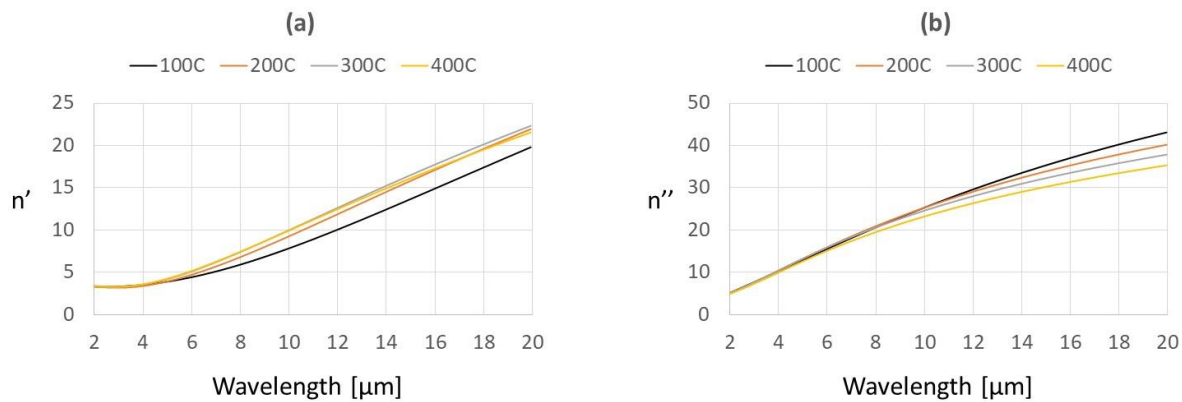

Fig. 4.6. Temperature-dependent refractive index of a Cr film sputtered at 400°C. (a): real part. (b): imaginary part. The fit MSE was lower than 2.21 for all samples.

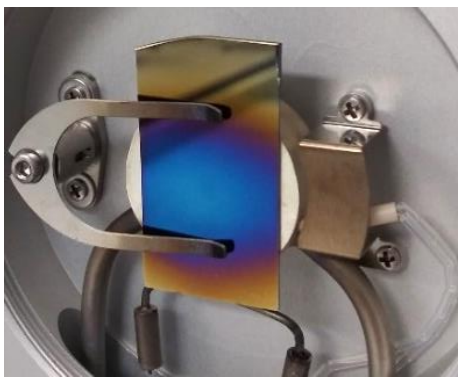

Fig. 4.7. Oxidized Cr surface after ellipsometric measurement at 600°C in $N_2$ ambient.

### 4.2.3. Ir

(The results and discussion in this section are also published in Ref. [74]) An [Ir ~220nm, Ti ~10nm] layer was deposited on a Si(100) substrate in a sputtering chamber. Ti was used as a buffer layer between the Ir and substrate. Sputtering parameters for Ir were: DC gun power 160W and Ar pressure 10 mTorr. Sputtering parameters for Ti were: DC gun power 200W and Ar pressure 4 mTorr. Argon (Ar) was the sputtering ion. The chamber pressure before Ar was introduced into the chamber was 4.3 × 10$^{-6}$ Torr. All layers were deposited at room temperature. After deposition, half the substrate was annealed at 700°C for 10 min in N$_2$ ambient. Fig. 4.8 shows atomic force microscopy (AFM) and scanning electron microscopy (SEM) images of as-deposited and annealed Ir surfaces. Their grazing incidence x-ray diffraction (GIXRD) results are shown in Fig. 4.9.

The temperature-dependent refractive index of Ir films is plotted in Fig. 4.10 –data values are provided in Appendix I. Ellipsometric measurements were performed in N$_2$ ambient, and no oxidation of Ir was detected across all measured temperatures. Fig. 4.11 shows dispersion models used to fit the optical data of each sample at room temperature – extended details are provided in Fig. A2 in Appendix I. For Drude parameters, the film resistivity measured by a four-point probe at room temperature was used as a starting point. The surface roughness of Ir films was also measured by AFM: 1.4 nm for the as-deposited and 2 nm for the annealed. To incorporate the optical effects of surface roughness, an "srough" layer was added on top of Ir when fitting the optical data in WVASE. This layer mixes Ir with voids at 50:50 volume percent, known as Bruggeman Effective Medium Approximation. The surface roughnesses by AFM were used as the thickness of the srough layers for each sample. The inclusion of Ti under Ir in the fit model did not make a difference since Ir was optically thick. The fit MSE was lower than 1.33 for all 14 measurements in Fig. 4.10.

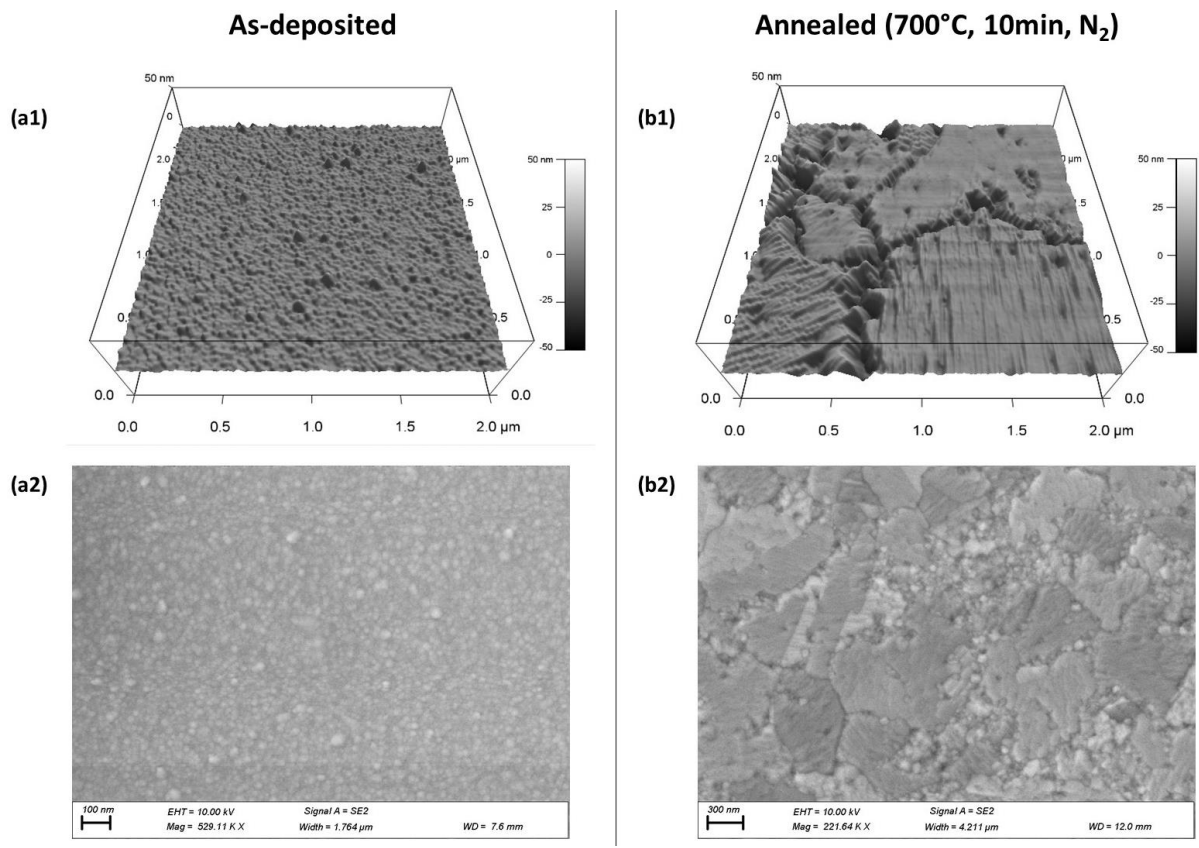

Fig. 4.8. AFM (top) and SEM (bottom) images of Ir surfaces. Larger grains are formed in the annealed sample. Annealing parameters were: 10 min at 700°C in $N_2$ ambient. These results are also published in Ref. [74] (open access).

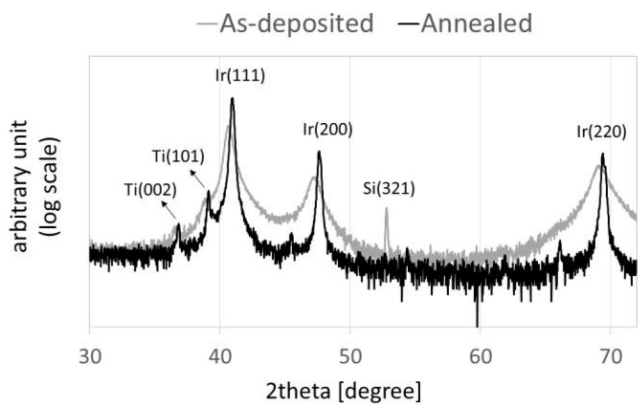

Fig. 4.9. GIXRD peaks of Ir films [29]. The sharper peaks of the annealed Ir are due to higher crystallinity (larger grains). The Si(321) peak is due to the substrate, single crystalline Si, which happened to be aligned for the peak. The XRD peaks and corresponding crystal orientations of Ir and Ti thin films are well-identified in Refs. [75] and [76]. The figure is also published in Ref. [74] (open access).

---

[29] I acknowledge John McElearney, a peer graduate student, for providing me with the XRD data based on which I was able to analyze the Ir films.

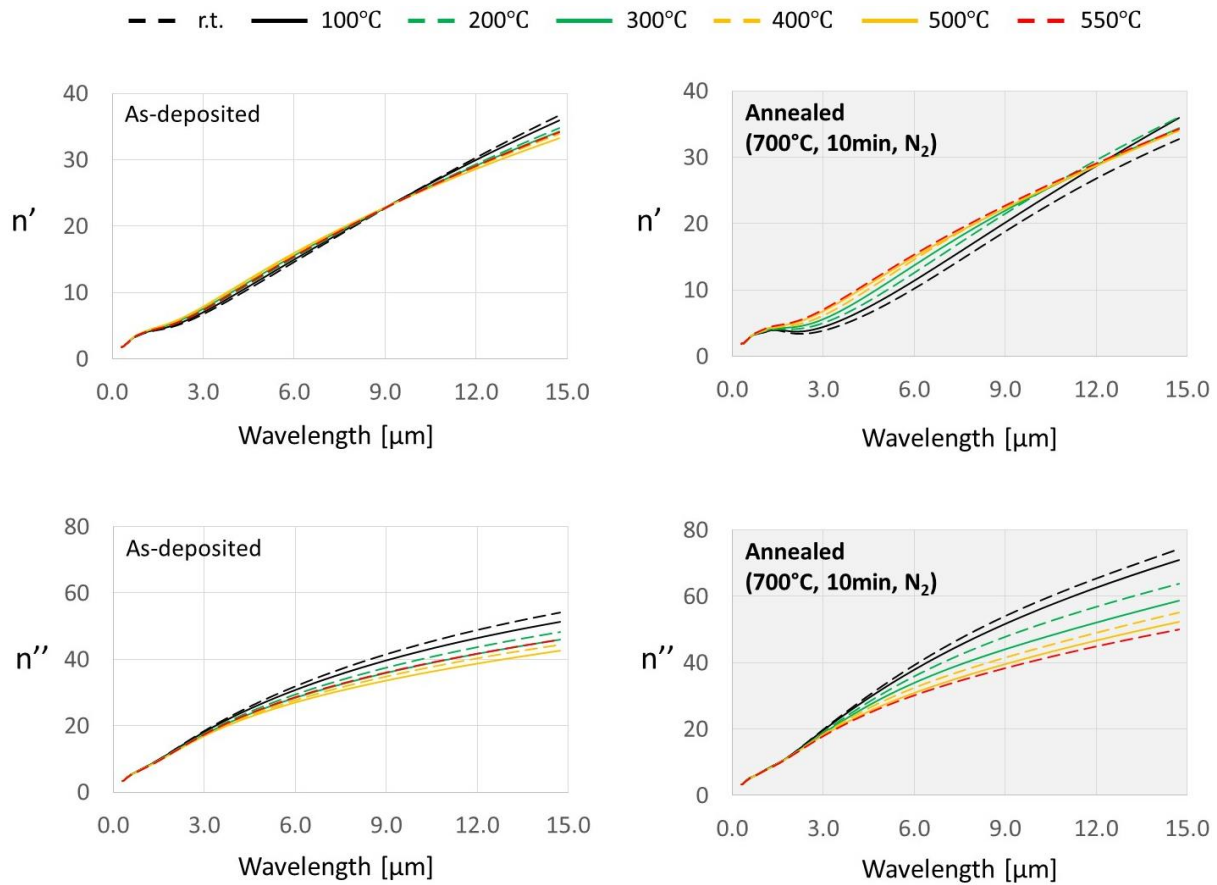

Fig. 4.10. Temperature-dependent refractive index of sputtered Ir films. Annealing parameters are written in the plot area. All Ir samples were deposited at room temperature. These results are also published in Ref. [74] (open access).

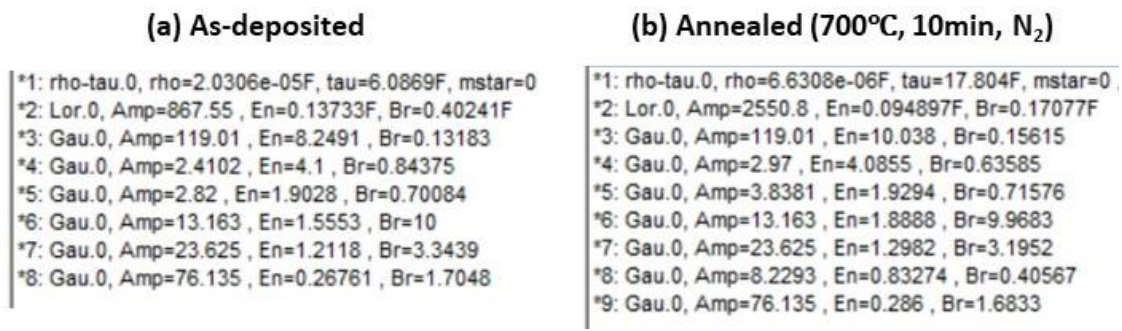

Fig. 4.11. Dispersion models used to fit ellipsometric data of Ir films at room temperature. The "rho-tau" is Drude model. The "Lor" and "Gau" are Lorentz and Gaussian models, respectively. The figure is also published in Ref. [74] (open access).

### 4.2.4. Si (undoped)

Three samples of a thin Si layer were fabricated by PECVD or sputtering. The refractive index of all samples was measured at room temperature, shown in Fig. 4.12(a), where the imaginary part is zero. The Cauchy's dispersion model was used for data fitting. Some deposition parameters and Cauchy's coefficients are shown in Table 4.1. Additional PECVD parameters were: SiH$_4$ 50 sccm, and Ar 600 sccm, and high frequency power 52W (Sample A). Additional sputtering parameters were: DC power 150W (Sample B) or DC power 200W (Sample C) and chamber pressure before deposition ~3 × 10$^{-5}$ Torr [30]. As seen in Fig. 4.12(a), Sample A has the highest refractive index, suggesting highest density. The lower refractive index of Samples B and C, which is due to voids in the films, indicates lower density. Fig. 4.12(b) shows a photograph of Sample B deposited on a silicon substrate. The surface tint is due to optical interference; the deposited Si film has a different refractive index than that of the Si substrate.

Table 4.1. Some deposition and data fit parameters for thin Si films. The Cauchy's third coefficient C is zero for all samples. Ar was the sputtering atom.

| Sample Name | Deposition Method | Deposition Temperature [°C] | Chamber Pressure [mTorr] | Cauchy's Coefficient A | Cauchy's Coefficient B | MSE | Substrate | Note |
|---|---|---|---|---|---|---|---|---|
| A | PECVD | 200 | 1000 | 3.28 | 0.3 | 45 | fused silica | |
| B | Sputter | r.t. | 8 (Ar pressure) | 1.67 | 0.08 | 3.7 | Si | Plasma didn't ignite at Ar pressure 4 mTorr |
| C | Sputter | 400 | 4 (Ar pressure) | 3.07 | 0.04 | 2.7 | Si | |

---

[30] The two sputtered samples have more than one differing parameters: sputtering power, temperature, and Ar pressure. I note that the goal of Si deposition here was not to study the effects of deposition parameters.

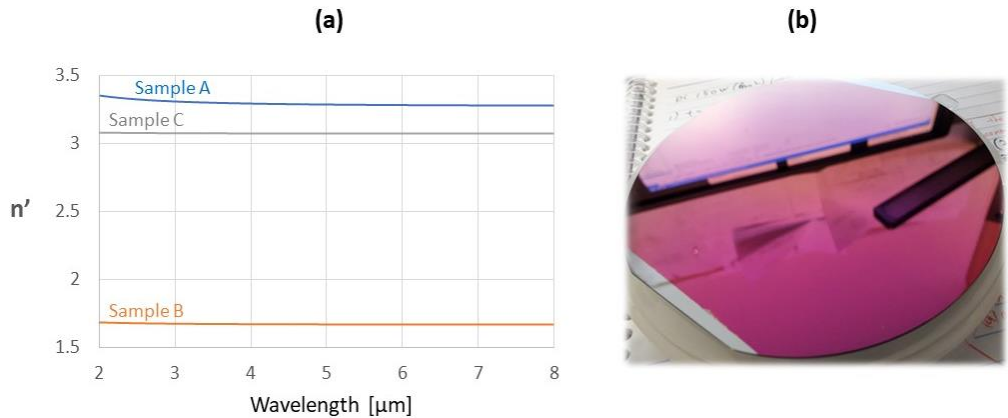

Fig. 4.12. (a): Refractive index (real part) of thin Si films in Table 4.1. The refractive index was measured at room temperature. (b): A photograph of Sample B deposited on a silicon substrate.

**4.2.5. Zr**

A Zr layer thicker than 200 nm was sputter-deposited on a silicon substrate at various temperatures up to 400°C. Deposition parameters were: DC power 200W, Ar 4mTorr, and chamber pressure before deposition ~$2.5 \times 10^{-5}$ Torr. As-deposited samples are shown in Fig. 4.13(a) – (c). For Zr films deposited at 200°C and lower, their surfaces had a metallic tint, and there was no color difference between samples. For the Zr film deposited at 400°C, however, its surface appeared in a different color, Fig. 4.13(c), which is attributed to oxidation due to residual oxygens in the chamber. This indicates that sputtering Zr at 400°C should be avoided if no oxidation is required.

The ellipsometric parameters of all Zr films were measured at room temperature, shown in Fig. 4.13(d) and (e). The films deposited at 100°C and 200°C exhibit almost identical optical constants, which are different from that of the film deposited at room temperature. This can be interpreted as that significant diffusion of Zr occurred at 100°C, as Zr has a higher diffusion coefficient [3]. The room temperature refractive indices of Zr films deposited at 100°C and 200°C are shown in Fig. 4.14.

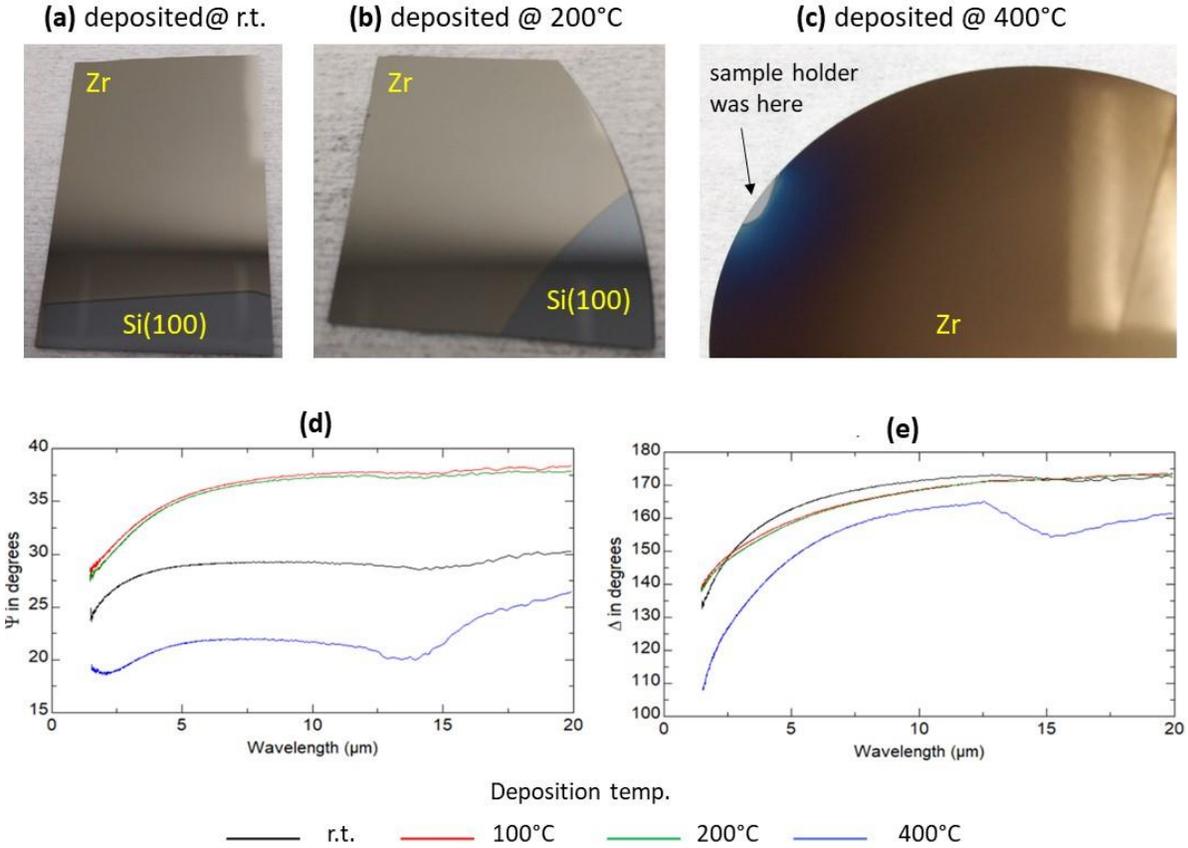

Fig. 4.13. As-deposited Zr samples. (a) – (c): photographs of Zr films. In (c), a blue tint is seen around where the sample holder was that was pressing the substrate. This color difference is attributed to oxidation; temperature is higher because the substrate makes a better contact with the substrate holder/heater due to the pressure from the sample holder. (d) and (e): ellipsometric parameters measured at room temperature (incidence angle 70°).

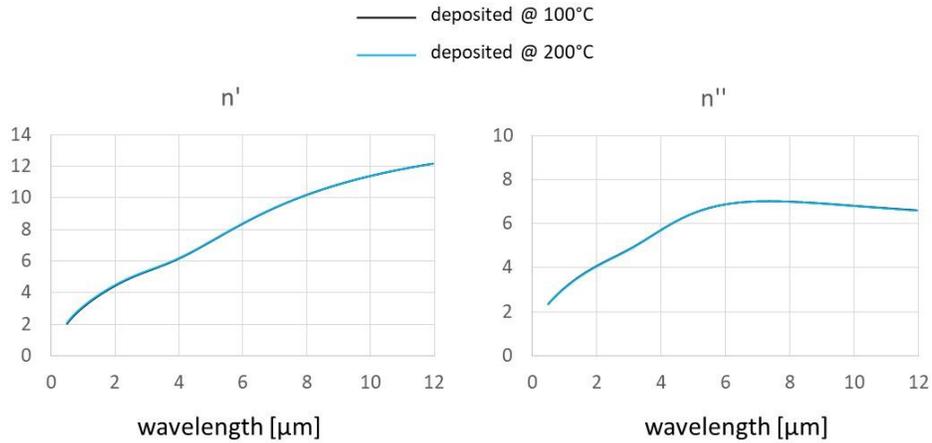

Fig. 4.14. Room temperature refractive index of sputtered Zr films. The MSE was lower than 5.4 for both samples.

## 4.3. Delamination of Thin Films

### 4.3.1. Motivation

TPV emitters must be both mechanically and chemically stable at high temperatures – i.e. no delamination or oxidation. Thin films of various materials were heated to a higher temperature to test their stability. This section reports delamination or oxidation results of the tested films.

### 4.3.2. Si (doped)

A thin layer of boron-doped p-type Si, p-Si(B), was deposited on various $SiO_2$-based substrates by PECVD. The deposition temperature was higher than room temperature. Table 4.2 shows delamination results of as-deposited films. For films deposited on a soda lime glass, the thinner layer (200 nm) did not delaminate while the thicker layer (> 500 nm) resulted in delamination. This can be interpreted as that the thicker layer caused higher stress at the interface. For all films thicker than 500 nm, the films deposited on Borofloat 33 and fused silica did not delaminate while the films on a soda lime glass and a quartz substrate delaminated. These results are attributed to larger CTE differences between the film and substrate [31]. Fig. 4.15 shows delaminated Si films at room temperature after deposition, where buckling delamination patterns are observed, due to a compressive stress in the film [77]. All samples thicker than 500 nm delaminated after annealing at a higher temperature. Fig. 4.16 shows delaminated films on a Borofloat 33 and a fused silica substrate after annealing.

Table 4.2. Delamination results of as-deposited p-Si(B) films. All films were deposited by PECVD. $CTE_{sub}$ and $CTE_{Si}$ are the coefficients of thermal expansion of the substrate and Si, respectively. $CTE_{Si} = 2.56 \times 10^{-6}$/K is used [63]. Y: yes and N: no.

| Film Thickness | Substrate | $CTE_{sub}$ [$10^{-6}$/K] | $CTE_{sub} - CTE_{Si}$ [$10^{-6}$/K] | Delaminated at Room Temperature? |
|---|---|---|---|---|
| > 500 nm | Soda lime glass | 9 [78] | 6.4 | Y |
| 200 nm | Soda lime glass | 9 [78] | 6.4 | N |
| > 500 nm | Quartz | 7.1 – 13.2 [79] | 4.5 – 10.6 | Y |
| > 500 nm | Borosilicate (Borofloat 33) | 3.25 [78] | 0.7 | N |
| > 500 nm | Fused silica | 0.55 [80] | -2 | N |

---

[31] The dominant chemical component of all substrates is $SiO_2$. For this reason, the bonding strength between the film and substrate is assumed to be similar for all samples.

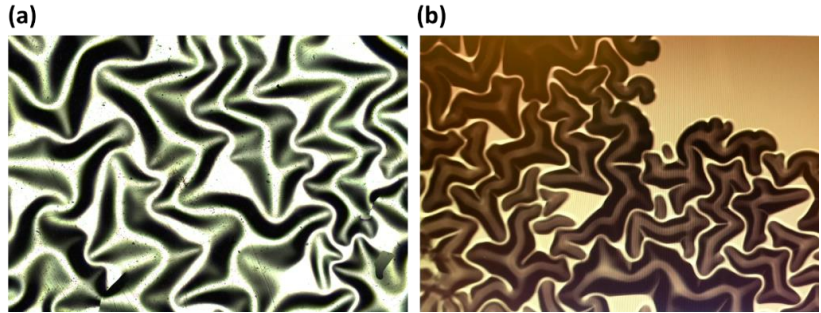

Fig. 4.15. Optical microscope images of delaminated p-Si(B) films on a soda lime glass (a) and a quartz substrate (b). Both films are thicker than 500 nm. Buckling delamination patterns are observed in both films.

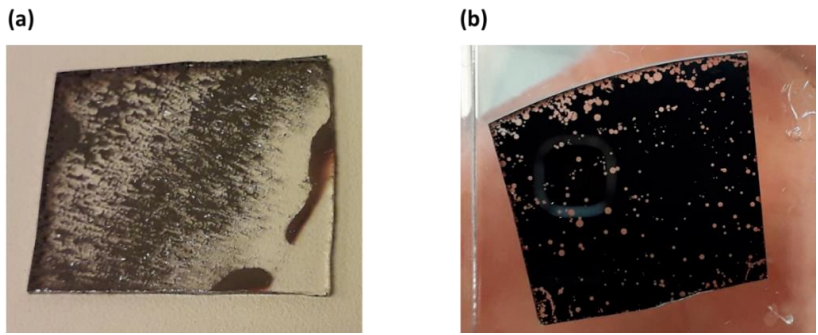

Fig. 4.16. Delaminated p-Si(B) films on a Borofloat 33 (a) and a fused silica (b) substrate after annealing.

### 4.3.3. Metals

Various metal thin films were deposited. To represent layer structures, I will write the element symbol of the top layer on the leftmost and the element symbol of the substrate material on the rightmost in brackets. Numbers in the brackets are thicknesses in nm. For example, Fig. 4.17 shows the layer structure of [Au40/Cr30/Si]. I will also denote the deposition method of each layer in parentheses within the brackets: [Au40(e-beam)/Cr30(e-beam)/Si] for example. The "e-beam" stands for electron-beam evaporator.

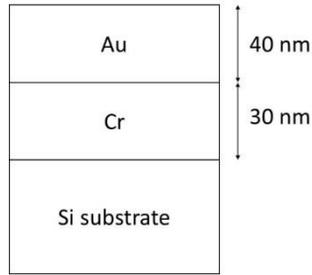

Fig. 4.17. Layer structure of [Au40/Cr30/Si].

### a. Au/Ti/Si

A [Au200(e-beam)/Ti10(e-beam)/Si] structure was deposited for a potential use of high-reflectance mirror in the emission measurement setup in Fig. 4.1. Ti was used as a buffer/adhesion layer due to a large CTE difference and poor adhesion between Au and Si [32]. However, the Au layer delaminated off the Ti after deposition, shown in Fig. 4.18. The Ti was still bonded to the substrate.

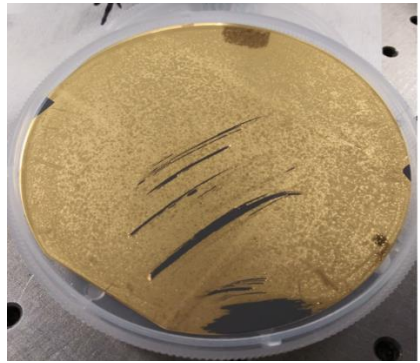

Fig. 4.18. Delamination Au in the [Au200(e-beam)/Ti10(e-beam)/Si] structure.

---

[32] Ti is known for its high toughness and elongation [114], which makes it a good buffer material.

**b. Au/Cr/Si**

A [Au40(e-beam)/Cr30(e-beam)/Si] structure was deposited, which was stable at room temperature after deposition. This structure was also thermally stable at up to 350°C, tested on a hot plate in air, maintaining a smooth surface. The surface became rough at 400°C. Due to its thermal stability at up to 350°C, [Au40(e-beam)/Cr30(e-beam)/Si] was used as a high-reflectance mirror in the emission measurement setup in Fig. 4.1.

**c. Ir/Cr/Si**

An [Ir25(sputtered)/Cr150(sputtered)/Si] structure was deposited. The structure was stable in air with a smooth surface at up to 400°C, shown in Fig. 4.19(a). The temperature ramp up rate in the heating test was as fast as ~200°C/min. At 620°C (893K), the structure did not delaminate, but the Ir surface discolored, which is attributed to oxidation [81]. The discolored Ir surface is shown in Fig. 4.19 (b).

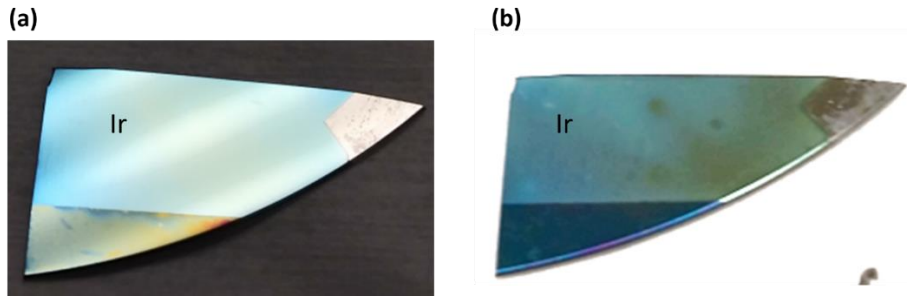

Fig. 4.19. An [Ir25(sputtered)/Cr150(sputtered)/Si] sample after heating test at 400°C (a) and at 620°C (b) in air.

**d. Ir/Si and Ir/Ti/Si**

An Ir layer, thinner than 100 nm, deposited on a silicon substrate without a buffer layer resulted in delamination after deposition, shown in Fig. 4.20(a). The delamination is due to a higher Young's modulus of Ir [3], which creates higher stresses at the interface according to Eq. 2.67. Thus, Ti was used as a buffer layer, and an [Ir240(sputtered)/Ti(a few nm, sputtered)/Si] was deposited. This structure was stable at 700°C (973K) in N₂ ambient and resulted in destruction at 1000°C (1273K) in vacuum. These results are shown in Fig. 4.20(b) and (c).

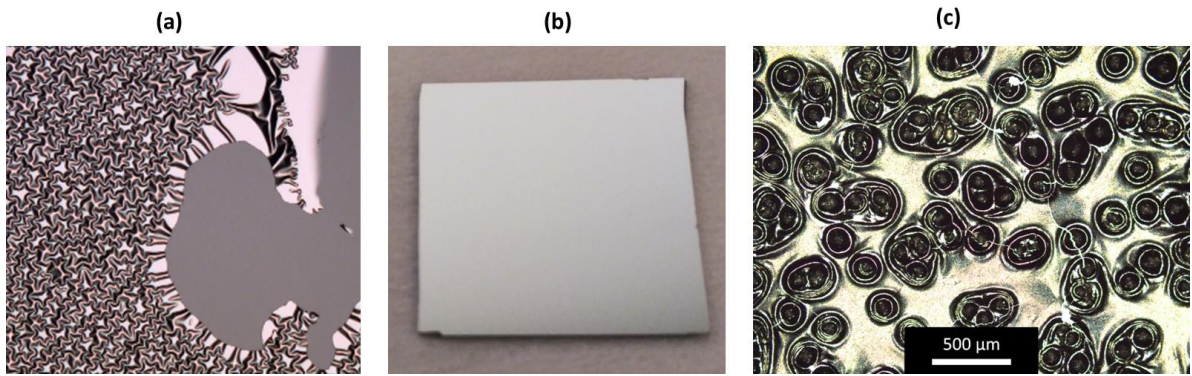

Fig. 4.20. Surfaces of Ir films. (a): An optical microscope image of an as-deposited [Ir≤100(sputtered)/Si]. (b): A photograph of an [Ir240(sputtered)/Ti(a few nm, sputtered)/Si] sample after annealing (700°C, 10min, N₂). (c): An optical microscope image of an [Ir240(sputtered)/Ti(a few nm, sputtered)/Si] sample after annealing (1000°C, vacuum). All layers were deposited at room temperature.

**e. Pt/Ti/Si**

A [Pt200(e-beam)/Ti10(e-beam)/Si] structure was deposited, where Ti is a buffer layer. This structure was deposited for its potential use as a reference emission material to normalize voltage signals from the emitter's radiation. However, the structure delaminated at 527°C (800K) in air, shown in Fig. 4.21.

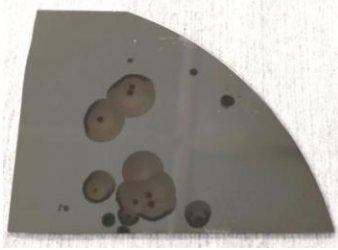

Fig. 4.21. Delamination of [Pt200(e-beam)/Ti10(e-beam)/Si].

**f. Ti/Cr/Ti/Si**

[Ti(a few nm, sputtered)/Cr185(sputtered)/Ti(a few nm, sputtered)/Si] was deposited. The structure was stable at 527°C (800K) in air. Fig. 4.22 shows the sample after the heating test, where surface tints are due to oxidation.

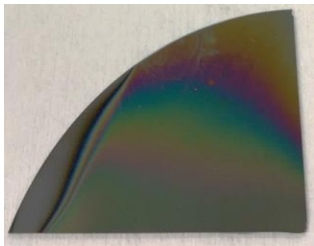

Fig. 4.22. A [Ti(a few nm, sputtered)/Cr185(sputtered)/Ti(a few nm, sputtered)/Si] sample after annealing in air at 527°C (800K).

## 4.4. Emitter: Initial Designs

### 4.4.1. Metamaterial Emitter #1 (doped Si)

**a. Motivation**

Optical responses of metamaterials (MMs) depend on their constituent materials' refractive index [82]–[84]. One can manipulate the refractive index of semiconductor materials, such as doped Si, by changing doping concentration [85]. Thus, MM emitters built with semiconductor materials provide an additional degree of freedom to engineer their emission properties. This section reports an MM emitter that is based on boron-doped p-type Si, p-Si(B).

**b. Results**

Fig. 4.23 shows simulated results of a p-Si(B)-based MM emitter at normal incidence at room temperature. The doping concentration is denoted by $N$. The refractive index of p-Si(B) was generated using an empirical equation in Ref. [85] [33], assuming Drude dispersion. The relevant dispersion parameters are:

$$\text{Eq. 4.1} \quad (a)\ \varepsilon_\infty = 11.74 \quad (b)\ \omega_p = \sqrt{\frac{Nq^2}{m_{eff}\varepsilon_o}} \quad (c)\ \gamma_D = \frac{\rho N q^2}{m_{eff}}$$

$\omega_p$ is the plasma frequency. $m_{eff}$ is the effective electron mass ($m_{eff}$ = 0.168$m_e$ is used, where $m_e$ is the electron mass). $\gamma_D$ is the damping (or collision) frequency. $\rho$ is the resistivity. The results of Fig. 4.23 indicate that absorbance (or absorptivity) of the MM emitter can be engineered by changing the doping concentration as well as structure dimensions. The selective emission is attributed to a combined resonance of electric and magnetic fields [10]. Fig. 4.24 shows fabricated cylindrical structures of p-Si(B) with similar dimensions to those in Fig. 4.23.

---

[33] Equation 3.8 in the reference is relevant.

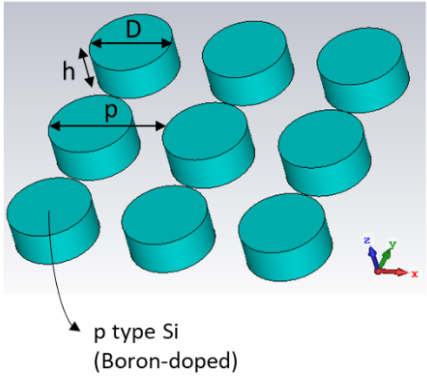
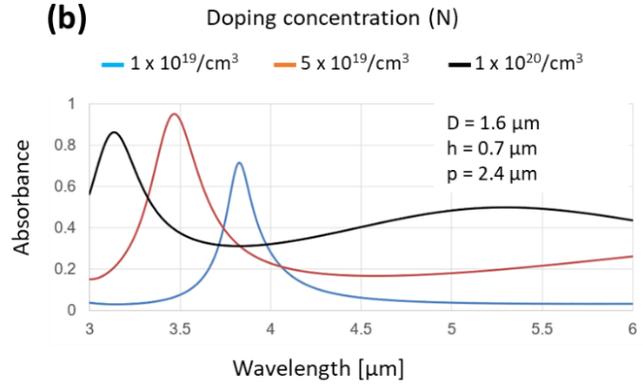
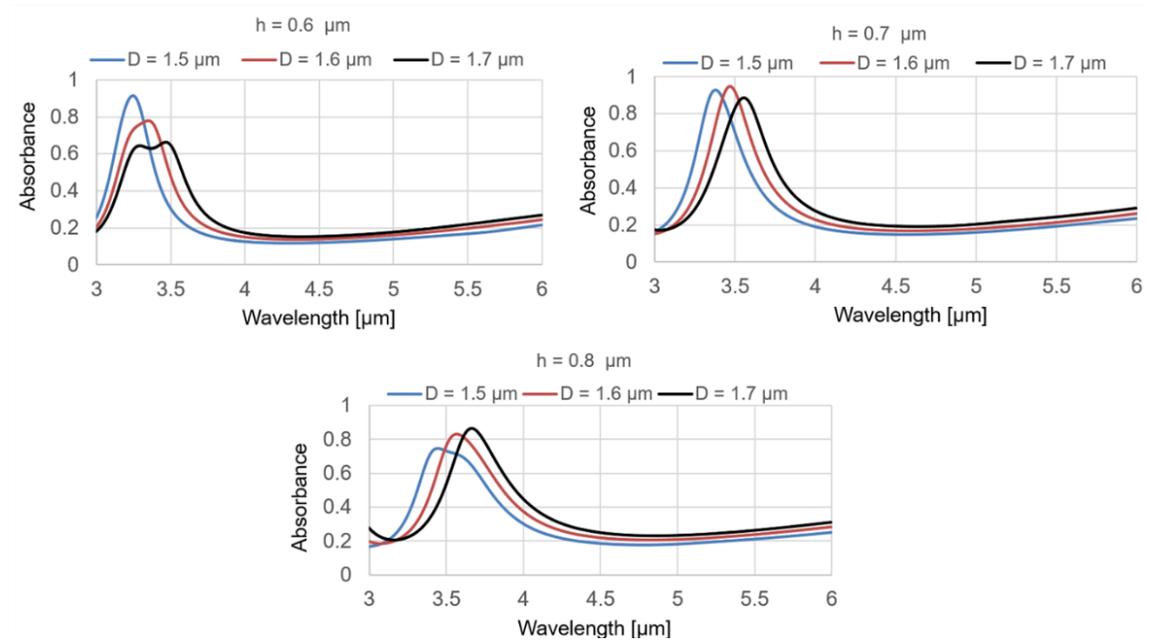

Fig. 4.23. Simulated results of the p-Si(B)-based MM emitter. (a): nine unit cells. (b): absorbance vs. doping concentration. (c): absorbance vs. unit cell dimensions. The figure is also published in Ref. [10] (open access).

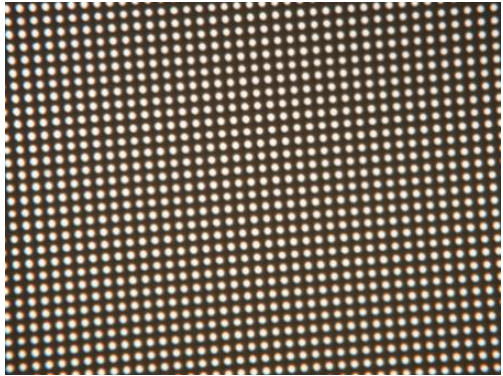

Fig. 4.24. Optical microscope image of fabricated cylindrical structures of p-Si(B) on a fused silica substrate (scale not provided) [34]. The p-Si(B) was deposited by PECVD. The doping concentration was much lower than $1 \times 10^{19}/cm^3$.

**c. Challenges**

The emitter can be fabricated using photolithography, as shown in Fig. 4.24. However, it was difficult to dope boron at $1 \times 10^{19}/cm^3$ or higher using PECVD. Another challenge was delamination of the p-Si(B) layer when it was heated for carrier distribution.

**4.4.2. Metamaterial Emitter #2 (Ir and HfO$_2$)**

**a. Motivation**

Ir and HfO$_2$ have a higher melting point and good oxidation resistance [3]. This makes them a good material candidate for TPV emitters. This section reports a MM emitter based on Ir and HfO$_2$.

---

[34] I acknowledge Emily Carlson, a peer graduate student, for guiding me on fabrication in the cleanroom.

### b. Results

Fig. 4.25 shows simulated results of the emitter at normal incidence at room temperature. It is observed that the absorbance can be engineered by changing the width ($w$) of the top Ir and HfO$_2$ layers. The absorption peak is due to a plasmonic resonance between Ir and HfO$_2$ layers [67], [86]. The resonance does not occur when the top layers' width is equal to the period of the structure; $w = p$. This is because the top Ir layer covers the whole surface, and the emitter becomes just like a mirror.

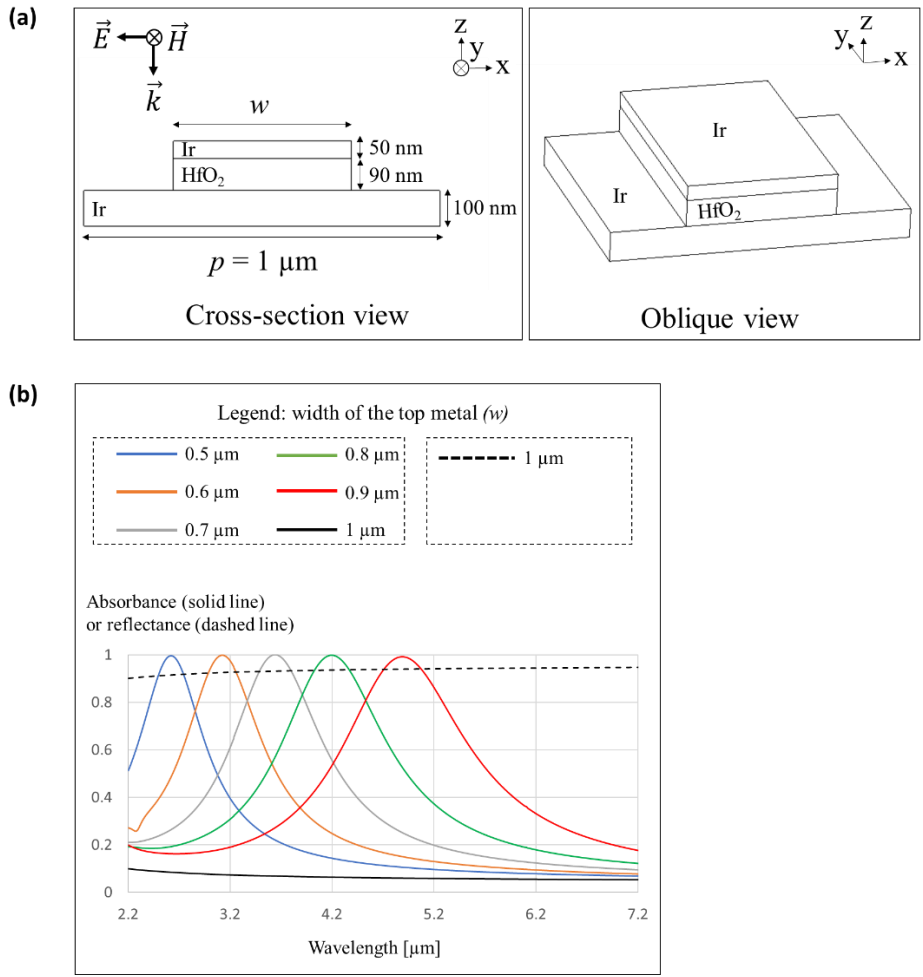

Fig. 4.25. Simulated results of the Ir and HfO$_2$-based MM emitter. (a): emitter unit cell and polarization of the incident light. $p$ is the period. (b): absorbance vs. top metal width ($w$). The figure is also published in Ref. [67].

c. Challenges

The emitter in Fig. 4.25 can be fabricated by electron beam lithography. However, it is generally highly time-consuming to create surface patterns on a larger area (>> 1 cm$^2$) using such process. This came as a challenge because large-area emitters are needed to produce higher TPV power output –a typical TPV power output density is on the order of 1W/cm$^2$ [4], [5], [66], [87].

### 4.4.3. 1D Emitter (undoped Si, Ir, and Cr)

a. Motivation

Since TPV power output is proportional to emitter size [88], large-area emitters are required to produce higher power output. Compared to MM emitters, emitters without two/three-dimensional surface patterns, as introduced in Section 1.2.2, can be fabricated over large areas easily. This section reports large-area emitters without such surface patterns. This type of emitters will be referred to as "1D emitters" in this work.

b. Results

Fig. 4.26 shows the emitter structure and its absorbance spectrum measured at room temperature. The Si (undoped) was deposited by PECVD. The Ir and Cr were sputtered at room temperature. The selective absorption properties are due to optical interference effects between Si and metals [33]. The peak-absorption (or peak-emission) wavelengths do not vary much with incidence angle. This suggests that the emitter can potentially maximize TPV power output by having photovoltaic cells surrounding it.

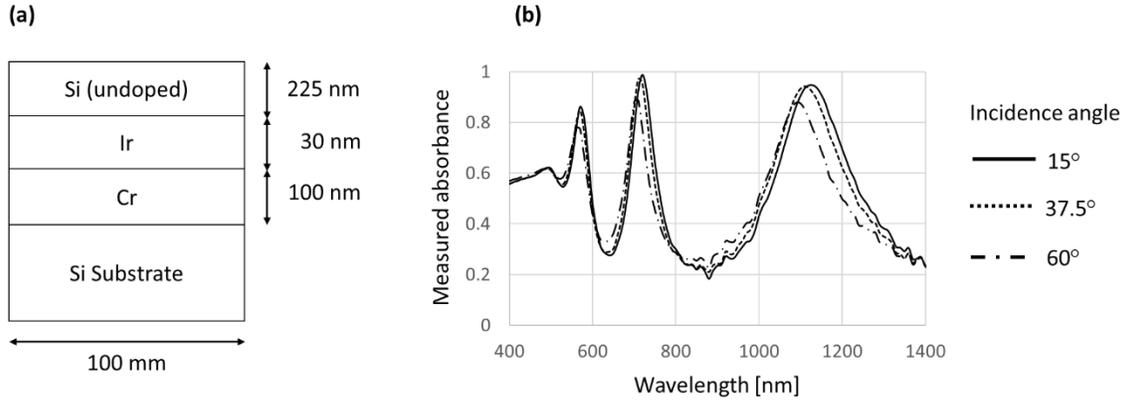

Fig. 4.26. 1D emitter (initial design). (a): Schematic cross-section. (b): Absorbance measured at room temperature with unpolarized light. The figure is also published in Ref. [33] (reprinted with permission).

c. Challenges

There were two challenges associated with the emitter design in Fig. 4.26. The first is that its peak-emission wavelengths are in the visible to near-infrared range. To maximize radiation energy from the emitter, the emitter needs to be at higher than 2600K. For example, the radiation energy from a blackbody peaks at the wavelength of 1.1 µm at 2635K, according to Eq. 2.16. This was a problem because the melting point of Si is below 2635K. Another challenge was delamination of the Si layer when heated up.

## 4.5. Emitter: Final Design

**4.5.1. Motivation**

To overcome challenges of Initial Design emitters in Section 4.4, the 1D emitter structure in Section 4.4.3 was chosen for further study. Fig. 4.27 shows the Final Design emitter. To prevent delamination of Si, various approaches were made. First, Ir was removed since it causes higher stresses at the Si/Ir interface due to its higher Young's modulus. Second, Ti was used as a buffer layer between Si and Cr. Lastly, Si was deposited by sputtering instead of PECVD. All layers were sputtered at 400°C for higher material density. A higher density is needed because if the material density changes during TPV operation due to annealing effects, then it may lead to changes in materials' refractive index as well as emitter's emissivity. Also, Si was deposited to a higher thickness, 740 nm, to bring the emissivity peaks farther out to the mid-infrared so that the emitter operates at a temperature much lower than Si's melting point.

**4.5.2. Emissivity**

Fig. 4.27(c) shows emissivity measured at 802K at normal incidence in air. The emitter was at 802K for less than 70 min during the measurement. Details about how the radiation was normalized are given in Appendix J. The emissivity peak is at the wavelength of ~3.75 µm. Theoretical emissivity curves were generated in WVASE based on the measured refractive index of each layer at near 802K. The theoretical emissivity curve in blue corresponds to the emitter before oxidation. The theoretical emissivity curve in red

corresponds to the emitter after oxidation. The difference between the two theoretical curves suggests that a 40 nm of Si is consumed to oxidation.

The bandwidth of the measured curve is wider than that of the theoretical curve in red. <u>This may be because of shorter optical coherence length and/or ununiform surface oxidation.</u> The coherence length of thermally radiated waves is smaller, but the ellipsometer software generates theoretical curves assuming infinitely long coherence length. The ununiform oxidation of the emitter surface, due to temperature gradient, which will be discussed in Fig. 4.30, could also have produced a difference in bandwidth.

Furthermore, Fig. 4.28 shows emissivity measured at higher than 802K. It is evident that the emissivity spectrum shifts to shorter wavelengths as temperature increases. This is because of the thinning of the top Si layer due to oxidation. Also, the spectrum bandwidth tends to increase as temperature increases. This can be attributed to ununiform surface oxidation.

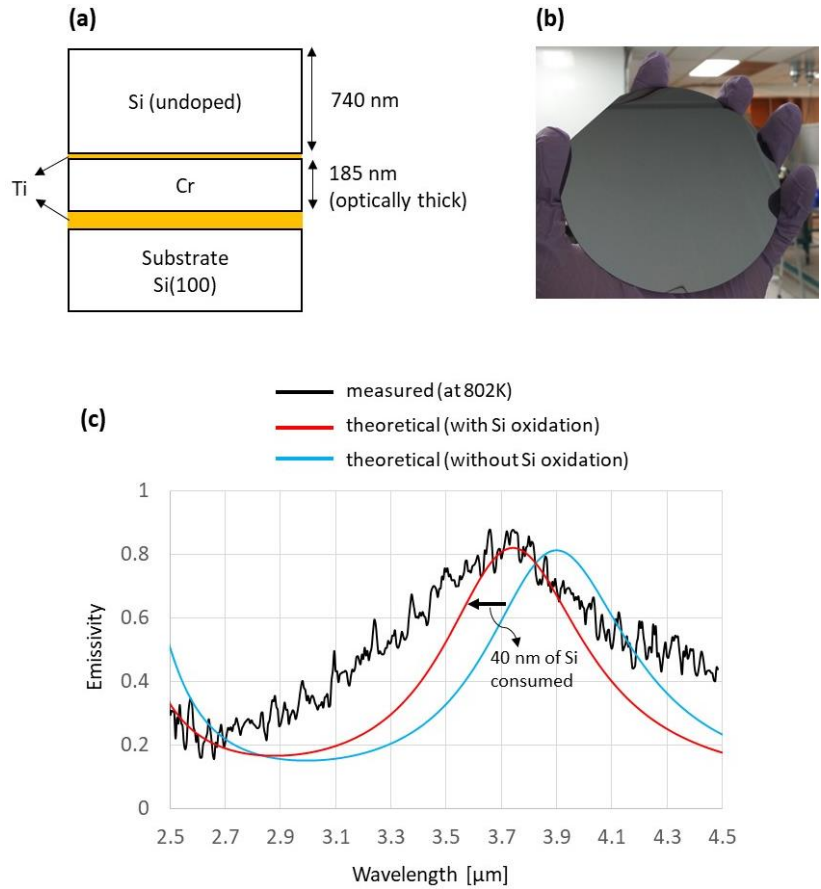

Fig. 4.27. Results of Final Design emitter. (a): schematic cross-section of an as-fabricated emitter. The thickness of the upper Ti layer is a few nm, deposition time 1 min. The lower Ti is 5 times thicker than the upper Ti, deposition time 5 min. (b): a photograph of an as-fabricated emitter. (c): emissivity measured at 802K at normal incidence in air. 802K was the temperature at the center of the emitter's top surface. The figure is also published in Ref. [88] (open access).

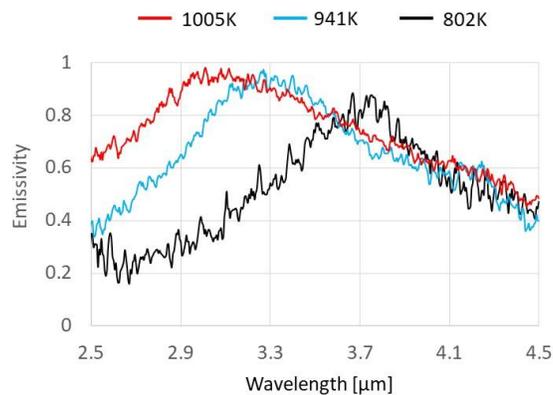

Fig. 4.28. Emissivity of Final Design emitter measured at higher than 802K at normal incidence in air. The temperatures in the legend refer to the temperature at the center of the emitter's top surface. The figure is also published in Ref. [89].

### 4.5.3. Spectral Efficiency

The spectral efficiency of a TPV emitter ($\eta_{SP}$) is given by [90]:

$$\text{Eq. 4.2} \qquad \eta_{SP} = \frac{\int_0^{\lambda_c} \epsilon_{emitter} M_{\lambda,bb}\, d\lambda}{\int_0^{\infty} \epsilon_{emitter} M_{\lambda,bb}\, d\lambda}$$

$$= \frac{\int_0^{\lambda_c} M_{\lambda,emitter}\, d\lambda}{\int_0^{\infty} M_{\lambda,emitter}\, d\lambda}$$

, where $\lambda_c$: cutoff wavelength, $\epsilon_{emitter}$: (spectral) emissivity of the emitter, $M_{\lambda,bb}$: spectral exitance of a blackbody, and $M_{\lambda,emitter}$: spectral exitance of the emitter. The measured wavelength range in Fig. 4.27(c) was not wide enough to precisely compute the spectral efficiency based on Eq. 4.2. For this reason, theoretical emissivity of a non-oxidized emitter was used for calculation, shown in Fig. 4.29(a). The normalized spectral exitance of the non-oxidized emitter at 802K is also shown in Fig. 4.29(b), where the spectral efficiency is 56%. Note that 4.26 μm was chosen as the cutoff wavelength ($\lambda_c$).

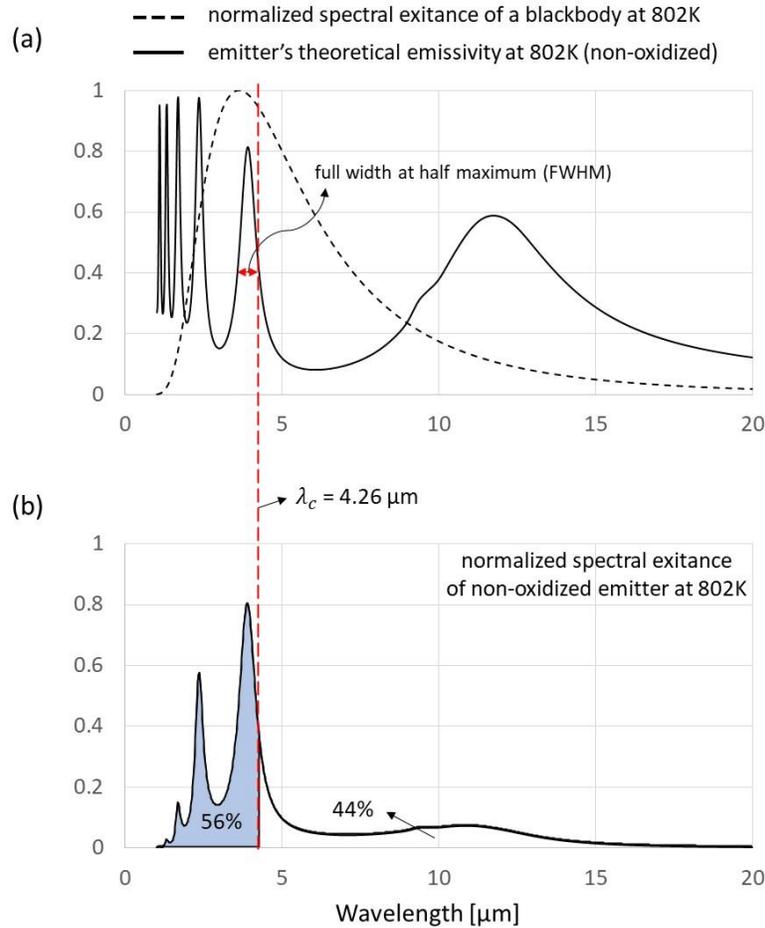

Fig. 4.29. Normalized radiation properties of a blackbody and a non-oxidized Final Design emitter.

### 4.5.4. Discussion on Thermal Stability

To further characterize surface oxidation of the emitter, an as-fabricated sample was annealed in air for 3h 14min. During annealing, the emitter was at 773K (at the center of the top surface) and 733K (at the edge). The emitter surface after annealing is shown in Fig. 4.30(a), where rings with different colors are seen. This is because of different oxidation rates of Si due to temperature gradient –see the heater in Fig. 4.2. After annealing, absorbance was measured at each ring at room temperature by a spectroscopic ellipsometer. The absorbance measurement spots are shown in Fig. 4.30(b), and the measured absorbance in Fig. 4.30(c). The spectrum at Spot 1 is located at shorter wavelengths than those at Spot 2 and Spot 3. Likewise, the spectrum at Spot 2 is at shorter wavelengths than that of Spot 3. Fig. 4.30(d) shows schematic cross-sections of the oxidized emitter at each Spot. The annealing temperature and oxide thickness for each Spot are shown in Table 4.3.

The results of Fig. 4.30 and Table 4.3 suggest significant Si oxidation. The oxidation rate is observed to be much faster than that of bulk Si in typical wet oxidation processes: where it is only a few nm of Si that would be consumed to oxidation at ~800K ~3 hours [91]. The accelerated Si oxidation may be due to tensile stress, also see Appendix K, and lower film density. Tensile stress is created in the Si layer because Cr and Ti have a larger CTE than the Si's. The tensile stress opens Si's atomic lattices and helps oxygen atoms diffuse more easily. Moreover, sputtered Si films have a lower density than bulk Si's due to a larger content of voids –discussion on Sample C in Fig. 4.12 is relevant. Since oxygen molecules can move more easily through voids, it can cause faster oxidation.

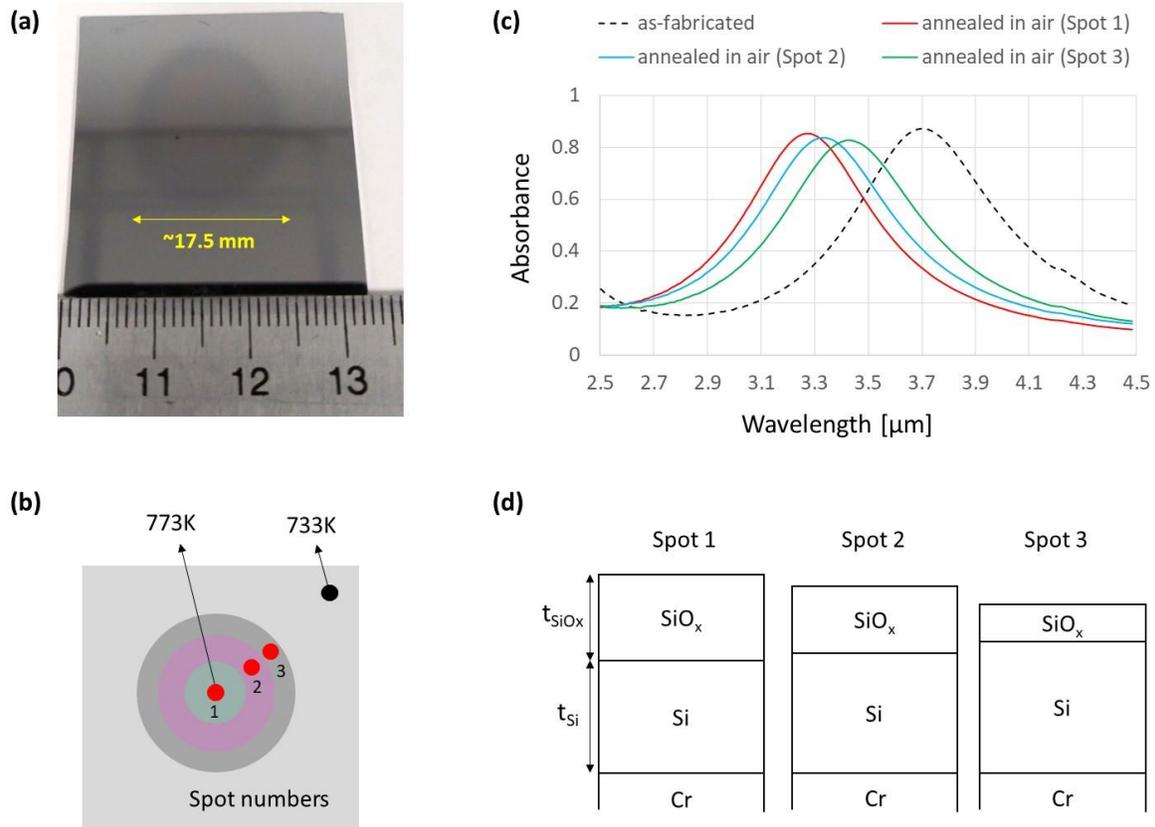

Fig. 4.30. Results of the Final Design emitter after annealing in air. (a): A photograph of the annealed emitter. The grid patterns on the surface are a reflection of the ceiling. (b): Absorbance measurement spots. Annealing temperatures were higher in the order of: Spot 1 (773K) > Spot 2 > Spot 3 > edge (733K). The ring colors correspond to those in (a). (c): Absorbance for unpolarized light measured at room temperature at near normal incidence (32°). (d): Schematic cross-sections of the oxidized emitter (Ti adhesion layers are not drawn). The figure is also published in Ref. [89].

Table 4.3. Emitter oxidation at different Spots. The thickness $t_{SiOx}$ includes both native and thermally grown oxides. $t_{Si}$ is the thickness of the remaining Si. The ellipsometric data were fit assuming a typical ratio between the thicknesses of consumed Si and thermally grown oxide; 0.44 (consumed Si) : 1 (silicon oxide).

|  | Annealed Emitter (Spot 1) | Annealed Emitter (Spot 2) | Annealed Emitter (Spot 3) | As-fabricated Emitter |
|---|---|---|---|---|
| Annealing Temperature | 773K | lower than Spot 1's | Lower than Spot 2's higher than 733K | - |
| $t_{SiOx}$ | 257 nm | 218 nm | 180 nm | 7 nm (native oxide) |
| $t_{Si}$ | 630 nm | 647 nm | 664 nm | 740 nm |
| Fit MSE | 6.7 | 3.5 | 4.4 | 2 |

Furthermore, another as-fabricated emitter sample was annealed at 802K in $N_2$ ambient to compare with the sample annealed in air. Grazing incidence x-ray diffraction (GIXRD) and ellipsometric analysis were performed after annealing. Fig. 4.31 shows GIXRD and ellipsometric results of as-fabricated and annealed samples in $N_2$ ambient. It is evident that there is no appreciable difference between the samples, suggesting that the emitter is thermally stable in $N_2$ (or nonoxidizing) ambient. Fig. 4.32 shows optical microscope images of three emitter samples; as-fabricated, annealed in air, and annealed in $N_2$. The surface of the annealed sample in air is more porous although its annealing temperature, ~733K, was lower than that of the one annealed in $N_2$ ambient, 802K.

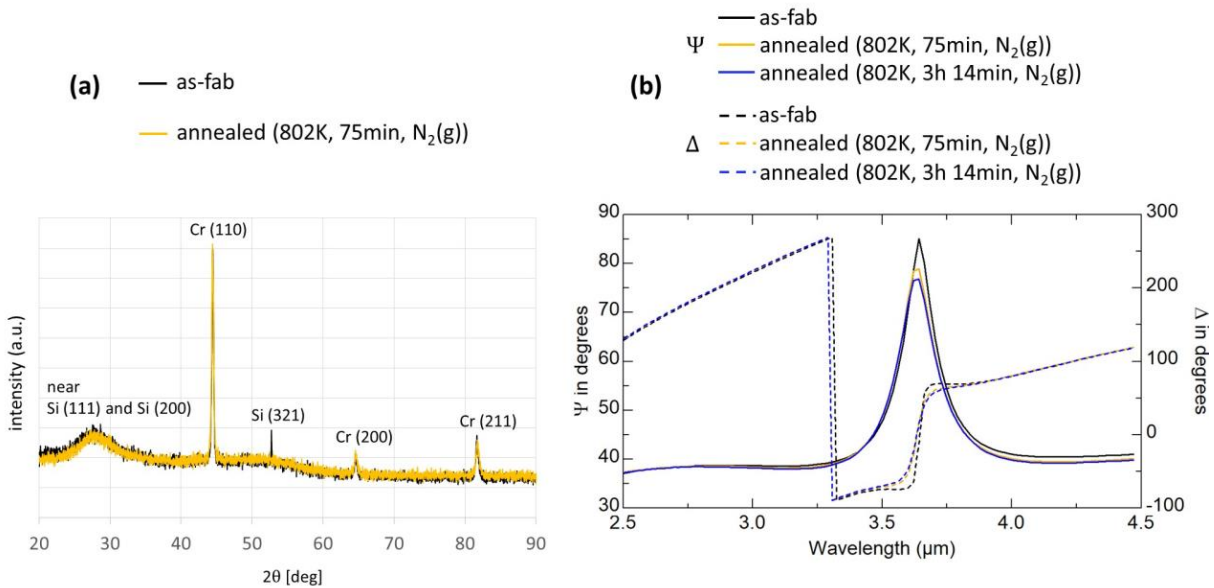

Fig. 4.31. Grazing incidence x-ray diffraction (GIXRD) (a) and ellipsometric (b) results of as-fabricated and annealed samples. Annealing parameters are notated in the legend. For example, "802K, 75min, $N_2(g)$" means the sample was annealed at 802K for 75min in $N_2$ ambient. Both GIXRD and ellipsometric measurements were performed at room temperature. Based on XRD results, Cr is crystalline, and Si is amorphous/nanocrystalline. The XRD peak corresponding to Si (321) is from the substrate, which was a single-crystalline Si wafer. The Si(321) peak appears only in the as-fabricated sample because the sample happened to be aligned for the peak when it was mounted in the instrument. The incidence angle of light was 70° for ellipsometric measurements.

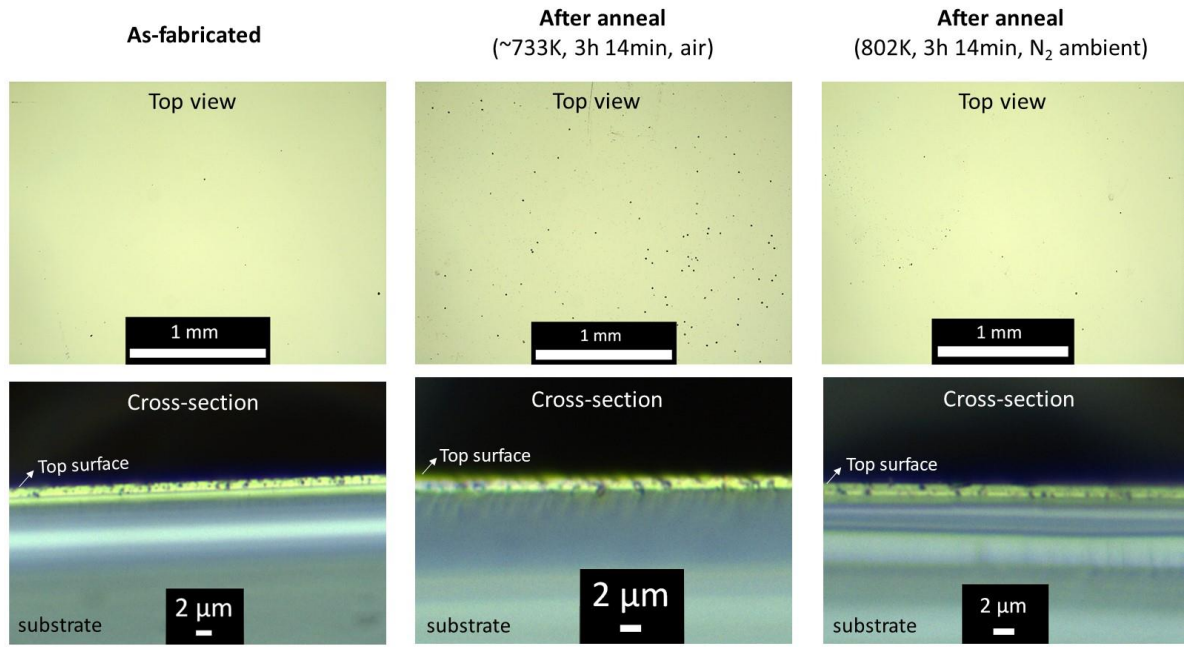

Fig. 4.32. Optical microscope images of as-fabricated and annealed emitter samples. Annealing parameters are shown on top of images.

## 4.6. Extended Discussion of Final Design Emitter: Effects of Surface Coating

### 4.6.1. Motivation

For practical deployment of TPV emitters in air, the emitters must be resistant to oxidation. However, the emission spectrum of Final Design emitter in Section 4.5 was severely impacted by oxidation. One approach I took to improve its oxidation stability was to coat the structure with protective materials. This section reports thermal stability of the emitter with various surface coatings.

### 4.6.2. Results

A similar emitter to Final Design emitter in Section 4.5 was fabricated. The stacking order of layers in this emitter was the same as Final Design emitter, but each layer was slightly thinner [35]. The as-fabricated emitter was cleaved into smaller pieces for surface coating with different materials. The thickness of the coating materials was chosen such that it doesn't significantly impact the emissivity spectrum; based on simulation in WVASE. After coating, each sample was annealed at ~808K for 1hr in air using the setup in Fig. 4.1.

Table 4.4 provides information about the coating layers and evaluates surface quality of each sample after annealing. Fig. 4.33 shows surfaces of coated samples before and after annealing. The samples that have a $Al_2O_3$ or $HfO_2$ coating on top of the Si layer maintained a mirror-like surface after annealing; evaluated by naked eyes. The samples without a $Al_2O_3$ or $HfO_2$ coating did not maintain mirror-like surface. Fig. 4.34 shows surface cracks and pores in the uncoated and 1150 nm $SiO_2$-coated emitters after annealing. Fig. 4.35 shows microstructures of the [1150 nm $SiO_2$/20 nm $Al_2O_3$]-coated emitter after annealing. The surface of this sample appears to be smooth without pores or cracks. This suggests that the 20 nm $Al_2O_3$ (ALD-deposited) between $SiO_2$ and Si helps the structure maintain smooth surface. However, even a $SiO_2$ coating as thick as 1150 nm did not prevent oxidation of Si, as can be seen in Fig. 4.33 and Fig. 4.36.

---

[35] This emitter was intended to be an exact reproduction of Final Design emitter. Therefore, sputtering parameters that were used for Final Design emitter were used. However, deposited layers came out of the chamber with ~80% of the intended thicknesses. This can be due to changes in the shape of the sputter targets, as this emitter was fabricated a year after the Final Design emitter was fabricated.

Table 4.4. Coated emitter samples. The surface quality after annealing (~808K, 1hr, air) was evaluated by naked eyes. The surface color changes are attributed to Si oxidation. r.t. is room temperature. Y: yes and N: no.

| Coating Material | Thickness | Deposition Method | Surface Color Changed after Anneal? | Surface Mirror-like after Anneal? |
|---|---|---|---|---|
| No coat | - | - | Y | Y [a] |
| $Al_2O_3$ | 20 nm | ALD | Y | Y |
| $WO_x$ | 20 nm | Sputtered @ r.t. | Y | N |
| $HfO_2$ | 20 nm | ALD | Y | Y |
| Au | 5.5 nm | e-beam @ r.t. | Y | N |
| $SiO_2$ | 1150 nm | PECVD | Difficult to tell | N |
| $SiO_2/Al_2O_3$ | $SiO_2$:1150 nm (top) $Al_2O_3$: 20 nm | $SiO_2$: PECVD $Al_2O_3$: ALD | Y | Y |

a: Surface was still mirror-like on the area where the annealing temperature was ~808K. However, it was not mirror-like on the area where the temperature was slightly higher, due to temperature gradient. This is seen in Fig. 4.33.

**Before anneal**

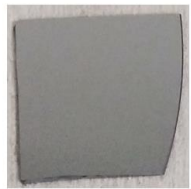
as-fabricated emitter

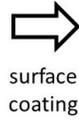
surface coating

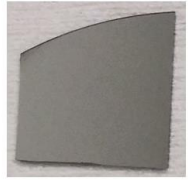
20nm Al$_2$O$_3$

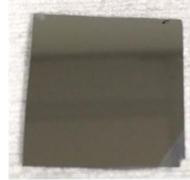
20nm WO$_x$

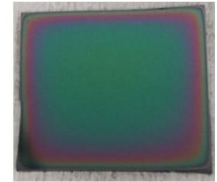
1150nm SiO$_2$

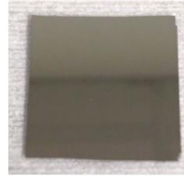
20nm HfO$_2$

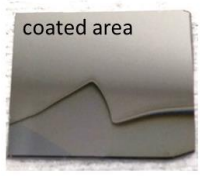
coated area
5.5nm Au

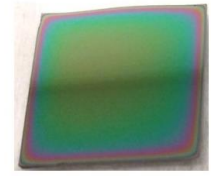
1150nm SiO$_2$
/20nm Al$_2$O$_3$

**After anneal**
(~808K, 1hr, air)

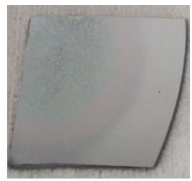
no coat

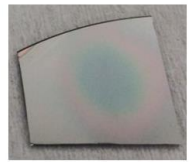
20nm Al$_2$O$_3$

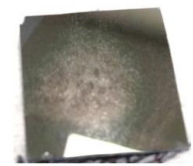
20nm WO$_x$

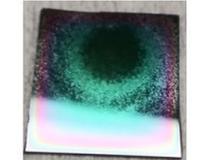
1150nm SiO$_2$

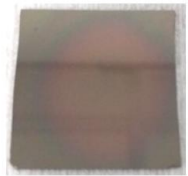
20nm HfO$_2$

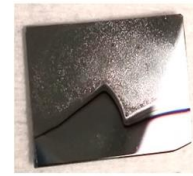
5.5nm Au

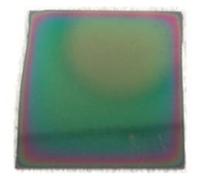
1150nm SiO$_2$
/20nm Al$_2$O$_3$

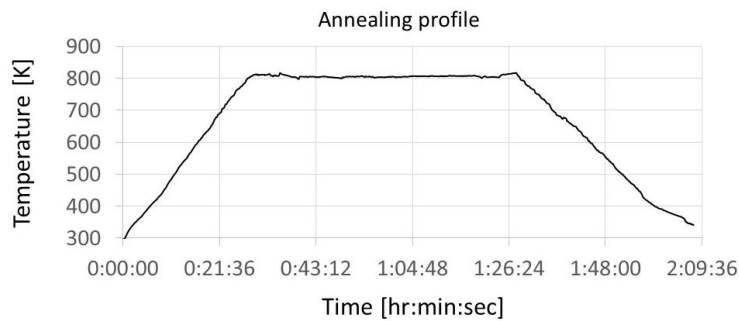

Fig. 4.33. Uncoated and coated emitters' surfaces before and after annealing (~808K, 1hr, air). The coating thickness and material are shown under each picture. Surfaces may appear in different colors depending on lighting or viewing angle.

## (a) Uncoated emitter

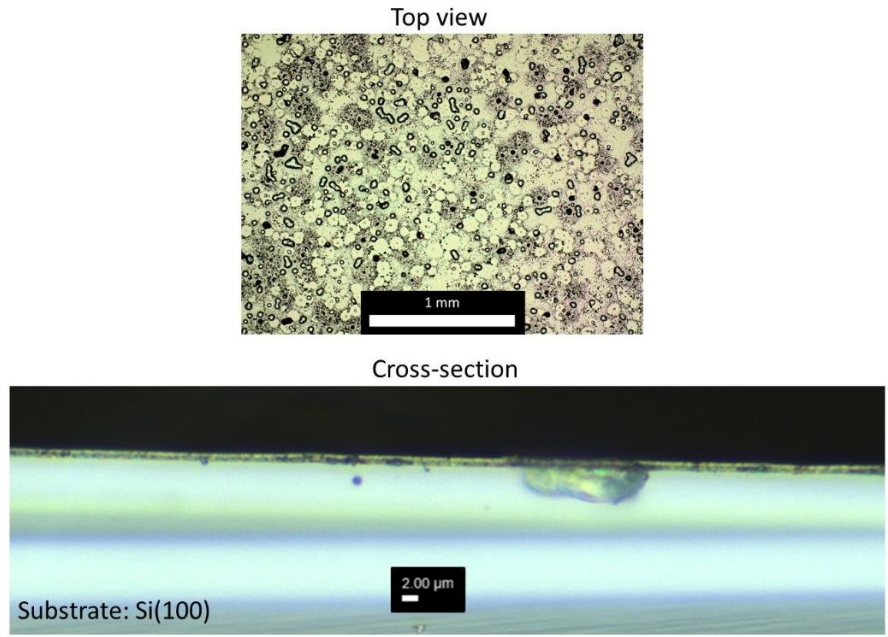

## (b) 1150 nm SiO$_2$-coated emitter

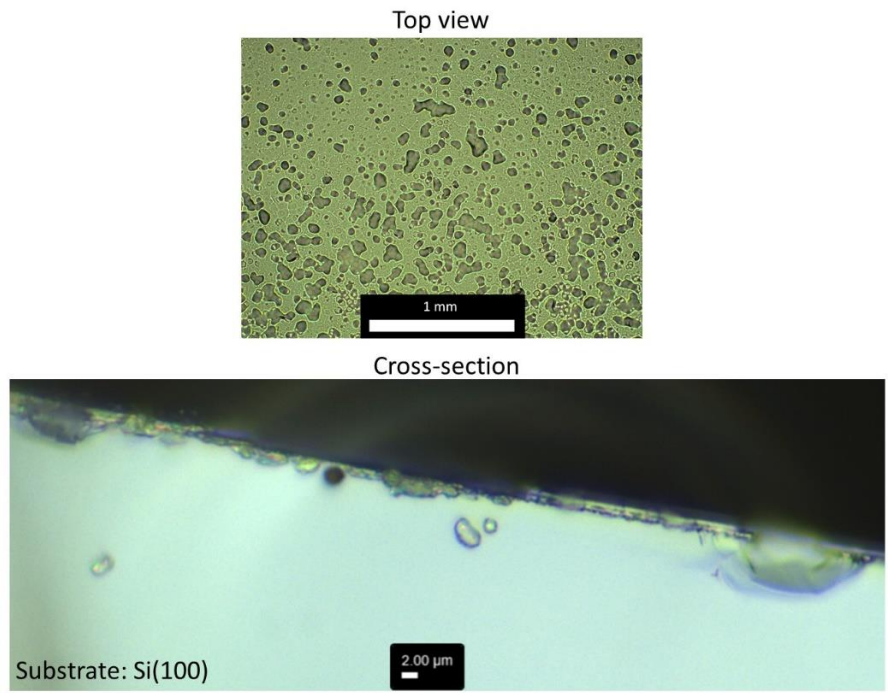

Fig. 4.34. Optical microscope images of the uncoated emitter and a 1150 nm SiO$_2$-coated emitter after annealing (~808K, 1hr, air).

## [1150 nm SiO₂/20 nm Al₂O₃]-coated emitter

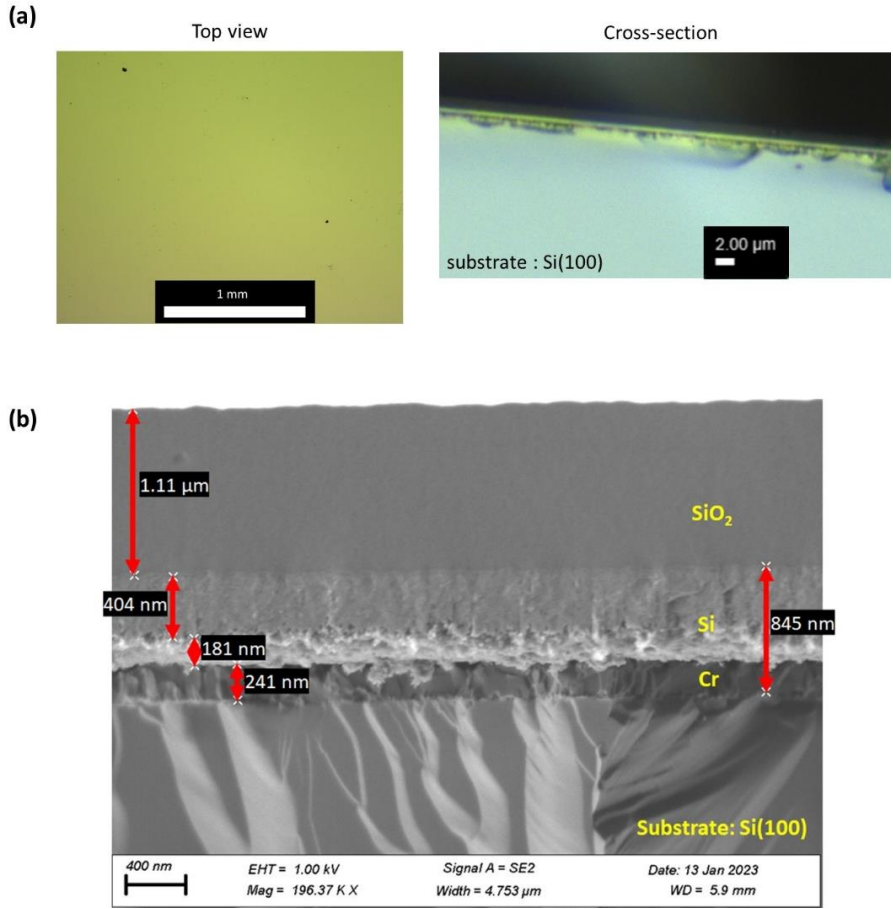

Fig. 4.35. Microstructure of the [1150 nm SiO$_2$/20 nm Al$_2$O$_3$]-coated emitter after annealing (~808K, 1hr, air). (a): Optical microscope images. (b): SEM image.

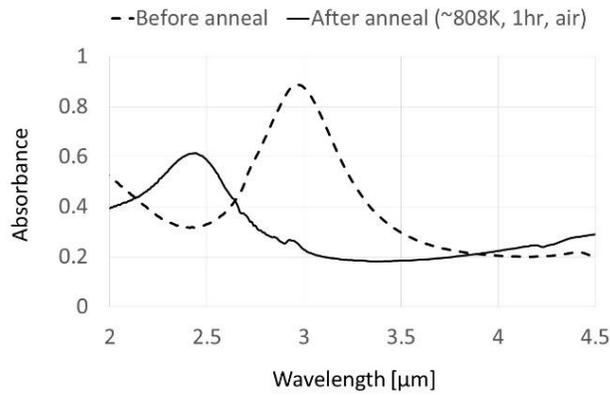

Fig. 4.36. Absorbance of the [1150 nm SiO$_2$/20 nm Al$_2$O$_3$]-coated emitter before and after annealing (~808K, 1hr, in air). The absorbance was measured at room temperature at near-normal incidence (32°).

# 5. CONCLUSION

1) In this work, TPV emitters built with multiple materials were studied. While delamination is a major challenge in such emitters at high temperatures, their mechanical stability can be improved with an adhesion layer. For example, Ti improves adhesion between Ir and Si substrate, as discussed in Section 4.3.3. However, as discussed in Section 2.4, numerous factors determine delamination of a layer: such as film thicknesses, stacking order, materials, temperature, temperature change rate, and so on. Therefore, an emitter which consists of multiple materials must be carefully designed for their stability.

2) Since TPV power output density is typically around 1W/cm$^2$ [88], large-area emitters are essential to generate higher power output [92]–[97]. Compared to MM emitters, 1D emitters can be fabricated over large areas with ease, enabling mass production. Thus, 1D emitters provide advantages for realizing TPV applications with higher power output.

3) The emitters should be resistant to oxidation. However, in Section 4.6, it was observed that even a SiO$_2$ surface coating as thick as 1150 nm did not prevent oxidation. Thus, applications of the 1D emitter in Section 4.5 may be limited to vacuum or non-oxidizing environments.

4) Fabrication of MM emitters typically requires electron beam lithography, which is a highly time-consuming process, due to their smaller feature sizes. However, at relatively lower temperatures, such as car engines, target wavelengths are longer –i.e. Wien's displacement law. This means that the feature sizes of a MM emitter can be much larger such that they can be fabricated by photolithography, a much faster fabrication process. Therefore, with fabrication ease and optical property tunability, MM emitters can be highly beneficial for lower-temperature TPVs.

# 6. FUTURE WORK

1) Indium tin oxide (ITO)

One way to create emitters stable against oxidation is to build them with oxide materials. ITO is an oxide and has a melting point greater than 1500°C. It also has some magnitude of extinction coefficients that can cause optical energy loss, thus producing optical emission, at infrared wavelengths. Fig. 6.1 shows the refractive index of a 150 nm-thick ITO film. The ITO was sputter-deposited on an R-cut sapphire substrate, which is also an oxide, at 500°C and then was annealed at 800°C for 10 min in oxidizing atmosphere ($N_2$ 780 sccm, $O_2$ 220 sccm, total pressure 8 Torr). Moreover, the ITO layer did not delaminate and maintained a flat, smooth surface even after a rapid thermal testing with the ramp up rate of ~800°C/min. Therefore, ITO with sapphire substrate can potentially provide a way to realizing thermally robust TPV emitters in air.

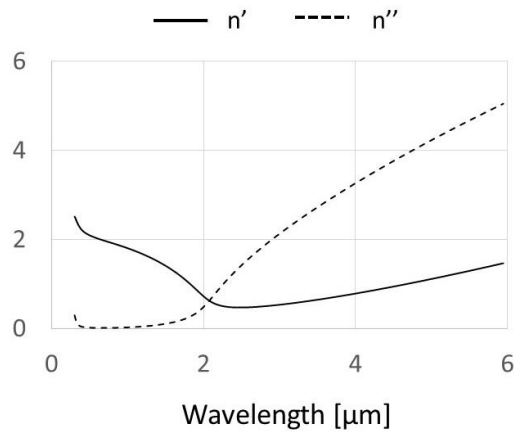

Fig. 6.1. Refractive index of indium tin oxide (ITO) after annealing. n' and n'' are real and imaginary parts, respectively.

2) Iridium (Ir)

It was reported that Ir does not oxidize up to at least 673K in air –it starts to oxidize between 673K and 873K [81]. Only a 5 nm-thick iridium oxide ($IrO_2$) was formed after annealing a thin Ir film at 873K for 1 hr in air [81]. This is significantly slower than the oxidation rate of Si in the Final Design emitter in Section 4.5. Thus, Ir-based MM emitters with fabrication ease are good candidates for practical TPV applications. Fig. 6.2 shows a MM emitter based on Ir and $HfO_2$ with larger features that can be fabricated by photolithography. This structure can be further studied for oxidation effects.

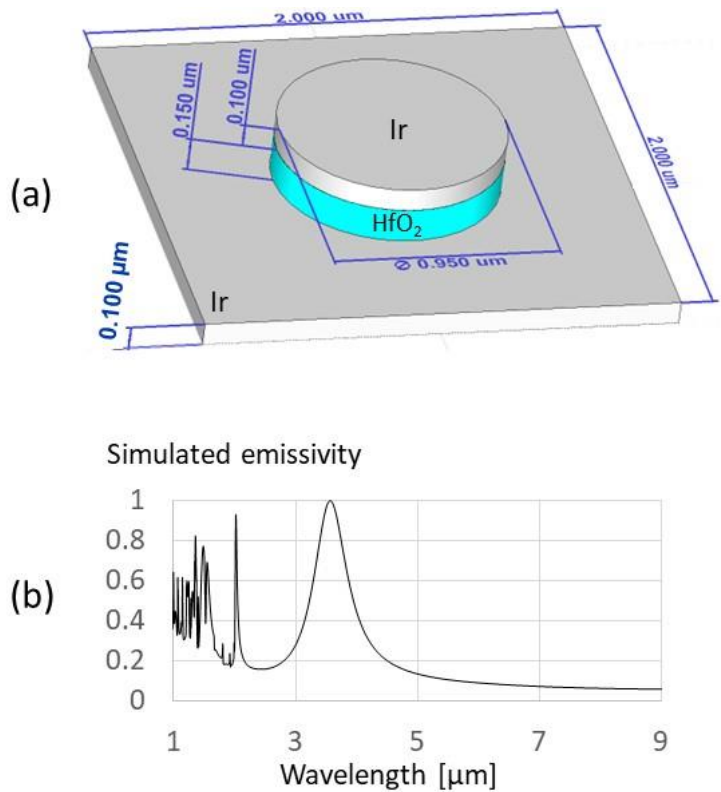

Fig. 6.2. An Ir-based metamaterial emitter with larger feature sizes. (a): unit cell. (b): simulated emissivity at normal incidence. The unit cell boundary conditions combined with Floquet modes were used to incorporate high-order diffractions at wavelengths smaller than 2 µm, which is the period of the unit cell.

# Appendix

### A. Charge Densities in Isotropic Media

Eq. 2.29(a) describes the divergence of electric fields in isotropic media, where the dielectric permittivity $\varepsilon$ is a scalar constant. From Eq. 2.29(a),

Eq. A1
$$\frac{\rho_b + \rho_f}{\varepsilon_o} = \frac{\rho_f}{\varepsilon}$$

By multiplying $\varepsilon$ to both sides of Eq. A1, we obtain $\varepsilon_r(\rho_b + \rho_f) = \rho_f$. Since $\varepsilon_r = 1 + \chi_e$, it follows that

Eq. A2
$$\rho_b = -\left(\frac{\chi_e}{1+\chi_e}\right)\rho_f$$

, where $\chi_e > 0$. Thus, Eq. A2 indicates the followings:

1. If $\rho_b = 0$ or $\rho_f = 0$, then $\rho_b = \rho_f = 0$.
2. $\rho_b$ and $\rho_f$ have opposite signs when they both are non-zero.
3. $|\rho_b| < |\rho_f|$ when they both are non-zero.

## B. Sinusoidal Electric and Magnetic Fields

Suppose a plane wave electric field:

Eq. A3
$$\vec{E} = \vec{E}_o e^{i(\vec{k}\cdot\vec{r}-\omega t)}$$
$$= (E_{ox}, E_{oy}, E_{oz}) e^{i(\vec{k}\cdot\vec{r}-\omega t)}$$
$$= \left(E_{ox} e^{i(\vec{k}\cdot\vec{r}-\omega t)}, E_{oy} e^{i(\vec{k}\cdot\vec{r}-\omega t)}, E_{oz} e^{i(\vec{k}\cdot\vec{r}-\omega t)}\right)$$
$$= (E_x, E_y, E_z)$$

, where the propagation vector $\vec{k} = (k_x, k_y, k_z)$ and the position vector $\vec{r} = (x, y, z)$ in the Cartesian coordinate. From $\nabla \times \vec{E} = -\frac{\partial \vec{B}}{\partial t}$ in Eq. 2.19(c),

Eq. A4
$$\nabla \times \vec{E} = \left(\frac{\partial}{\partial y} E_z - \frac{\partial}{\partial z} E_y\right)\hat{x} - \left(\frac{\partial}{\partial x} E_z - \frac{\partial}{\partial z} E_x\right)\hat{y} + \left(\frac{\partial}{\partial x} E_y - \frac{\partial}{\partial y} E_x\right)\hat{z}$$
$$= (ik_y E_z - ik_z E_y)\hat{x} - (ik_x E_z - ik_z E_x)\hat{y} + (ik_x E_y - ik_y E_x)\hat{z}$$
$$= i\left([k_y E_z - k_z E_y], -[k_x E_z - k_z E_x], [k_x E_y - k_y E_x]\right)$$
$$= i\vec{k} \times \vec{E}$$
$$= i e^{i(\vec{k}\cdot\vec{r}-\omega t)} (\vec{k} \times \vec{E}_o)$$
$$= -\frac{\partial \vec{B}}{\partial t}$$

Using the expression $\frac{\partial \vec{B}}{\partial t} = -i e^{i(\vec{k}\cdot\vec{r}-\omega t)} (\vec{k} \times \vec{E}_o)$ above, the magnetic field can be obtained as

Eq. A5
$$\vec{B} = \int \left(\frac{\partial \vec{B}}{\partial t}\right) \partial t$$
$$= -i \left[\int e^{i(\vec{k}\cdot\vec{r}-\omega t)} \partial t\right] (\vec{k} \times \vec{E}_o)$$

$$= -i\left[\frac{1}{-i\omega}e^{i(\vec{k}\cdot\vec{r}-\omega t)}\right](\vec{k}\times\vec{E}_o)$$

$$= \frac{1}{\omega}(\vec{k}\times\vec{E}_o)e^{i(\vec{k}\cdot\vec{r}-\omega t)}$$

$$= \vec{B}_o e^{i(\vec{k}\cdot\vec{r}-\omega t)} = \frac{1}{\omega}(\vec{k}\times\vec{E})$$

The divergence of magnetic fields is always zero; $\nabla\cdot\vec{B}=0$. This can be checked using the magnetic field expression derived from the electric field above. That is,

Eq. A6  $\nabla\cdot\vec{B} = \nabla\cdot\vec{B}_o e^{i(\vec{k}\cdot\vec{r}-\omega t)}$

$$= \nabla\cdot\frac{1}{\omega}(\vec{k}\times\vec{E})$$

$$= \frac{1}{\omega}\left(\frac{\partial}{\partial x},\frac{\partial}{\partial y},\frac{\partial}{\partial z}\right)\cdot\left([k_y E_z - k_z E_y], -[k_x E_z - k_z E_x], [k_x E_y - k_y E_x]\right)$$

$$= \frac{1}{\omega}\left(\frac{\partial}{\partial x}[k_y E_z - k_z E_y] - \frac{\partial}{\partial y}[k_x E_z - k_z E_x] + \frac{\partial}{\partial z}[k_x E_y - k_y E_x]\right)$$

$$= \frac{1}{\omega}\left([ik_x k_y E_z - ik_x k_z E_y] - [ik_x k_y E_z - ik_y k_z E_x] + [ik_x k_z E_y - ik_y k_z E_x]\right)$$

$$= 0$$

## C. Complex-Valued Speed of Light

Suppose an expression of complex-valued speed of light ($\tilde{v}$) as $\frac{1}{\tilde{v}^2} = \varepsilon_{eff}\mu$. Now, consider the "real" speed of light (or phase velocity) that is given by $v = \frac{1}{Re(\sqrt{\varepsilon_{eff}\mu})}$. The $\tilde{v}$ and $v$ are compared below:

Eq. A7
$$Re(\tilde{v}) = Re\left(\frac{1}{\sqrt{\varepsilon_{eff}\mu}}\right)$$

$$\neq \frac{1}{Re(\sqrt{\varepsilon_{eff}\mu})} = v$$

$$\rightarrow Re(\tilde{v}) \neq v$$

Therefore, the real part of the complex-valued speed of light is not equal to the phase velocity. Typically, the real and imaginary parts of complex-valued electromagnetic properties have their physical meanings. For example, the real part of refractive index relates to the phase velocity and the imaginary part to the optical energy loss. However, as can be seen in Eq. A7, the real part of the complex-valued speed of light does not relate to the actual speed of light. Indeed, the relationship between $\tilde{v}$ and $v$ is:

Eq. A8
$$v = \frac{1}{Re(\sqrt{\varepsilon_{eff}\mu})}$$

$$= \frac{1}{Re\left(\frac{1}{\tilde{v}}\right)}$$

## D. Speed of Light in Vacuum

According to Eq. 2.30, the lefthand and righthand sides of Eq. 2.43 can be written as $\nabla_v^2 \vec{B} = (\nabla_s^2 B_x, \nabla_s^2 B_y, \nabla_s^2 B_z)$ and $\varepsilon_{eff}\mu \frac{\partial^2 \vec{B}}{\partial t^2} = \varepsilon_{eff}\mu \left(\frac{\partial^2 B_x}{\partial t^2}, \frac{\partial^2 B_y}{\partial t^2}, \frac{\partial^2 B_z}{\partial t^2}\right)$, respectively. Similarly, for electric fields: $\nabla_v^2 \vec{E} = (\nabla_s^2 E_x, \nabla_s^2 E_y, \nabla_s^2 E_z)$ and $\varepsilon_{eff}\mu \frac{\partial^2 \vec{E}}{\partial t^2} = \varepsilon_{eff}\mu \left(\frac{\partial^2 E_x}{\partial t^2}, \frac{\partial^2 E_y}{\partial t^2}, \frac{\partial^2 E_z}{\partial t^2}\right)$, according to Eq. 2.44. Thus, we obtain the following relationships:

Eq. A9

$$\nabla_s^2 B_x = \varepsilon_{eff}\mu \frac{\partial^2 B_x}{\partial t^2}$$

$$\nabla_s^2 B_y = \varepsilon_{eff}\mu \frac{\partial^2 B_y}{\partial t^2}$$

$$\nabla_s^2 B_z = \varepsilon_{eff}\mu \frac{\partial^2 B_z}{\partial t^2}$$

Eq. A10

$$\nabla_s^2 E_x = \varepsilon_{eff}\mu \frac{\partial^2 E_x}{\partial t^2}$$

$$\nabla_s^2 E_y = \varepsilon_{eff}\mu \frac{\partial^2 E_y}{\partial t^2}$$

$$\nabla_s^2 E_z = \varepsilon_{eff}\mu \frac{\partial^2 E_z}{\partial t^2}$$

For magnetic fields, taking the real parts of the lefthand and righthand sides of Eq. A9 for vacuum:

Eq. A 11

$$Re[\nabla_s^2 B_{axis}] = Re\left[\varepsilon_{eff}\mu \frac{\partial^2 B_{axis}}{\partial t^2}\right]$$

$$= \varepsilon_o \mu_o \frac{\partial^2 Re[B_{axis}]}{\partial t^2}$$

$$= \nabla_s^2 Re[B_{axis}]$$

$$\rightarrow \nabla_s^2 B_x' = \varepsilon_o \mu_o \frac{\partial^2 B_x'}{\partial t^2}$$

$$\rightarrow \nabla_s^2 B_y' = \varepsilon_o \mu_o \frac{\partial^2 B_y'}{\partial t^2}$$

$$\rightarrow \nabla_s^2 B_z' = \varepsilon_o \mu_o \frac{\partial^2 B_z'}{\partial t^2}$$

, where $B_{axis}' = Re[B_{axis}]$. The subscript "axis" is used instead of x, y, or z. Now, for electric fields:

Eq. A12

$$Re[\nabla_s^2 E_{axis}] = Re\left[\varepsilon_{eff} \mu \frac{\partial^2 E_{axis}}{\partial t^2}\right]$$

$$= \varepsilon_o \mu_o \frac{\partial^2 Re[E_{axis}]}{\partial t^2}$$

$$= \nabla_s^2 Re[E_{axis}]$$

$$\rightarrow \nabla_s^2 E_x' = \varepsilon_o \mu_o \frac{\partial^2 E_x'}{\partial t^2}$$

$$\rightarrow \nabla_s^2 E_y' = \varepsilon_o \mu_o \frac{\partial^2 E_y'}{\partial t^2}$$

$$\rightarrow \nabla_s^2 E_z' = \varepsilon_o \mu_o \frac{\partial^2 E_z'}{\partial t^2}$$

, where $E_{axis}' = Re[E_{axis}]$. It follows that the propagation speed of both electric and magnetic fields along any axis is the same as $c = \sqrt{\frac{1}{\varepsilon_o \mu_o}} \approx 3 \times 10^8 \, m/s$. Therefore, the speed of light in vacuum is independent of propagation direction or oscillation direction of the fields.

### E. Metal Adhesion with and without DI Water Rinse of Substrate

Fig. A1 shows sputtered Zr on a Si(100) substrate after rapid thermal annealing. The temperature ramp up rate was ~1000 °C/min in $N_2$ ambient; which was intentionally fast for delamination testing. Prior to deposition, the substrate of each sample was cleaned by the following method:

- Sample A: sonication for 3 min in Acetone → rinse with isopropyl alcohol (IPA) → $N_2$ blow.

- Sample B: sonication for 3 min in Acetone → rinse with IPA → rinse with deionized (DI) water → $N_2$ blow → dry at 140°C for 2min on a hot plate.

While delamination is seen in both samples, Sample A delaminated more aggressively. These results suggest that the rinse with DI water improves adhesion.

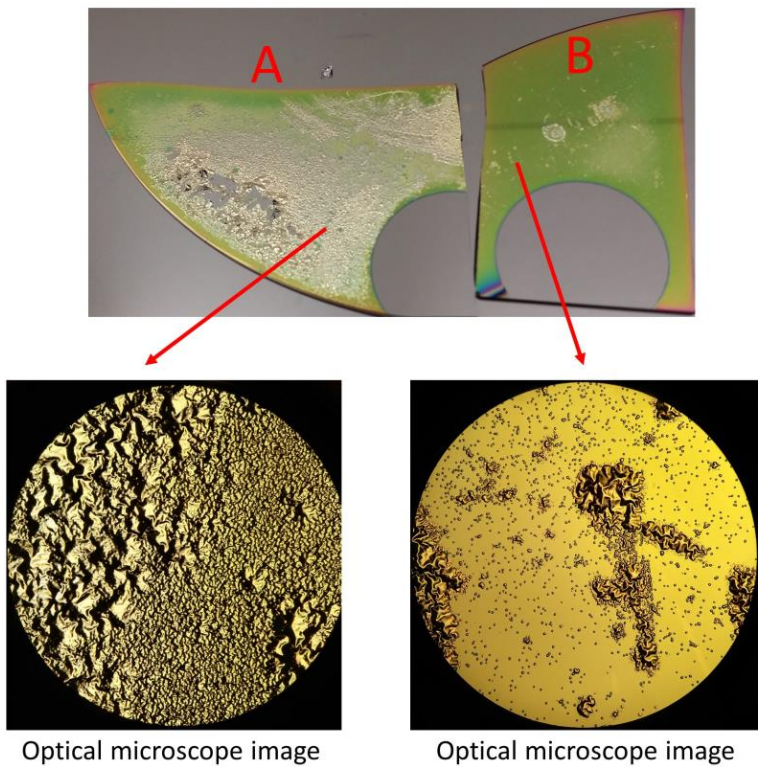

Fig. A1. Sputtered Zr on a Si(100) substrate after rapid thermal annealing.

## F. Emission Measurement Apparatus: Component Diagram

It is assumed that thermally radiated waves from the emitter in the below setup can be approximated as a plane wave (large L). Black arrows represent data signal flow.

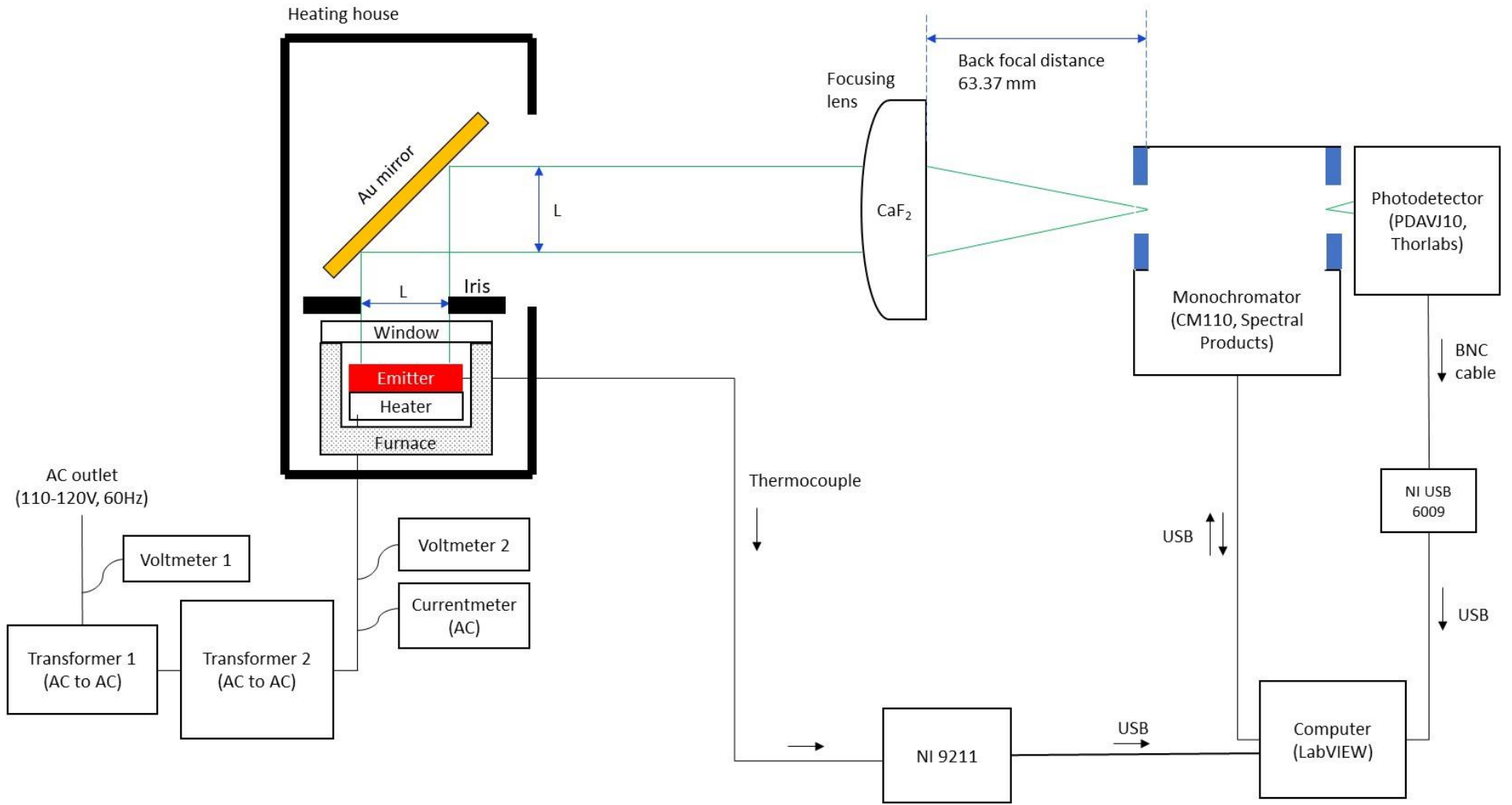

# G. Emission Measurement Apparatus: LabVIEW Code for Emission Measurement

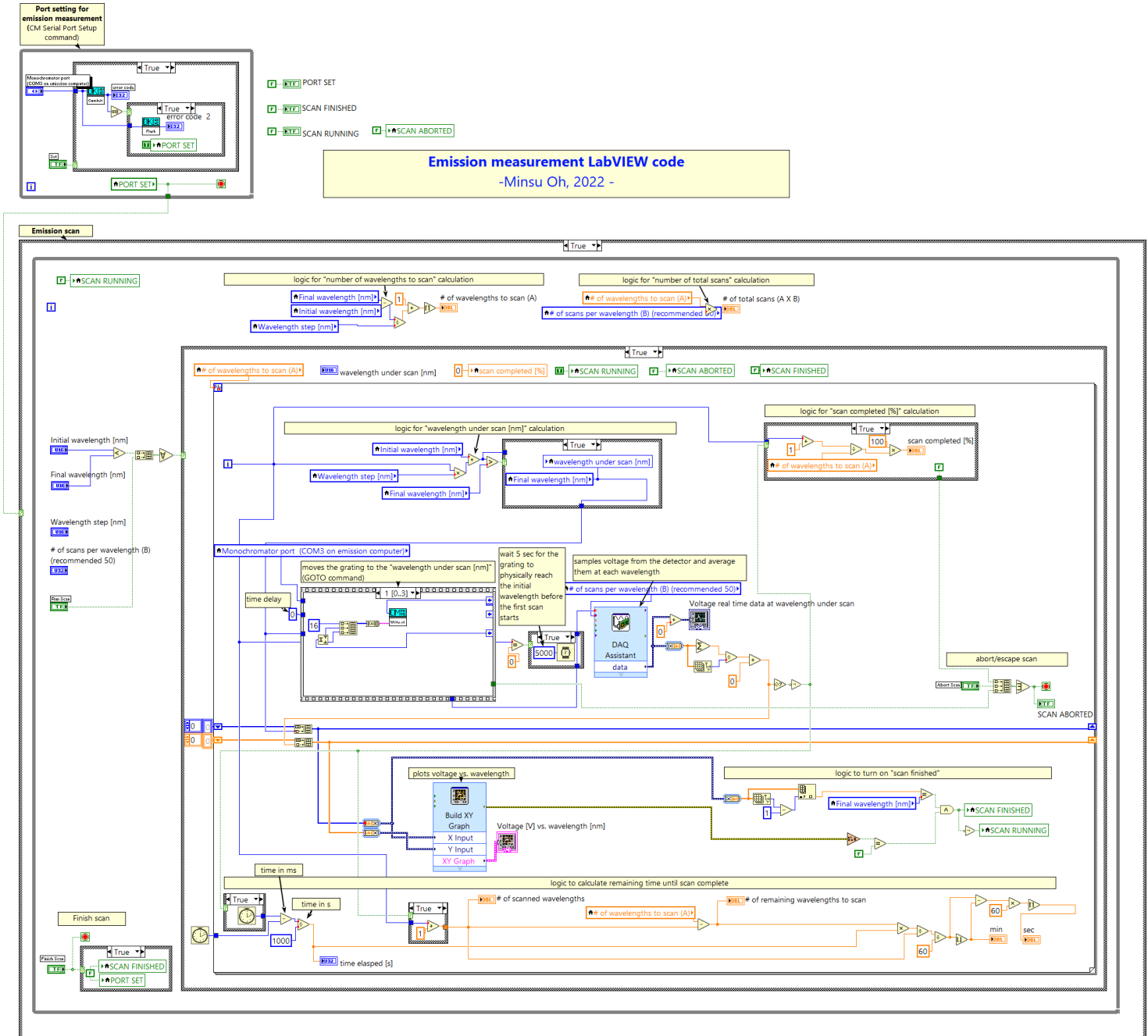

## H. Emission Measurement Apparatus: LabVIEW Code for Temperature Monitoring

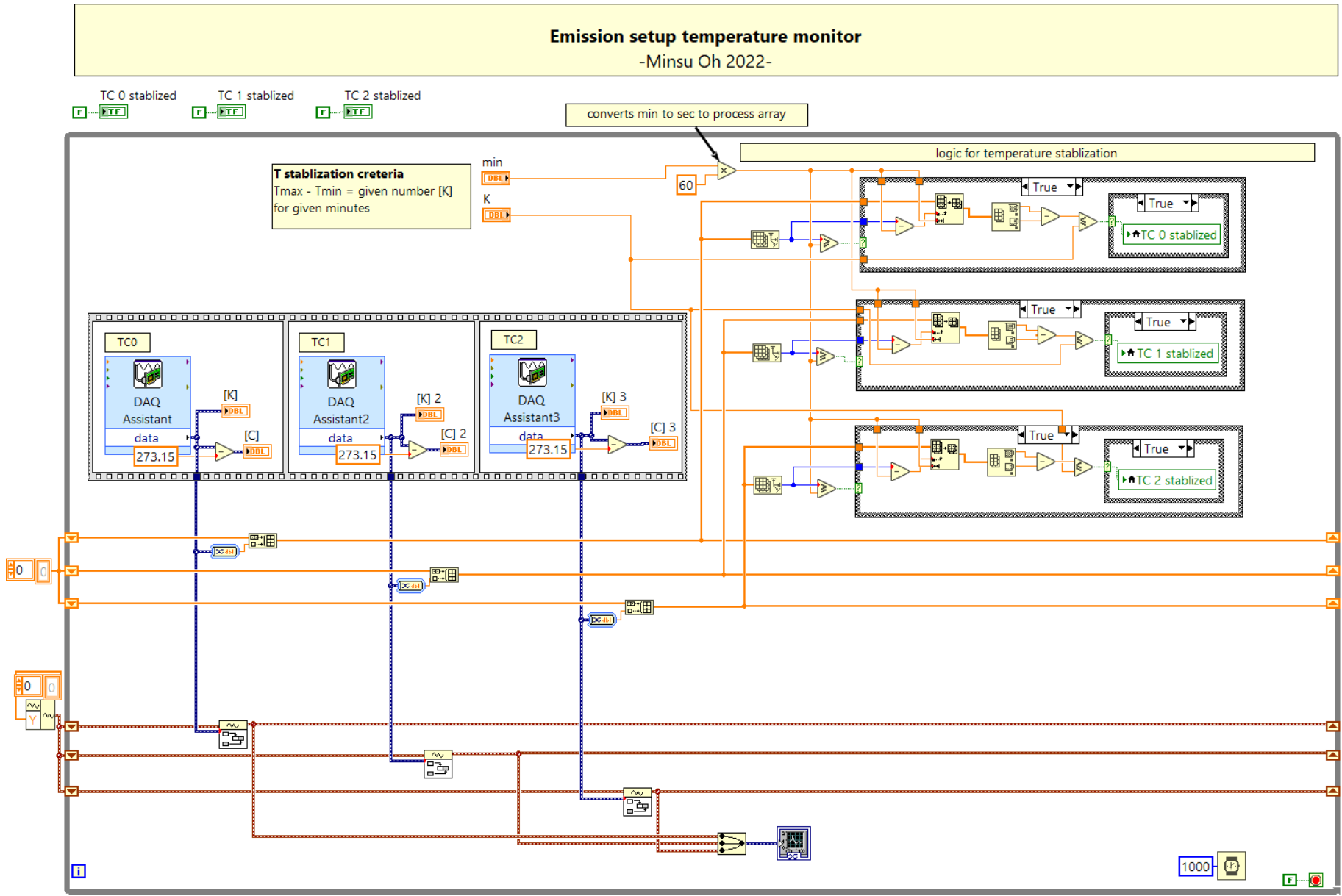

# I. Optical Properties of Sputtered Iridium (Ir) Films

Since the Ir films were optically thick, the modeled results were independent of their thickness. The below results are also published in Ref. [74].

Table A1. Temperature-dependent refractive index of as-deposited Ir. The photon energy in the unit of eV is given in the column of "eV". The n and k are real and imaginary parts of the refractive index, respectively. "r.t." stands for room temperature. A much higher resolution, smaller than 0.01eV, was used for analysis.

|      | r.t. |       | 100°C |       | 200°C |       | 300°C |       | 400°C |       | 500°C |       | 550°C |       |
|------|------|-------|-------|-------|-------|-------|-------|-------|-------|-------|-------|-------|-------|-------|
| eV   | n    | k     | n     | k     | n     | k     | n     | k     | n     | k     | n     | k     | n     | k     |
| 4.13 | 1.69 | 3.31  | 1.68  | 3.30  | 1.68  | 3.30  | 1.68  | 3.30  | 1.68  | 3.30  | 1.67  | 3.30  | 1.67  | 3.33  |
| 4.03 | 1.74 | 3.32  | 1.73  | 3.32  | 1.74  | 3.32  | 1.74  | 3.32  | 1.74  | 3.33  | 1.73  | 3.33  | 1.74  | 3.34  |
| 3.93 | 1.77 | 3.34  | 1.77  | 3.34  | 1.78  | 3.33  | 1.78  | 3.34  | 1.78  | 3.34  | 1.77  | 3.35  | 1.78  | 3.35  |
| 3.83 | 1.79 | 3.36  | 1.79  | 3.37  | 1.80  | 3.36  | 1.80  | 3.36  | 1.81  | 3.37  | 1.80  | 3.37  | 1.81  | 3.36  |
| 3.73 | 1.79 | 3.40  | 1.79  | 3.41  | 1.81  | 3.40  | 1.82  | 3.40  | 1.82  | 3.41  | 1.82  | 3.41  | 1.81  | 3.39  |
| 3.63 | 1.79 | 3.47  | 1.80  | 3.47  | 1.81  | 3.46  | 1.82  | 3.46  | 1.83  | 3.46  | 1.83  | 3.46  | 1.81  | 3.45  |
| 3.53 | 1.80 | 3.55  | 1.80  | 3.55  | 1.82  | 3.53  | 1.83  | 3.53  | 1.84  | 3.53  | 1.85  | 3.53  | 1.82  | 3.53  |
| 3.43 | 1.81 | 3.64  | 1.82  | 3.64  | 1.84  | 3.62  | 1.85  | 3.62  | 1.86  | 3.62  | 1.86  | 3.61  | 1.84  | 3.62  |
| 3.33 | 1.84 | 3.74  | 1.85  | 3.74  | 1.87  | 3.72  | 1.88  | 3.71  | 1.89  | 3.71  | 1.89  | 3.71  | 1.87  | 3.71  |
| 3.23 | 1.88 | 3.85  | 1.89  | 3.84  | 1.91  | 3.82  | 1.92  | 3.81  | 1.93  | 3.81  | 1.93  | 3.81  | 1.92  | 3.82  |
| 3.13 | 1.93 | 3.96  | 1.94  | 3.95  | 1.96  | 3.92  | 1.97  | 3.92  | 1.97  | 3.92  | 1.98  | 3.91  | 1.97  | 3.92  |
| 3.03 | 1.98 | 4.06  | 1.99  | 4.06  | 2.01  | 4.03  | 2.02  | 4.02  | 2.03  | 4.02  | 2.04  | 4.01  | 2.03  | 4.02  |
| 2.93 | 2.04 | 4.17  | 2.05  | 4.16  | 2.07  | 4.14  | 2.08  | 4.13  | 2.09  | 4.13  | 2.10  | 4.12  | 2.09  | 4.12  |
| 2.83 | 2.10 | 4.28  | 2.11  | 4.28  | 2.13  | 4.24  | 2.14  | 4.24  | 2.15  | 4.23  | 2.16  | 4.22  | 2.15  | 4.22  |
| 2.73 | 2.16 | 4.40  | 2.18  | 4.39  | 2.20  | 4.36  | 2.21  | 4.35  | 2.21  | 4.35  | 2.23  | 4.34  | 2.22  | 4.33  |
| 2.63 | 2.23 | 4.53  | 2.24  | 4.52  | 2.26  | 4.48  | 2.27  | 4.47  | 2.28  | 4.47  | 2.30  | 4.45  | 2.28  | 4.44  |
| 2.53 | 2.30 | 4.67  | 2.31  | 4.65  | 2.33  | 4.61  | 2.35  | 4.60  | 2.35  | 4.60  | 2.37  | 4.58  | 2.36  | 4.57  |
| 2.43 | 2.38 | 4.81  | 2.39  | 4.80  | 2.41  | 4.76  | 2.43  | 4.74  | 2.43  | 4.74  | 2.45  | 4.72  | 2.44  | 4.71  |
| 2.33 | 2.48 | 4.97  | 2.49  | 4.96  | 2.51  | 4.91  | 2.52  | 4.90  | 2.53  | 4.89  | 2.54  | 4.87  | 2.53  | 4.86  |
| 2.23 | 2.59 | 5.13  | 2.60  | 5.12  | 2.62  | 5.07  | 2.63  | 5.05  | 2.64  | 5.05  | 2.65  | 5.02  | 2.64  | 5.01  |
| 2.13 | 2.72 | 5.29  | 2.73  | 5.27  | 2.75  | 5.23  | 2.75  | 5.21  | 2.76  | 5.21  | 2.77  | 5.18  | 2.76  | 5.16  |
| 2.03 | 2.85 | 5.44  | 2.86  | 5.42  | 2.88  | 5.38  | 2.89  | 5.36  | 2.89  | 5.36  | 2.91  | 5.33  | 2.89  | 5.31  |
| 1.93 | 2.97 | 5.59  | 3.00  | 5.57  | 3.01  | 5.52  | 3.02  | 5.51  | 3.03  | 5.51  | 3.04  | 5.48  | 3.02  | 5.46  |
| 1.83 | 3.09 | 5.74  | 3.12  | 5.73  | 3.13  | 5.68  | 3.15  | 5.67  | 3.15  | 5.67  | 3.17  | 5.64  | 3.14  | 5.61  |
| 1.73 | 3.19 | 5.92  | 3.23  | 5.90  | 3.25  | 5.85  | 3.27  | 5.84  | 3.28  | 5.84  | 3.30  | 5.81  | 3.26  | 5.77  |
| 1.63 | 3.29 | 6.12  | 3.33  | 6.10  | 3.36  | 6.04  | 3.38  | 6.04  | 3.39  | 6.04  | 3.42  | 6.00  | 3.37  | 5.96  |
| 1.53 | 3.40 | 6.37  | 3.44  | 6.34  | 3.47  | 6.28  | 3.49  | 6.27  | 3.51  | 6.27  | 3.54  | 6.22  | 3.48  | 6.19  |
| 1.43 | 3.51 | 6.65  | 3.55  | 6.61  | 3.59  | 6.55  | 3.62  | 6.54  | 3.65  | 6.54  | 3.67  | 6.49  | 3.60  | 6.45  |
| 1.33 | 3.64 | 6.97  | 3.69  | 6.93  | 3.73  | 6.87  | 3.76  | 6.85  | 3.79  | 6.85  | 3.81  | 6.79  | 3.73  | 6.75  |
| 1.23 | 3.78 | 7.34  | 3.83  | 7.30  | 3.87  | 7.23  | 3.92  | 7.21  | 3.95  | 7.21  | 3.97  | 7.14  | 3.88  | 7.10  |
| 1.13 | 3.92 | 7.77  | 3.97  | 7.72  | 4.03  | 7.64  | 4.08  | 7.62  | 4.13  | 7.62  | 4.14  | 7.55  | 4.03  | 7.51  |
| 1.03 | 4.05 | 8.29  | 4.12  | 8.23  | 4.19  | 8.14  | 4.25  | 8.12  | 4.31  | 8.12  | 4.32  | 8.04  | 4.18  | 8.01  |
| 0.93 | 4.18 | 8.94  | 4.26  | 8.88  | 4.35  | 8.77  | 4.43  | 8.74  | 4.50  | 8.73  | 4.51  | 8.65  | 4.34  | 8.62  |
| 0.83 | 4.31 | 9.80  | 4.42  | 9.72  | 4.53  | 9.59  | 4.63  | 9.54  | 4.72  | 9.52  | 4.73  | 9.44  | 4.52  | 9.43  |
| 0.73 | 4.49 | 10.96 | 4.63  | 10.86 | 4.76  | 10.69 | 4.89  | 10.61 | 5.00  | 10.58 | 5.03  | 10.48 | 4.76  | 10.51 |
| 0.63 | 4.81 | 12.56 | 4.97  | 12.42 | 5.14  | 12.19 | 5.30  | 12.08 | 5.44  | 12.01 | 5.48  | 11.90 | 5.14  | 11.99 |
| 0.53 | 5.39 | 14.78 | 5.60  | 14.57 | 5.80  | 14.26 | 6.00  | 14.09 | 6.18  | 13.97 | 6.24  | 13.83 | 5.82  | 14.03 |
| 0.43 | 6.50 | 17.91 | 6.78  | 17.59 | 7.03  | 17.15 | 7.27  | 16.88 | 7.48  | 16.67 | 7.57  | 16.48 | 7.07  | 16.86 |
| 0.33 | 8.61 | 22.40 | 8.96  | 21.87 | 9.26  | 21.21 | 9.56  | 20.75 | 9.78  | 20.40 | 9.94  | 20.11 | 9.36  | 20.81 |
| 0.22 | 13.35| 30.15 | 13.74 | 29.12 | 14.04 | 27.91 | 14.34 | 27.06 | 14.53 | 26.44 | 14.74 | 25.85 | 14.23 | 27.19 |
| 0.12 | 25.26| 44.20 | 25.25 | 42.06 | 24.96 | 39.68 | 24.90 | 37.97 | 24.83 | 36.84 | 24.76 | 35.50 | 24.94 | 38.00 |
| 0.08 | 36.70| 54.20 | 35.99 | 51.30 | 34.83 | 48.25 | 34.20 | 46.00 | 33.83 | 44.52 | 33.18 | 42.76 | 34.19 | 45.90 |

Table A2. Temperature-dependent refractive index of annealed Ir (700°C, 10min, N$_2$). The photon energy in the unit of eV is given in the column of "eV". The n and k are real and imaginary parts of the refractive index, respectively. "r.t." stands for room temperature. A much higher resolution, smaller than 0.01eV, was used for analysis.

|      | r.t.  |       | 100°C |       | 200°C |       | 300°C |       | 400°C |       | 500°C |       | 550°C |       |
| ---- | ----- | ----- | ----- | ----- | ----- | ----- | ----- | ----- | ----- | ----- | ----- | ----- | ----- | ----- |
| eV   | n     | k     | n     | k     | n     | k     | n     | k     | n     | k     | n     | k     | n     | k     |
| 4.13 | 1.82  | 3.40  | 1.84  | 3.40  | 1.86  | 3.38  | 1.87  | 3.38  | 1.87  | 3.39  | 1.84  | 3.43  | 1.84  | 3.44  |
| 4.03 | 1.89  | 3.38  | 1.91  | 3.38  | 1.92  | 3.37  | 1.93  | 3.37  | 1.93  | 3.38  | 1.91  | 3.43  | 1.91  | 3.44  |
| 3.93 | 1.92  | 3.35  | 1.93  | 3.36  | 1.95  | 3.36  | 1.96  | 3.36  | 1.96  | 3.38  | 1.95  | 3.42  | 1.95  | 3.43  |
| 3.83 | 1.90  | 3.35  | 1.92  | 3.36  | 1.95  | 3.36  | 1.96  | 3.37  | 1.97  | 3.39  | 1.97  | 3.42  | 1.97  | 3.43  |
| 3.73 | 1.87  | 3.38  | 1.89  | 3.39  | 1.94  | 3.40  | 1.95  | 3.40  | 1.97  | 3.42  | 1.96  | 3.45  | 1.96  | 3.46  |
| 3.63 | 1.84  | 3.45  | 1.87  | 3.47  | 1.92  | 3.46  | 1.94  | 3.46  | 1.96  | 3.47  | 1.95  | 3.50  | 1.95  | 3.51  |
| 3.53 | 1.83  | 3.55  | 1.86  | 3.57  | 1.91  | 3.54  | 1.94  | 3.54  | 1.95  | 3.55  | 1.95  | 3.58  | 1.95  | 3.59  |
| 3.43 | 1.85  | 3.67  | 1.87  | 3.68  | 1.92  | 3.64  | 1.95  | 3.63  | 1.96  | 3.64  | 1.96  | 3.67  | 1.96  | 3.68  |
| 3.33 | 1.88  | 3.78  | 1.91  | 3.79  | 1.95  | 3.74  | 1.97  | 3.74  | 1.99  | 3.75  | 1.99  | 3.78  | 1.99  | 3.78  |
| 3.23 | 1.93  | 3.89  | 1.95  | 3.90  | 1.99  | 3.85  | 2.01  | 3.84  | 2.03  | 3.85  | 2.03  | 3.88  | 2.03  | 3.89  |
| 3.13 | 1.98  | 4.00  | 2.00  | 4.01  | 2.04  | 3.96  | 2.06  | 3.95  | 2.07  | 3.96  | 2.08  | 3.99  | 2.08  | 4.00  |
| 3.03 | 2.03  | 4.11  | 2.06  | 4.11  | 2.09  | 4.06  | 2.11  | 4.06  | 2.13  | 4.06  | 2.13  | 4.09  | 2.14  | 4.10  |
| 2.93 | 2.09  | 4.22  | 2.12  | 4.22  | 2.15  | 4.17  | 2.17  | 4.16  | 2.19  | 4.17  | 2.19  | 4.20  | 2.20  | 4.20  |
| 2.83 | 2.15  | 4.33  | 2.18  | 4.34  | 2.21  | 4.28  | 2.23  | 4.27  | 2.25  | 4.28  | 2.26  | 4.31  | 2.27  | 4.31  |
| 2.73 | 2.21  | 4.45  | 2.24  | 4.45  | 2.27  | 4.40  | 2.29  | 4.39  | 2.31  | 4.39  | 2.32  | 4.42  | 2.33  | 4.42  |
| 2.63 | 2.28  | 4.58  | 2.31  | 4.58  | 2.34  | 4.52  | 2.36  | 4.51  | 2.38  | 4.52  | 2.39  | 4.54  | 2.40  | 4.54  |
| 2.53 | 2.35  | 4.72  | 2.39  | 4.72  | 2.41  | 4.66  | 2.44  | 4.65  | 2.45  | 4.65  | 2.46  | 4.67  | 2.47  | 4.67  |
| 2.43 | 2.44  | 4.88  | 2.47  | 4.87  | 2.50  | 4.81  | 2.52  | 4.79  | 2.53  | 4.79  | 2.54  | 4.81  | 2.55  | 4.81  |
| 2.33 | 2.55  | 5.03  | 2.58  | 5.02  | 2.60  | 4.96  | 2.62  | 4.94  | 2.63  | 4.94  | 2.63  | 4.96  | 2.64  | 4.96  |
| 2.23 | 2.67  | 5.18  | 2.70  | 5.18  | 2.72  | 5.11  | 2.73  | 5.09  | 2.74  | 5.09  | 2.74  | 5.12  | 2.75  | 5.11  |
| 2.13 | 2.81  | 5.32  | 2.83  | 5.32  | 2.85  | 5.25  | 2.86  | 5.23  | 2.86  | 5.24  | 2.87  | 5.28  | 2.87  | 5.27  |
| 2.03 | 2.94  | 5.45  | 2.96  | 5.45  | 2.98  | 5.38  | 2.99  | 5.37  | 2.99  | 5.38  | 3.00  | 5.43  | 3.01  | 5.42  |
| 1.93 | 3.06  | 5.57  | 3.09  | 5.58  | 3.10  | 5.51  | 3.11  | 5.50  | 3.12  | 5.52  | 3.13  | 5.57  | 3.14  | 5.57  |
| 1.83 | 3.15  | 5.70  | 3.19  | 5.71  | 3.21  | 5.64  | 3.22  | 5.64  | 3.23  | 5.67  | 3.25  | 5.72  | 3.27  | 5.72  |
| 1.73 | 3.23  | 5.85  | 3.28  | 5.86  | 3.29  | 5.79  | 3.32  | 5.80  | 3.34  | 5.83  | 3.36  | 5.88  | 3.38  | 5.89  |
| 1.63 | 3.29  | 6.04  | 3.35  | 6.04  | 3.36  | 5.98  | 3.40  | 5.99  | 3.43  | 6.02  | 3.46  | 6.08  | 3.49  | 6.08  |
| 1.53 | 3.35  | 6.28  | 3.41  | 6.27  | 3.43  | 6.21  | 3.48  | 6.21  | 3.52  | 6.24  | 3.56  | 6.31  | 3.59  | 6.31  |
| 1.43 | 3.41  | 6.56  | 3.47  | 6.55  | 3.51  | 6.49  | 3.57  | 6.49  | 3.61  | 6.51  | 3.67  | 6.59  | 3.71  | 6.59  |
| 1.33 | 3.47  | 6.90  | 3.54  | 6.88  | 3.60  | 6.82  | 3.66  | 6.81  | 3.71  | 6.83  | 3.80  | 6.92  | 3.84  | 6.91  |
| 1.23 | 3.54  | 7.30  | 3.61  | 7.28  | 3.70  | 7.21  | 3.76  | 7.19  | 3.82  | 7.21  | 3.94  | 7.29  | 4.00  | 7.28  |
| 1.13 | 3.64  | 7.79  | 3.71  | 7.76  | 3.81  | 7.65  | 3.88  | 7.64  | 3.95  | 7.66  | 4.10  | 7.71  | 4.18  | 7.70  |
| 1.03 | 3.77  | 8.34  | 3.84  | 8.32  | 3.93  | 8.17  | 4.01  | 8.16  | 4.09  | 8.20  | 4.27  | 8.21  | 4.35  | 8.18  |
| 0.93 | 3.89  | 8.95  | 3.97  | 8.94  | 4.03  | 8.80  | 4.14  | 8.80  | 4.25  | 8.84  | 4.42  | 8.81  | 4.51  | 8.77  |
| 0.83 | 3.88  | 9.67  | 4.01  | 9.69  | 4.09  | 9.60  | 4.25  | 9.59  | 4.40  | 9.63  | 4.55  | 9.61  | 4.66  | 9.54  |
| 0.73 | 3.70  | 10.75 | 3.90  | 10.76 | 4.09  | 10.70 | 4.32  | 10.67 | 4.52  | 10.69 | 4.70  | 10.70 | 4.82  | 10.61 |
| 0.63 | 3.45  | 12.49 | 3.74  | 12.46 | 4.10  | 12.31 | 4.40  | 12.24 | 4.67  | 12.23 | 4.95  | 12.25 | 5.09  | 12.13 |
| 0.53 | 3.37  | 15.16 | 3.77  | 15.06 | 4.25  | 14.73 | 4.66  | 14.58 | 5.03  | 14.51 | 5.47  | 14.46 | 5.64  | 14.30 |
| 0.43 | 3.74  | 19.16 | 4.29  | 18.94 | 4.87  | 18.37 | 5.44  | 18.07 | 5.95  | 17.88 | 6.57  | 17.62 | 6.80  | 17.38 |
| 0.33 | 4.99  | 25.27 | 5.79  | 24.78 | 6.52  | 23.89 | 7.35  | 23.26 | 8.06  | 22.78 | 8.79  | 22.10 | 9.07  | 21.69 |
| 0.22 | 8.96  | 36.58 | 10.13 | 35.38 | 11.23 | 33.65 | 12.38 | 32.05 | 13.23 | 30.80 | 13.77 | 29.36 | 14.07 | 28.63 |
| 0.12 | 21.55 | 57.99 | 22.99 | 55.43 | 24.19 | 50.90 | 24.35 | 46.82 | 24.60 | 44.16 | 24.51 | 42.03 | 24.84 | 40.54 |
| 0.08 | 32.76 | 74.23 | 35.96 | 70.86 | 36.02 | 63.72 | 34.46 | 58.72 | 34.24 | 55.22 | 34.08 | 52.32 | 34.25 | 49.89 |

Fig. A2. Ellipsometric data fit parameters for sputtered Ir films at room temperature. (a): as-deposited film. (b): annealed film (700℃, 10min, $N_2$).

## J. Method of Emission Signal Normalization

(The below method and discussions are also published in Ref. [88]) In the setup in Fig. 4.1, the radiated beam from the sample undergoes multiple reflections between the window and sample, illustrated in Fig. A3 below. The total power that is reflected from the Au mirror ($P_{sum}$) is given by

Eq. A13
$$P_{sum} = P_{rad}T_w(1 + R_wR_s)R_{Au}$$

, where it assumes that most power is in the first and second beams that transmit the window. $P_{rad}$ is the thermally radiated power from the sample. $R_w$ and $T_w$ are the reflectance and transmittance of the window, respectively. $R_s$ is the reflectance of the sample. $R_{Au}$ is the reflectance of the Au mirror. The power that enters the photodetector ($P_{in}$) is linearly proportional to the power reflected from the mirror ($P_{sum}$). The output voltage ($V$) from the photodetector used in this work (PDAVJ10, Thorlabs) is also linearly proportional to the input power ($P_{in}$) at a given wavelength. Thus, the output voltage is linearly proportional to $P_{rad}$ by the relationship:

Eq. A14
$$P_{rad} \propto \frac{V}{(1+R_wR_s)\,T_w\,R_{Au}}$$

It follows that the emitter's emissivity ($\epsilon_{emitter}$) can be obtained by

Eq. A15
$$\epsilon_{emitter} = \frac{P_{rad,emitter}}{P_{rad,ref}}\epsilon_{ref}$$
$$= \left(\frac{V_{emitter}}{V_{ref}}\right)\left(\frac{1+R_wR_{ref}}{1+R_wR_{emitter}}\right)\epsilon_{ref}$$

, where the subscripts "emitter" and "ref" are used for the emitter and reference material, respectively. The absorbance of the reference material characterized by spectroscopic ellipsometry was used as its emissivity.

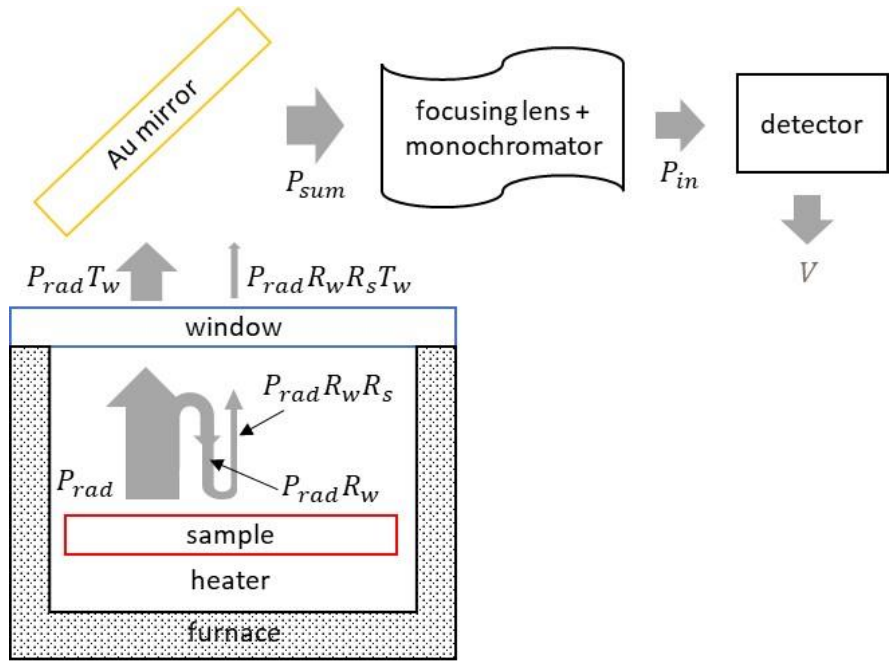

Fig. A3. Radiation path and power flow. This figure is also published in Ref. [88] (open access).

## K. Microstructure of an Annealed Final Design Emitter

Fig. A4(a) is an optical microscope image (dark field) of the cross-section of a Final Design emitter sample in Section 4.5 that was annealed in air. In the figure, needle-like patterns are seen ~100 µm into the depth of the substrate, which are attributed to tensile stress. The temperature profile for annealing is also provided in Fig. A4(b).

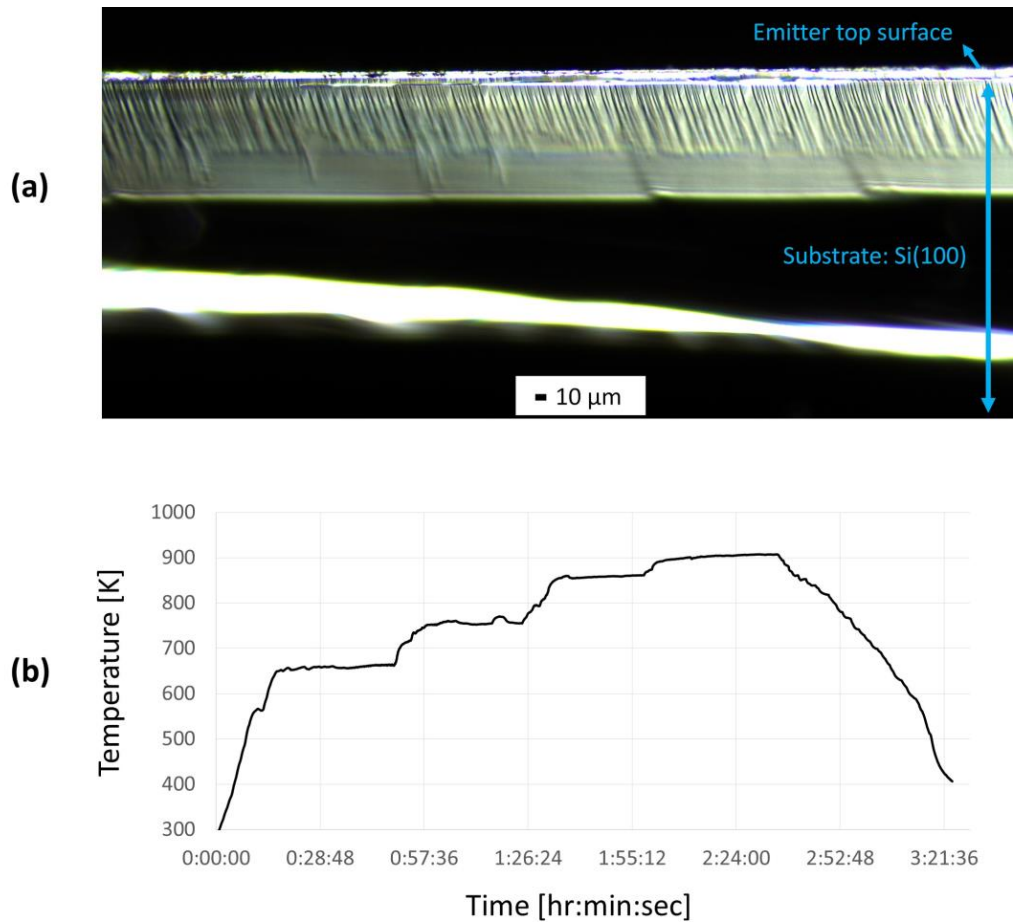

Fig. A4. A Final Design emitter sample that was annealed in air. (a): a dark field optical microscope image of the emitter cross-section. (b): temperature profile for annealing.

# References

[1]  "Estimated U.S. Energy Consumption," 2021. [Online]. Available: https://flowcharts.llnl.gov/.

[2]  U.S. Energy Information Administration, "Electric Power Monthly," 2023. [Online]. Available: https://www.eia.gov/electricity/monthly/.

[3]  M. Oh, J. McElearney, A. Lemire, and T. E. Vandervelde, "Material candidates for thermally robust applications of selective thermophotovoltaic emitters," *Phys. Rev. Mater.*, vol. 6, no. 11, p. 110201, Nov. 2022, doi: 10.1103/PhysRevMaterials.6.110201.

[4]  M. Suemitsu, T. Asano, T. Inoue, and S. Noda, "High-Efficiency Thermophotovoltaic System That Employs an Emitter Based on a Silicon Rod-Type Photonic Crystal," *ACS Photonics*, vol. 7, no. 1, pp. 80–87, 2020, doi: 10.1021/acsphotonics.9b00984.

[5]  D. N. Woolf *et al.*, "High-efficiency thermophotovoltaic energy conversion enabled by a metamaterial selective emitter," *Optica*, vol. 5, no. 2, p. 213, 2018, doi: 10.1364/optica.5.000213.

[6]  Y. X. Yeng, W. R. Chan, V. Rinnerbauer, J. D. Joannopoulos, M. Soljačić, and I. Celanovic, "Performance analysis of experimentally viable photonic crystal enhanced thermophotovoltaic systems," *Opt. Express*, vol. 21, no. S6, p. A1035, 2013, doi: 10.1364/oe.21.0a1035.

[7]  A. Fiorino, L. Zhu, D. Thompson, R. Mittapally, P. Reddy, and E. Meyhofer, "Nanogap near-field thermophotovoltaics," *Nat. Nanotechnol.*, vol. 13, no. 9, pp. 806–811, 2018, doi: 10.1038/s41565-018-0172-5.

[8]  W. E. S. W. A. Rashid, P. J. Ker, M. Z. Bin Jamaludin, M. M. A. Gamel, H. J. Lee, and N. B. A. Rahman, "Recent Development of Thermophotovoltaic System for Waste Heat Harvesting Application and Potential Implementation in Thermal Power Plant," *IEEE Access*, vol. 8, pp. 105156–105168, 2020, doi: 10.1109/ACCESS.2020.2999061.

[9]  N. A. Pfiester and T. E. Vandervelde, "Selective emitters for thermophotovoltaic applications," *Phys. Status Solidi Appl. Mater. Sci.*, vol. 214, no. 1, pp. 1–24, 2017, doi: 10.1002/pssa.201600410.

[10] M. Oh, E. Carlson, and T. E. Vandervelde, "Design of an All-Semiconductor Selective Metamaterial Emitter in the Mid-IR Regime with Larger Feature Sizes for Thermophotovoltaic Energy Conversion Applications," *J. Electron. Mater.*, vol. 49, no. 6, pp. 3504–3511, Jun. 2020, doi: 10.1007/s11664-020-07972-8.

[11] I. E. Titkov *et al.*, "Temperature - Dependent Internal Quantum Efficiency of Blue High - Brightness Light - Emitting Diodes," doi: 10.1109/JQE.2014.2359958.

[12] A. K. Baranwal *et al.*, "Thermal Degradation Analysis of Sealed Perovskite Solar Cell with Porous Carbon Electrode at 100 °C for 7000 h," *Energy Technol.*, vol. 7, no. 2, pp. 245–252, 2019, doi: doi.org/10.1002/ente.201800572.

[13] A. Datas and A. Martí, "Thermophotovoltaic energy in space applications: Review and future potential," *Sol. Energy Mater. Sol. Cells*, vol. 161, no. December 2016, pp. 285–296, 2017, doi: 10.1016/j.solmat.2016.12.007.

[14] E. Hecht and A. R. Ganesan, *Optics*, 5th ed. Pearson, 2020.

[15]     A. Vijayakumar and S. Bhattacharya, *Design and Fabrication of Diffractive Optical Elements with MATLAB*. SPIE Pres, 2017.

[16]     C. M. Watts, X. Liu, and W. J. Padilla, "Metamaterial electromagnetic wave absorbers," *Adv. Mater.*, vol. 24, no. 23, 2012, doi: 10.1002/adma.201200674.

[17]     W. Stork, N. Streibl, H. Haidner, and P. Kipfer, "Artificial distributed-index media fabricated by zero-order gratings," *Opt. Lett.*, vol. 16, no. 24, p. 1921, Dec. 1991, doi: 10.1364/OL.16.001921.

[18]     A. Zhan, S. Colburn, R. Trivedi, T. K. Fryett, C. M. Dodson, and A. Majumdar, "Low-Contrast Dielectric Metasurface Optics," *ACS Photonics*, vol. 3, no. 2, pp. 209–214, Feb. 2016, doi: 10.1021/acsphotonics.5b00660.

[19]     H. Wakatsuchi, D. F. Sievenpiper, and C. Christopoulos, "Designing flexible and versatile metamaterial absorbers," *IEEE Electromagn. Compat. Mag.*, vol. 5, no. 2, pp. 76–82, 2016, doi: 10.1109/MEMC.0.7543955.

[20]     J. Valentine *et al.*, "Three-dimensional optical metamaterial with a negative refractive index," *Nature*, vol. 455, no. 7211, pp. 376–379, Sep. 2008, doi: 10.1038/nature07247.

[21]     X. Liu, C. Lan, B. Li, Q. Zhao, and J. Zhou, "Dual band metamaterial perfect absorber based on artificial dielectric 'molecules,'" *Sci. Rep.*, vol. 6, pp. 2–7, 2016, doi: 10.1038/srep28906.

[22]     E. Arbabi, S. M. Kamali, A. Arbabi, and A. Faraon, "Full-Stokes Imaging Polarimetry Using Dielectric Metasurfaces," *ACS Photonics*, vol. 5, no. 8, pp. 3132–3140, Aug. 2018, doi: 10.1021/acsphotonics.8b00362.

[23]     G. Liu *et al.*, "Near-unity, full-spectrum, nanoscale solar absorbers and near-perfect blackbody emitters," *Sol. Energy Mater. Sol. Cells*, vol. 190, no. 2–3, pp. 20–29, Feb. 2019, doi: 10.1016/j.solmat.2018.10.011.

[24]     S. M. Rudolph, W. S. Wall, S. K. Hong, and K. L. Morgan, "Broadband switching nonlinear metamaterial," in *2014 IEEE Antennas and Propagation Society International Symposium (APSURSI)*, Jul. 2014, pp. 1228–1229, doi: 10.1109/APS.2014.6904941.

[25]     P. Genevet, F. Capasso, F. Aieta, M. Khorasaninejad, and R. Devlin, "Recent advances in planar optics: from plasmonic to dielectric metasurfaces," *Optica*, vol. 4, no. 1, p. 139, 2017, doi: 10.1364/optica.4.000139.

[26]     S. Zhang, W. Fan, K. J. Malloy, S. R. J. Brueck, N. C. Panoiu, and R. M. Osgood, "Near-infrared double negative metamaterials," *Opt. Express*, vol. 13, no. 13, p. 4922, 2005, doi: 10.1364/OPEX.13.004922.

[27]     C. Shemelya *et al.*, "Stable high temperature metamaterial emitters for thermophotovoltaic applications," *Appl. Phys. Lett.*, vol. 104, no. 20, 2014, doi: 10.1063/1.4878849.

[28]     S. Jahani and Z. Jacob, "All-dielectric metamaterials," *Nat. Nanotechnol.*, vol. 11, no. 1, pp. 23–36, 2016, doi: 10.1038/nnano.2015.304.

[29]     K. Fan, J. Y. Suen, X. Liu, and W. J. Padilla, "All-dielectric metasurface absorbers for uncooled terahertz imaging," *Optica*, vol. 4, no. 6, p. 601, 2017, doi: 10.1364/optica.4.000601.

[30]     P. Nagpal, S. E. Han, A. Stein, and D. J. Norris, "Efficient low-temperature thermophotovoltaic emitters from metallic photonic crystals," *Nano Lett.*, vol. 8, no. 10, pp. 3238–3243, 2008, doi:

10.1021/nl801571z.

[31] V. Rinnerbauer *et al.*, "Recent developments in high-temperature photonic crystals for energy conversion," *Energy Environ. Sci.*, vol. 5, no. 10, pp. 8815–8823, 2012, doi: 10.1039/c2ee22731b.

[32] T. F. Krauss and R. M. D. La Rue, "Photonic crystals in the optical regime — past, present and future," *Prog. Quantum Electron.*, vol. 23, no. 2, pp. 51–96, Mar. 1999, doi: 10.1016/S0079-6727(99)00004-X.

[33] M. Oh and T. E. Vandervelde, "Wafer-area selective emitters based on optical interference," *MRS Adv.*, Jun. 2021, doi: 10.1557/s43580-021-00076-3.

[34] H. Fujiwara and R. W. Collins, *Spectroscopic Ellipsometry for Photovoltaics Volume 1: Fundamental Principles and Solar Cell Characterization*, vol. 212. Cham: Springer, 2018.

[35] T. Inoue, M. De Zoysa, T. Asano, and S. Noda, "Realization of narrowband thermal emission with optical nanostructures," *Optica*, vol. 2, no. 1, p. 27, 2015, doi: 10.1364/optica.2.000027.

[36] V. Sapritsky and A. Prokhorov, *Blackbody Radiometry Volume 1: Fundamentals*. Springer, 2020.

[37] S. M.Stewart and R. B. Johnson, *Blackbody Radiation: A History of Thermal Radiation Computational Aids and Numerical Methods*. CRC Press, 2017.

[38] "Planck's radiation law, Encyclopaedia Britannica." https://www.britannica.com/science/Plancks-radiation-law (accessed Sep. 21, 2022).

[39] W. M. S. J. L. Jeff Sanny, *University Physics Volume 3*. OpenStax, 2022.

[40] L. J. Klein, H. F. Hamann, Y.-Y. Au, and S. Ingvarsson, "Coherence properties of infrared thermal emission from heated metallic nanowires," *Appl. Phys. Lett.*, vol. 92, no. 21, p. 213102, May 2008, doi: 10.1063/1.2936835.

[41] S. Ingvarsson, L. Klein, Y.-Y. Au, J. A. Lacey, and H. F. Hamann, "Enhanced thermal emission from individual antenna-like nanoheaters," *Opt. Express*, vol. 15, no. 18, p. 11249, 2007, doi: 10.1364/OE.15.011249.

[42] D. J. Griffiths, *Introduction to electrodynamics*, 4th ed. Pearson, 2016.

[43] D. Fleisch, *A student's guide to Maxwell's equations*. Cambridge University Press, 2008.

[44] M. Oh, "Study of Cu/SiO2/Cu Metamaterials: Design, Simulation, Fabrication, Testing, and Optical Applications," Rose-Hulman Institute of Technology, 2017.

[45] A. Yariv and P. Yeh, *Optical Waves in Crystals: Propagation and Control of Laser Radiation.* Wiley & Sons, Inc., 1984.

[46] J. Peatross and M. Ware, *Physics of Light and Optics*. Brigham Young University, 2015.

[47] P. Moon and D. E. Spencer, "The meaning of the vector Laplacian," *J. Franklin Inst.*, vol. 256, no. 6, pp. 551–558, Dec. 1953, doi: 10.1016/0016-0032(53)91160-0.

[48] E. Kreyszig, *Advanced Engineering Mathematics*, 9th ed. John Wiley & Sons, Inc., 2006.

[49] J. D. Kraus and K. R. Carver, *Electromagnetics*, 2nd ed. McGraw-Hilll, 1973.

[50] M. F. Iskander, *Electromagneitc Fields and Waves*, 2nd ed. Waveland Press Inc., 2013.

[51] J. Krupka, J. Breeze, A. Centeno, N. Alford, T. Claussen, and L. Jensen, "Measurements of Permittivity, Dielectric Loss Tangent, and Resistivity of Float-Zone Silicon at Microwave Frequencies," *IEEE Trans. Microw. Theory Tech.*, vol. 54, no. 11, pp. 3995–4001, Nov. 2006, doi: 10.1109/TMTT.2006.883655.

[52] C. A. Balanis, *Advanced Engineering Electromagnetics*, 2nd ed. John Wiley & Sons, Inc., 2012.

[53] S. J. Orfanidis, *Electromagnetic Waves and Antennas*. 2016.

[54] F. L. Pedrotti, L. M. Pedrotti, and L. S. Pedrotti, *Introduction to Optics*, 3rd ed. Addison-Wesley, 2006.

[55] H. Wang *et al.*, "Extended Drude Model for Intraband-Transition-Induced Optical Nonlinearity," *Phys. Rev. Appl.*, vol. 11, no. 6, p. 064062, 2019, doi: 10.1103/PhysRevApplied.11.064062.

[56] H. Y. Li *et al.*, "Analysis of the Drude model in metallic films," *Appl. Opt.*, vol. 40, no. 34, p. 6307, 2001, doi: 10.1364/AO.40.006307.

[57] M. A. Ordal, R. J. Bell, R. W. Alexander, L. L. Long, and M. R. Querry, "Optical properties of fourteen metals in the infrared and far infrared: Al, Co, Cu, Au, Fe, Pb, Mo, Ni, Pd, Pt, Ag, Ti, V, and W," *Appl. Opt.*, vol. 24, no. 24, p. 4493, 1985, doi: 10.1364/AO.24.004493.

[58] A. Moridi, H. Ruan, L. C. Zhang, and M. Liu, "Residual stresses in thin film systems: Effects of lattice mismatch, thermal mismatch and interface dislocations," *Int. J. Solids Struct.*, vol. 50, no. 22–23, pp. 3562–3569, Oct. 2013, doi: 10.1016/j.ijsolstr.2013.06.022.

[59] D. M. Broadway *et al.*, "Achieving zero stress in iridium, chromium, and nickel thin films," *EUV X-ray Opt. Synerg. between Lab. Sp. IV*, vol. 9510, no. May 2015, p. 95100E, 2015, doi: 10.1117/12.2180641.

[60] A. Carter and S. Elhadj, "Modulus of Elasticity and Thermal Expansion Coefficient of ITO Film," Livermore, CA (United States), Jun. 2016. doi: 10.2172/1325877.

[61] C. Iliescu *et al.*, "Residual stress in thin films PECVD depositions: A review," *J. Optoelectron. Adv. Mater.*, vol. 13, no. 4, pp. 387–394, 2011.

[62] Y. S. Touloukian, *Thermal Expansion: Metallic Elements and Alloys (Thermophysical Properties of Matter, Vol. 12)*. IFI/Plenum, 1975.

[63] G. K. White and M. L. Minges, "Thermophysical properties of some key solids: An update," *Int. J. Thermophys.*, vol. 18, no. 5, pp. 1269–1327, Sep. 1997, doi: 10.1007/BF02575261.

[64] M. Oh and T. Vandervelde, "Bridging the gaps between different sign conventions of Fresnel reflection coefficients towards a universal form," in *Midwest Symposium on Circuits and Systems*, 2020, vol. 2020August, pp. 707–713, doi: 10.1109/MWSCAS48704.2020.9184440.

[65] C. Wu *et al.*, "Spectrally selective chiral silicon metasurfaces based on infrared Fano resonances," *Nat. Commun.*, vol. 5, no. May, pp. 1–9, 2014, doi: 10.1038/ncomms4892.

[66] D. Woolf, J. Hensley, J. G. Cederberg, D. T. Bethke, A. D. Grine, and E. A. Shaner, "Heterogeneous metasurface for high temperature selective emission," *Appl. Phys. Lett.*, vol. 105, no. 8, pp. 1–5, 2014, doi: 10.1063/1.4893742.

[67] M. Oh, E. S. Carlson, and T. E. Vandervelde, "Coupled resonance via localized surface plasmon

[68] J. C. G. de Sande, C. N. Afonso, J. L. Escudero, R. Serna, F. Catalina, and E. Bernabéu, "Optical properties of laser-deposited a-Ge films: a comparison with sputtered and e-beam-deposited films," *Appl. Opt.*, vol. 31, no. 28, p. 6133, Oct. 1992, doi: 10.1364/AO.31.006133.

polaritons in Iridium-based refractory metamaterials," *Comput. Mater. Sci.*, vol. 197, no. May, p. 110598, 2021, doi: 10.1016/j.commatsci.2021.110598.

[69] N. A. Pfiester, K. A. Grossklaus, M. A. Stevens, and T. E. Vandervelde, "Effect of microstructure on the optical properties of sputtered iridium thin films," *Opt. Mater. Express*, vol. 10, no. 4, p. 1120, 2020, doi: 10.1364/ome.390421.

[70] B. N. Chapman, "Thin-film adhesion," *J. Vac. Sci. Technol.*, vol. 11, no. 1, pp. 106–113, Jan. 1974, doi: 10.1116/1.1318537.

[71] N. Laegreid and G. K. Wehner, "Sputtering Yields of Metals for Ar + and Ne + Ions with Energies from 50 to 600 ev," *J. Appl. Phys.*, vol. 32, no. 3, pp. 365–369, Mar. 1961, doi: 10.1063/1.1736012.

[72] H. Klumbies *et al.*, "Thickness dependent barrier performance of permeation barriers made from atomic layer deposited alumina for organic devices," *Org. Electron.*, vol. 17, pp. 138–143, Feb. 2015, doi: 10.1016/j.orgel.2014.12.003.

[73] B. Cui, "Microfabrication and thin film technology (Chapter 9 Thin film deposition III, NE 343 Lecture Note)," *University of Waterloo*. https://ece.uwaterloo.ca/~bcui/NE_343.html (accessed Dec. 31, 2022).

[74] M. Oh, J. Mcelearney, and T. E. Vandervelde, "High-temperature electrical and optical properties of sputtered iridium at wavelengths 300 nm to 15 µm," *Opt. Mater. Express*, 2023.

[75] X. Yan, T. Ai, X. Su, Z. Wang, G. Sun, and P. Zhao, "Synthesis and Thermal Decomposition Mechanism Study of a Novel Iridium Precursor," *MATEC Web Conf.*, vol. 43, p. 01002, Feb. 2016, doi: 10.1051/matecconf/20164301002.

[76] N. J. Suliali *et al.*, "Ti thin films deposited by high-power impulse magnetron sputtering in an industrial system: Process parameters for a low surface roughness," *Vacuum*, vol. 195, p. 110698, Jan. 2022, doi: 10.1016/j.vacuum.2021.110698.

[77] H. H. Yu and J. W. Hutchinson, "Influence of substrate compliance on buckling delamination of thin films," *Int. J. Fract.*, vol. 113, no. 1, pp. 39–55, 2002, doi: https://doi.org/10.1023/A:1013790232359.

[78] S. Emami, J. Martins, L. Andrade, J. Mendes, and A. Mendes, "Low temperature hermetic laser-assisted glass frit encapsulation of soda-lime glass substrates," *Opt. Lasers Eng.*, vol. 96, no. May, pp. 107–116, Sep. 2017, doi: 10.1016/j.optlaseng.2017.04.006.

[79] Crystran, "Quartz Crystal (SiO2)." https://www.crystran.co.uk/optical-materials/quartz-crystal-sio2.

[80] B. Deng, Y. Shi, and F. Yuan, "Investigation on the structural origin of low thermal expansion coefficient of fused silica," *Materialia*, vol. 12, no. April, p. 100752, Aug. 2020, doi: 10.1016/j.mtla.2020.100752.

[81] S. Kohli, D. Niles, C. D. Rithner, and P. K. Dorhout, "Structural and optical properties of Iridium

films annealed in air," *Adv. X-ray Anal.*, vol. 45, no. c, pp. 352–358, 2002.

[82] V. Sanphuang, N. Ghalichechian, N. K. Nahar, and J. L. Volakis, "Reconfigurable THz Filters Using Phase-Change Material and Integrated Heater," *IEEE Trans. Terahertz Sci. Technol.*, vol. 6, no. 4, pp. 583–591, 2016, doi: 10.1109/TTHZ.2016.2560175.

[83] T. Cao, C. Wei, R. E. Simpson, L. Zhang, and M. J. Cryan, "Rapid phase transition of a phase-change metamaterial perfect absorber," *Opt. Mater. Express*, vol. 3, no. 8, p. 1101, Aug. 2013, doi: 10.1364/OME.3.001101.

[84] S. G. Carrillo *et al.*, "A Nonvolatile Phase-Change Metamaterial Color Display," *Adv. Opt. Mater.*, vol. 7, no. 18, p. 1801782, Sep. 2019, doi: 10.1002/adom.201801782.

[85] T. E. Tiwald, "Measurement of free carriers in silicon and silicon carbide using infrared ellipsometry," University of Nebraska - Lincoln, 1999.

[86] M. Oh, E. S. Carlson, and T. E. Vandervelde, "Localized surface plasmon resonance in refractory metamaterials," in *Advanced Fabrication Technologies for Micro/Nano Optics and Photonics XIV*, Mar. 2021, vol. 11696, p. 40, doi: 10.1117/12.2581294.

[87] W. R. Chan *et al.*, "Toward high-energy-density, high-efficiency, and moderate-temperature chip-scale thermophotovoltaics," *Proc. Natl. Acad. Sci. U. S. A.*, vol. 110, no. 14, pp. 5309–5314, 2013, doi: 10.1073/pnas.1301004110.

[88] M. Oh, K. Grossklaus, and T. E. Vandervelde, "Large-area 1D selective emitter for thermophotovoltaic applications in the mid-infrared," *J. Vac. Sci. Technol. B*, vol. 41, no. 1, p. 012203, Jan. 2023, doi: 10.1116/6.0002198.

[89] M. Oh, K. Grossklaus, and T. E. Vandervelde, "1D Thermophotovoltaic Emitter: Performance Comparison in N2 Ambient and Air," in *2023 7th IEEE Electron Devices Technology & Manufacturing Conference (EDTM)*, Mar. 2023, pp. 1–3, doi: 10.1109/EDTM55494.2023.10103125.

[90] F. T. Maremi, N. Lee, G. Choi, T. Kim, and H. H. Cho, "Design of multilayer ring emitter based on metamaterial for thermophotovoltaic applications," *Energies*, vol. 11, no. 9, pp. 1–9, 2018, doi: 10.3390/en11092299.

[91] V. K. Bhat, M. Pattabiraman, K. N. Bhat, and A. Subrahmanyam, "The growth of ultrathin oxides of silicon by low temperature wet oxidation technique," *Mater. Res. Bull.*, vol. 34, no. 10–11, pp. 1797–1803, Jul. 1999, doi: 10.1016/S0025-5408(99)00158-0.

[92] L. M. Fraas, J. E. Avery, and Han Xiang Huang, "Thermophotovoltaics: heat and electric power from low bandgap 'solar' cells around gas fired radiant tube burners," in *Conference Record of the Twenty-Ninth IEEE Photovoltaic Specialists Conference, 2002.*, 2002, no. June, pp. 1553–1556, doi: 10.1109/PVSC.2002.1190909.

[93] T. Bauer, R. P. Forbes, and N. Pearsall, "The Potential of Thermophotovoltaic Heat Recovery for the Glass Industry," in *AIP Conference Proceedings*, 2003, vol. 653, pp. 101–110, doi: 10.1063/1.1539368.

[94] T. Bauer, I. Forbes, and N. Pearsall, "The potential of thermophotovoltaic heat recovery for the UK industry," *Int. J. Ambient Energy*, vol. 25, no. 1, pp. 19–25, 2004, doi: 10.1080/01430750.2004.9674933.

[95] L. M. Fraas, "Economic potential for thermophotovoltaic electric power generation in the steel industry," *2014 IEEE 40th Photovolt. Spec. Conf. PVSC 2014*, pp. 766–770, 2014, doi: 10.1109/PVSC.2014.6925031.

[96] G. Bianchi *et al.*, "Estimating the waste heat recovery in the European Union Industry," *Energy, Ecol. Environ.*, vol. 4, no. 5, pp. 211–221, Oct. 2019, doi: 10.1007/s40974-019-00132-7.

[97] US Department of Energy, *Waste Heat Recovery: Technology and Opportunities in U.S. industry*. US Department of Energy, 2008.

[98] J. Walker, *Fundamentals of Physics*, 10th ed. Wiley, 2018.

[99] A. Fiorino, "Near-Field Radiative Thermal Transport and Energy Conversion by," 2018.

[100] G. R. Bhatt *et al.*, "Integrated near-field thermo-photovoltaics for heat recycling," *Nat. Commun.*, vol. 11, no. 1, pp. 1–7, 2020, doi: 10.1038/s41467-020-16197-6.

[101] J. Holmes and C. Balanis, "Refraction of a uniform plane wave incident on a plane boundary between two lossy media," *IEEE Trans. Antennas Propag.*, vol. 26, no. 5, pp. 738–741, Sep. 1978, doi: 10.1109/TAP.1978.1141924.

[102] J. R. Reitz, F. J. Milford, and R. W. Christy, *Foundations of electromagnetic theory, Ch 18-4*, 4th ed. Addison-Wesley Publishing Company, 2009.

[103] M. Oh, "Complex Unit Vector for the Complex Wave Constant k in a Lossy Medium [Letters to the Editor]," *IEEE Antennas Propag. Mag.*, vol. 63, no. 1, pp. 117–118, 2021, doi: 10.1109/MAP.2020.3039796.

[104] M. Oh, "Polarization of Nonuniform Plane Waves: Transmission from a Lossless Medium into a Lossy Medium," *TechRxiv*, 2023, doi: https://doi.org/10.36227/techrxiv.21565683.

[105] P. B. Johnson and R. W. Christy, "Optical Constants of the Noble Metals," *Phys. Rev. B*, vol. 6, no. 12, pp. 4370–4379, Dec. 1972, doi: 10.1103/PhysRevB.6.4370.

[106] J. Als-Nielsen and D. McMorrow, *Elements of Modern X-ray Physics*, 2nd ed. Wiley, 2011.

[107] P. G. Bergmann, *Introduction to the theory of relativity*. Prentice-Hall, Inc., 1976.

[108] R. Y. Chiao, "Superluminal (but causal) propagation of wave packets in transparent media with inverted atomic populations," *Phys. Rev. A*, vol. 48, no. 1, pp. R34–R37, Jul. 1993, doi: 10.1103/PhysRevA.48.R34.

[109] D. Mugnai, A. Ranfagni, and R. Ruggeri, "Observation of Superluminal Behaviors in Wave Propagation," *Phys. Rev. Lett.*, vol. 84, no. 21, pp. 4830–4833, May 2000, doi: 10.1103/PhysRevLett.84.4830.

[110] "Faster than a Speeding Light Wave," *American Physical Society (APS Physics)*, 2000. https://physics.aps.org/story/v5/st23.

[111] F. H. Cocks and R. Gettliffe, "Total external X-ray reflection: A novel method of surface chemical analysis," *Mater. Lett.*, vol. 3, no. 4, pp. 133–136, Mar. 1985, doi: 10.1016/0167-577X(85)90144-2.

[112] D. R. Smith, S. Schultz, P. Markoš, and C. M. Soukoulis, "Determination of effective permittivity and permeability of metamaterials from reflection and transmission coefficients," *Phys. Rev. B*,

vol. 65, no. 19, p. 195104, Apr. 2002, doi: 10.1103/PhysRevB.65.195104.

[113] X. Chen, T. M. Grzegorczyk, B.-I. Wu, J. Pacheco, and J. A. Kong, "Robust method to retrieve the constitutive effective parameters of metamaterials," *Phys. Rev. E*, vol. 70, no. 1, p. 016608, Jul. 2004, doi: 10.1103/PhysRevE.70.016608.

[114] G. Hua and D. Li, "Electron work function: A novel probe for toughness," *Phys. Chem. Chem. Phys.*, vol. 18, no. 6, pp. 4753–4759, 2016, doi: 10.1039/c5cp04873g.